\documentclass[a4paper,11pt]{article}
\usepackage{jheppub}
\usepackage[utf8]{inputenc}
\usepackage{amsfonts}
\usepackage{latexsym}
\usepackage{bbold}
\usepackage{verbatim}
\usepackage{setspace}
\usepackage{ytableau}
\usepackage[colorinlistoftodos]{todonotes}

\usepackage{slashed}

\newcommand{\beq}{\begin{equation}}
\newcommand{\eeq}{\end{equation}}
\newcommand{\bc}{}
\newcommand{\nn}{\nonumber}

\newcommand{\mF}{\mathcal{F}}
\newcommand{\mD}{\mathcal{D}}

\newcommand{\mN}{\mathcal{N}}

\newcommand{\mS}{\mathcal{S}}

\newcommand{\p}{\partial}

\newcommand{\f}{\frac}

\newcommand{\al}{\alpha}
\newcommand{\be}{\beta}
\newcommand{\ga}{\gamma}       \newcommand{\Ga}{\Gamma}
\newcommand{\de}{\delta}       \newcommand{\De}{\Delta}
\newcommand{\ep}{\epsilon}

\newcommand{\et}{\eta}
\newcommand{\te}{\theta}

\newcommand{\ka}{\kappa}
\newcommand{\la}{\lambda}      

\newcommand{\rh}{\rho}

\newcommand{\si}{\sigma}       \newcommand{\Si}{\Sigma}

\newcommand{\ph}{\phi}         \newcommand{\Ph}{\Phi}
\newcommand{\ps}{\psi}         

\newcommand{\ch}{\chi}

\newcommand{\bph}{\bar{\ph}}
\newcommand{\bPh}{\bar{\Ph}}

\newcommand{\bF}{\bar{F}}

\newcommand{\mbp}{\mathbf{p}}
\newcommand{\bps}{\bar{\psi}}

\newcommand{\lan}{\langle}
\newcommand{\ran}{\rangle}

\newcommand{\maS}{\mathcal{S}}

\newcommand{\TB}{\mathcal{T}_B}
\newcommand{\TF}{\mathcal{T}_F}

\newcommand{\mO}{\mathcal{O}}
\def\vec[#1]{\boldsymbol{#1}}
\def\vecs[#1,#2]{\boldsymbol{{#1}_{#2}}}

\title{Scattering Amplitudes in $\mathcal{N} = 3$ Supersymmetric $SU(N)$ Chern-Simons-Matter Theory at Large $N$}

\author[a]{Karthik Inbasekar,}
\author[b]{Lavneet Janagal,}
\author[c]{and Ashish Shukla}

\affiliation[a]{Department of Physics, Ben-Gurion University of the Negev, Beer-Sheva 84105, Israel}
\affiliation[b]{School of Physics, Korea Institute for Advanced Studies, Seoul 02455, South Korea}
\affiliation[c]{Department of Physics \& Astronomy, University of Victoria \\ 3800 Finnerty Road, Victoria, British Columbia  V8P 5C2, Canada}

\emailAdd{inbaseka@post.bgu.ac.il}
\emailAdd{lavneet@kias.re.kr}
\emailAdd{ashish@uvic.ca}

\abstract{
We study $\mN=3$ supersymmetric Chern-Simons-matter theory coupled to matter in the fundamental representation of $SU(N)$.  In the \rq t Hooft large $N$ limit, we compute the exact $2\to 2$ scattering amplitudes of the fundamental scalar superfields to all orders in the \rq t Hooft coupling $\la$. 
Our computations are presented in $\mN=1$ superspace and make significant use of the residual $SO(2)_R$ symmetry in order to solve for the exact four-point correlator of the scalar superfields. By taking the on-shell limit, we are able to extract the exact $2\to 2$ scattering amplitudes of bosons/fermions in the symmetric, anti-symmetric and adjoint channels of scattering. We find that the scattering amplitude of the $\mathcal{N}=3$ theory in the planar limit is tree-level exact to all orders in the \rq t Hooft coupling $\la$. The result is consistent with the conjectured bosonization duality and is expected to have enhanced symmetry structures such as dual superconformal symmetry and Yangian symmetry.}

\arxivnumber{2001.02363}

\begin{document}
\maketitle

\newpage
\section{Introduction}
\label{intro}
Pure Chern-Simons theories in the absence of matter are topological and have no propagating degrees of freedom. The source free action\footnote{Here, $\ka$ is the level of the Chern-Simons action defined in dimensional regularization. Gauge invariance of the action \eqref{csp} requires that $\ka$ take integer values. The gauge field transforms in the adjoint representation of the gauge group $G$, which we will assume to be $SU(N)$ for the purpose of this paper. } 
\beq\label{csp}
S_{CS}=\f{\ka}{4\pi} \int d^3 x \, \text{Tr} \left(A \wedge dA+ \f{2}{3} A \wedge A \wedge A \right),
\eeq
where the trace runs over some compact gauge group $G$ leads to the equation of motion $F_{\mu\nu}=0$. The solutions are flat connections $A_\mu=g^{-1} \p_\mu g$, for $g \in G$. 
The only physical observables in this theory are n-point correlation functions of gauge invariant operators such as a Wilson loop $l$ in representation $R$, 
\beq
W_R(l) = \text{Tr} \left( \ \mathcal{P} \int \exp\left(\oint_l A\right)\right).
\eeq
When the theory is on a 3-sphere, the n-point correlation function formed from the product of Wilson loops around $n$ links is proportional to Knot polynomials \cite{Witten:1988hf}. However, addition of matter to the theory modifies the topological character dramatically and makes the theory highly relevant to several physical phenomena.  For instance, in the $U(1)$ theory interacting with a charged scalar, the equation of motion reads as 
\beq
\ep^{\mu\nu\rh} F_{\mu\nu}= \f{2\pi}{\ka} J^\rh.
\eeq
In components, this means that interactions of the Chern-Simons gauge field with matter attaches magnetic fluxes proportional to $\f{1}{\ka}$ to the particle quanta.  When such particles undergo an adiabatic exchange the multi-particle wave function picks up the well known Aharonov-Bohm phase $\nu\propto \f{1}{\ka}$ \cite{Aharonov:1959fk}. This phase of the wave-function is responsible for fractional statistics (anyonic).\footnote{In a non-Abelian theory, when particles in representation $R_1$ and $R_2$ are exchanged, the phase is proportional to the quadratic Casimirs in their representations,
\beq\label{anp}
\nu_{R_m}= \f{1}{\ka} (C_2(R_m)-C_2(R_1)-C_2(R_2)),
\eeq
where $R_m$ is a channel in the decomposition $R_1\otimes R_2 = \sum_m R_m$. In \S \ref{skin}, we will discuss the case of $G= SU(N)$ in detail. }
The anyonic character and flux attachment features of Chern-Simons interactions are essential ingredients in effective field theories that describe excitations about the ground state in quantum Hall effect (see for instance \cite{Tong:2016kpv}). Since Chern-Simons gauge fields change the nature of the statistics of particles, it is not entirely wild to think that perhaps Chern-Simons theories interacting with matter of one kind (say bosons) may be related to Chern-Simons theories interacting with matter of another kind (say fermions).

The first hint for such dual descriptions appeared from holographic studies in $AdS/CFT$, through the observation that higher-spin gravity theories in $AdS$ \footnote{Constructed and studied by Vasiliev in pre-$AdS/CFT$ era. See \cite{Giombi:2016ejx} for a comprehensive review and more references.} have a dual description in terms of $O(N)$ vector models living on the boundary of $AdS$ \cite{Klebanov:2002ja,Sezgin:2003pt,Leigh:2003gk}. The most general Vasiliev higher-spin gravity theory in 3+1 dimensions is characterized by a phase that is restricted only by parity invariance. The parity-even model (also known as Vasiliev Type A) is dual to the regular bosonic vector model (scaling dimension $\De=1$), while the parity-odd model (also known as Vasiliev Type B) is dual to the critical fermionic vector model ($\De=1$). Under RG flow due to the $\ph^4$ operator, the free boson theory flows to the Wilson-Fisher critical point ($\De=2$), while the critical fermion theory flows to the regular fermion ($\De=2$) due to the $\psi^4$ operator. In the bulk, the corresponding higher-spin gravity theories employ $\De=2$ boundary conditions.  Of course, it is not surprising that higher-spin symmetries exist in free field theories; when the higher-spin symmetries are exact the corresponding field theories are necessarily free \cite{Maldacena:2011jn}. 

There exists a one parameter deformation of this picture that generalizes it to a large class of vector models interacting via Chern-Simons gauge fields. From the higher-spin gravity point of view, relaxing parity invariance leads to a  general set of parity breaking higher-spin theories labeled by an infinite set of  parameters. These parameters holographically map to the \rq t Hooft coupling $\la=\f{N}{\ka}$ of the Chern-Simons vector models \cite{Giombi:2011kc,Aharony:2011jz}. The higher-spin currents in the field theories obey the conservation laws to the leading order in large $N$, in other words the symmetries are weakly broken by $\mO\left(\f{1}{N}\right)$ effects. This fact constrains the three-point functions of higher-spin currents in the fermionic and bosonic theories to have identical structures and leads to a map between the different parameters in these theories \cite{Aharony:2011jz,Maldacena:2012sf,GurAri:2012is}. Thus one is led to the conjecture that the regular boson/critical boson is dual to the critical fermion/regular fermion, respectively. 

The conjectured bosonization duality has successfully passed a battery of tests in the large $N$ limit, such as thermal partition functions \cite{Aharony:2012ns,Yokoyama:2013pxa,Takimi:2013zca,Jain:2013py}, correlation functions of spin $s$ currents \cite{Aharony:2012nh,Bedhotiya:2015uga,Gur-Ari:2015pca,Geracie:2015drf,Turiaci:2018dht,Yacoby:2018yvy,Inbasekar:2019wdw,Kalloor:2019xjb}, and $S$-matrices \cite{Jain:2014nza,Dandekar:2014era,Inbasekar:2015tsa,Yokoyama:2016sbx}. As we alluded to earlier, there is significant evidence that the bosonization duality is preserved along RG flows \cite{Minwalla:2015sca,Aharony:2018pjn}. Some of these flows lead to supersymmetric fixed points where the bosonization duality \cite{Jain:2012qi,Jain:2013gza,Aharony:2019mbc,Dey:2019ihe} is related to the already well known Giveon-Kutasov duality \cite{Benini:2011mf,Park:2013wta,Aharony:2013dha}.\footnote{It is important to emphasize that the Giveon-Kutasov duality is valid for any $N$ and $\ka$.}
Recent computations have also investigated the duality in the Higgsed phase \cite{Choudhury:2018iwf,Dey:2018ykx}, finite temperature, background fields and additional flavors (see \cite{Gur-Ari:2016xff,Ghosh:2019sqf,Halder:2019foo,Jensen:2019mga}). In particular, the puzzle of matching the dimensions of monopole operators \cite{Radicevic:2015yla} has led to the conjecture of the duality for finite $N,\ka$ \cite{Aharony:2015mjs}. This conjecture has led to several important developments in the field, for instance the well known particle-vortex duality can be derived from the finite $N,\ka$ version of the bosonization duality and leads to a web of dualities that has potential applications in condensed matter physics \cite{Seiberg:2016gmd,Karch:2016sxi,Hsin:2016blu,Cordova:2017kue,Metlitski_2017,Cordova:2018qvg}.

In this paper, we study scattering amplitudes in supersymmetric Chern-Simons theories. We will motivate our study as follows. As we saw earlier, Chern-Simons gauge field interacts with matter and gives it an anyonic character. A natural question to ask is whether the anyonic character is visible in the scattering amplitudes. In $2+1$ dimensions, the particle is stuck to the plane, it does not have the freedom of the extra dimension to escape various windings, and this leads to braiding. However, we do not yet know how to directly compute amplitudes with particles that have anyonic character. In Chern-Simons-matter theories, the particle in any given channel of scattering behaves like an Aharonov-Bohm particle with a phase given by \eqref{anp}. For example, in $SU(N)$ Chern-Simons-matter theories at large $N,\ka$ the symmetric, anti-symmetric and adjoint channels (see details in \S\ref{skin}) have their phases suppressed by $1/N$ at leading order in large $N$. Hence, these are ``non-anyonic channels" and the amplitude can be computed using well known methods in QFT. However, the singlet channel exhibits a finite Aharonov-Bohm phase and is effectively anyonic in the large $N$ limit, and cannot be computed directly.  Furthermore, the amplitudes in the symmetric, anti-symmetric and adjoint channels are $\mO\left(\f{1}{N}\right)$ and hence the unitarity equations are linear.\footnote{The unitarity equations read $i(T^\dagger-T)= TT^\dagger$. If $T$ is $\mO\left( \f{1}{N}\right)$, then the RHS of the unitarity equation is $\mO\left( \f{1}{N^2}\right)$.} On the other hand, the singlet channel amplitude is $\mO(\f{N}{\ka})$, and hence in the planar limit is effectively $\mO(\la)$. Thus in addition to being anyonic in character, the singlet channel also satisfies the full non-linear unitarity equation at any given order in $\f{1}{N}$. As the singlet channel cannot be directly computed using the rules of QFT, one may attempt to use the naive crossing rule in QFT to obtain the amplitude in the singlet channel from any of the non-anyonic channels. However, this leads to a tension with unitarity. In particular, unitarity of the scattering amplitudes in Chern-Simons-matter theories requires a modification of the rules of crossing symmetry \cite{Jain:2014nza,Inbasekar:2015tsa,Yokoyama:2016sbx}. These conjectured modifications are expected to be universal features of all Chern-Simons-matter theories. Although there have been several tests of this conjecture \cite{Jain:2014nza}, there has not been a direct proof yet.

There are further under-appreciated features of amplitudes in Chern-Simons-matter theories (especially with a high degree of supersymmetry). For $\mN=2$ supersymmetric Chern-Simons-matter theories, the $2\to 2$ scattering amplitude computed to all orders in the \rq t Hooft coupling remains tree-level exact (vanishing loop corrections to all orders) \cite{Inbasekar:2015tsa}. Furthermore, the amplitude enjoys dual superconformal symmetry \cite{Inbasekar:2017sqp} and Yangian symmetry \cite{Yangian} exact to all loops. Furthermore, using recursion relations, arbitrary $n$-point tree-level amplitudes have also been constructed \cite{Inbasekar:2017ieo}. It remains to be seen if the higher point amplitudes continue to enjoy such symmetries as well. It is possible that such hidden symmetries are special features of Chern-Simons-matter theories with a high degree of supersymmetry. For instance, n-point tree amplitudes in $\mathcal{N}=6$ supersymmetric Chern-Simons-matter theory or the ABJM theory (Aharony-Bergman-Jafferis-Maldacena, \cite{Aharony:2008ug}) enjoy invariance under dual superconformal symmetry and Yangian symmetry \cite{Gang:2010gy,Bargheer:2010hn}. 

While this is sufficient motivation for us to study the scattering amplitudes in supersymmetric Chern-Simons-matter theories, there is a significant technical roadblock as we go to higher supersymmetries. The exact computations performed in \cite{Inbasekar:2015tsa} use the supersymmetric light cone gauge in $\mN=1$ superspace to construct and solve the Dyson-Schwinger equations. For $\mN\geq 3$ supersymmetry there does not exist a superspace formalism, and the supersymmetric light cone gauge in $\mN=2$ superspace is not easy to handle. In practice one can try to formulate higher supersymmetric theories in $\mN=1$ superspace, and then apply the technology established in \cite{Inbasekar:2015tsa}. 

In this paper we initiate a program to compute exact amplitudes in higher supersymmetric theories in $\mN=1$ superspace. In particular we compute $2\to 2$ scattering amplitudes in $\mN=3$ supersymmetric $SU(N)$ Chern-Simons-matter theories coupled to fundamental matter at large $N,\ka$, to all orders in the \rq t Hooft coupling $\la=\f{N}{\ka}$. It is well known that the maximal supersymmetric extension of Chern-Simons theory coupled to matter in fundamental representation of $SU(N)$ has $\mathcal{N} = 3$ supersymmetry \cite{Kao:1995gf, Kapustin:1999ha, Gaiotto:2007qi, Chang:2012kt}.\footnote{For $\mN\geq 4$, supersymmetry requires matter to transform in bi-fundamental representation.}
The $\mathcal{N} = 3$ theory consists of a fundamental and an anti-fundamental matter multiplet, and a unique triplet mass deformation for the matter multiplets \cite{Cordova:2016xhm}.  The mass deformed theory with $\mN=3$ supersymmetry was formulated in $\mN=1$ superspace in \cite{Inbasekar:2019azv}. There have been some recent studies on $\mN=3$ theories, see for instance \cite{Buchbinder:2009dc} for studies on non-renormalization theorems and UV finiteness of supergraph perturbation theory, and \cite{Lehum:2019kzb} for effective potentials and RG flows. The mass deformation preserves $\mN=3$ supersymmetry but breaks the $R$ symmetry from $SU(2)_R \to U(1)_R$.  Since the mass parameter is a central charge that appears in the $\mN=3$ supersymmetry algebra, it is protected against quantum corrections by supersymmetry. The main reason to consider the mass parameter is to employ a manifestly supersymmetric IR regulator. 

 We set up a Dyson-Schwinger equation for the exact four-point correlator of the matter multiplets in $\mN=1$ superspace. The fundamental and anti-fundamental matter multiplets transform under two inequivalent one-dimensional representations of the $SO(2)_R$ symmetry in $\mN=1$ superspace.\footnote{In $2+1$ dimensions, a theory with $\mN$ superconformal symmetry has $SO(\mN)_R$ symmetry. Formulating the theory on $\mN=1$ superspace leaves the matter multiplets manifestly invariant under $SO(\mN-1)_R$ symmetry.}
A judicious application of this symmetry allows us to organize and solve the Dyson-Schwinger equation in different ``sectors" labeled by the $SO(2)_R$ charges.  We employ the supersymmetric light cone gauge for the computations and solve for the correlation function in the kinematic regime $q_{\pm}=0$. The off-shell four-point correlation function of matter fields is defined in \S\ref{org} - \S\ref{off4pt}. Our final results for the off-shell correlation function for matter multiplets are expressed in \eqref{solutionsN3ch}, \eqref{solpp} and \eqref{solmp}. The final result for the correlator is written in $\mN=1$ superspace and therefore it is not possible to visualize the $\mN=3$ supersymmetry.
 
By taking the on-shell limit of the off-shell four-point correlator we extract the $2\to2$ scattering amplitude. We find the $2\to 2$ bosonic/fermionic amplitude to be tree-level exact to all orders in $\la$ (no loop corrections). The final result can be expressed in terms of the function \eqref{bosf}-\eqref{ferf}
\beq\label{res}
\mathcal{F}(p,q,k,m) = \f{4\pi i}{\ka} \ep_{\mu\nu\rh} \f{q^\mu (p-k)^\nu (p+k)^\rh}{(p-k)^2} +\f{8m\pi}{\ka}.
\eeq
The result is in essence identical to that of the $2\to2$ amplitude computed to all orders in $\la$ in the $\mN=2$ theory \cite{Inbasekar:2015tsa}.  What is remarkable is that this is a second example in supersymmetric Chern-Simons-matter theories where the amplitude computed to all loops has vanishing loop corrections. It is highly likely that this amplitude also enjoys dual superconformal symmetry \cite{Inbasekar:2017sqp} and Yangian symmetry \cite{Yangian} exact to all loops. It may also be possible to use BCFW recursion relations to compute arbitrary $n$-point tree-level amplitudes \cite{Inbasekar:2017ieo} in the $\mN=3$ theory.

We observe that \eqref{res} admits a smooth $m\to 0$ limit, and one recovers the full $SU(2)_R$ symmetry.  In the covariant form, the results for the symmetric, anti-symmetric and adjoint channels can be conveniently expressed in terms of Mandelstam variables and are summarized in \S\ref{resT}. In these channels (as explained in \S\ref{skin} ) the unitarity equations are linear and thus Hermiticity guarantees unitarity. The special kinematic limit $q_\pm=0$ in which the correlation functions were computed allows direct extraction of the scattering amplitudes in the symmetric, anti-symmetric and adjoint channels. In the singlet channel, the kinematic regime $q_\pm=0$ renders the exchange momentum to be space-like, which is forbidden (see for instance fig.\ \ref{chann}).  Thus it is not possible to extract the singlet channel amplitude directly from our results \cite{Jain:2014nza,Inbasekar:2015tsa}. 

Nevertheless, the conjecture of \cite{Jain:2014nza} for the singlet channel should apply directly to the $\mN=3$ theory. In \S\ref{ampsing} we merely state the conjecture without proof. The conjectured $2\to 2$ bosonic/fermionic singlet amplitude for the $\mN=3$ theory takes the form 
\begin{align}\label{ampsingf}
\mathcal{S}_{B/F}^{S}(s,\te) & = 8\pi \sqrt{s} \cos(\pi\la) \de(\te) + \, i\f{\sin(\pi\la)}{\pi\la}  \mathcal{T}_{B/F}^{naive}(s,\te),\nn\\
\mathcal{T}_{B/F}^{naive}(s,\te) &= 4\pi i \la \sqrt{s}\cot\f{\te}{2}.
\end{align}
Here the function $\mathcal{T}_{B/F}^{naive}$ is the naive analytic continuation from either of symmetric, antisymmetric or adjoint channels. While the amplitude \eqref{ampsingf} already satisfies the non-linear $\mN=2$ unitarity conditions \cite{Inbasekar:2015tsa}, it would be satisfying to derive it from a manifestly $\mN=3$ on-shell formalism. The on-shell formalism is fairly well developed for even supersymmetries \cite{Elvang:2013cua}, for odd supersymmetries our attempts to proceed as in \cite{Inbasekar:2015tsa} have not been very fruitful.\footnote{For instance, by formulating the Ward identities in $\mN=1$ superspace, it can be shown that the most general $\mN=3$ superamplitude can be written in terms of two independent functions for the mass deformed theory and a single function for the massless theory. However, the relations between the various component amplitudes are quite cumbersome. } Writing the full superamplitude in a suitable basis should also enable construction of n-point tree amplitudes and to show dual superconformal symmetry \cite{Inbasekar:2017sqp,Inbasekar:2017ieo}. We hope to return to these issues in a future work.

The paper is organized as follows. We begin with an elementary introduction to scattering kinematics in $SU(N)$ Chern-Simons-matter theories in \S\ref{skin}, followed by the definition of the mass deformed $\mN=3$ theory in \S\ref{lagr}. In \S\ref{onrepN3} we cover the on-shell representation of mass deformed $\mN=3$ theory in $\mN=1$ superspace. We discuss the mode expansions in \S\ref{fmon}, asymptotic states and $S$-matrices in \S\ref{astate} (for the $\mN=3$ theory in the $\mN=1$ language). Following this we switch gears to the massless limit in \S\ref{N3co} and study the on-shell representation of the $\mN=3$ theory in a manifestly $SU(2)_R$ language. In particular, in \S\ref{scmpa} we relate the $S$-matrices in the $SU(2)_R$ language to the $S$-matrices in the $U(1)_R$ language (\S\ref{astate}). The \S\ref{exaccomp} is the crucial technical section of the paper. In \S\ref{exacpro} we give a quick derivation of the non-renormalization of the two-point function. Following this, we define the four-point correlator in \S\ref{org}-\S\ref{off4pt}. Furthermore, in \S\ref{inteqall}, we solve the Dyson-Schwinger equations for the exact four-point correlator to all orders in \rq t Hooft coupling $\la$ in the large $N,\ka$ limit. We then discuss the on-shell limit in \S\ref{onsmatrix} and describe our results for the all loop $2\to2$ scattering amplitude in \S\ref{ampchannCov}. We conclude in \S\ref{disco} with comments and discussion. Our notations and conventions are summarized in \S\ref{conv}, the $\mN=3$ supersymmetry transformations are given in \S\ref{N3susytransf}.  Some supplementary equations are given in \S\ref{termsineff}, and \S\ref{treeamp} is a summary of the tree level superamplitude in the theory. 

\section{Background}\label{bkg}
\subsection{Scattering kinematics in $SU(N)$ Chern-Simons-matter theories}
\label{skin}
In this section we briefly review the kinematics of $2\to 2$ $S$-matrices in $SU(N)$ Chern-Simons-matter theories (for a detailed exposition see \S 2.3 - \S 2.7 of \cite{Jain:2014nza}). Let us represent the quanta as
\beq
\left.\text{Particles} \ (\bf{N}) \ \ \equiv\ytableausetup{centertableaux,boxsize=1.3em}
\begin{ytableau}
\none &
\end{ytableau}\ , \ 
\text{Anti-Particles}  \ (\overline{\mathbf{N}})  \ \equiv \ytableausetup
 {centertableaux, boxsize=1.3em}
\begin{ytableau}
 \ \ \\
 \ \ \\
 \none[\vdots]\\
 \ \
\end{ytableau} \ \ \right\} N-1 
\eeq
It follows that particle-particle scattering ($P_i(p_1)+P_j(p_2)\to P_k(p_3)+P_l(p_4)$) can be decomposed as
\beq
\ytableausetup{centertableaux,boxsize=1.3em}
\begin{ytableau}
\none &
\end{ytableau} \quad\otimes \ytableausetup{centertableaux,boxsize=1.3em}
\begin{ytableau}
\none &
\end{ytableau} \quad = \quad \ytableausetup{centertableaux,boxsize=1.3em}
\begin{ytableau}
\ \ & \ \
\end{ytableau}\quad \oplus \quad \ytableausetup{centertableaux,boxsize=1.3em}
\begin{ytableau}
\ \ &\none\\
\ \
\end{ytableau}
\eeq
Note that $P_i(p)$ stands for a particle with color index $i$ carrying three-momentum $p$. Similarly, $A^j(p)$ will stand for an anti-particle with color index $j$ and three-momentum $p$. We refer to the symmetric channel of scattering in particle-particle scattering as $U_s$ and the anti-symmetric channel as $U_a$,
\beq
S_{\text{particle-particle}} = S_{U_s} \f{(\de_i^l\de_j^k+\de_i^k\de_j^l)}{2}+ S_{U_a} \f{(\de_i^l\de_j^k-\de_i^k\de_j^l)}{2} .
\eeq
Anti-particle anti-particle scattering is related to the above by $\mathcal{CPT}$. Similarly, particle--anti-particle scattering  ($P_i(p_1)+A^j(p_2)\to P_k(p_3)+A^l(p_4)$) can be decomposed into
\beq
\ytableausetup
 {centertableaux, boxsize=1.3em}
N-1\left\{\begin{ytableau}
 \ \ \\
 \ \ \\
 \none[\vdots]\\
 \ \
\end{ytableau}\right. \ \ \otimes
\ytableausetup{centertableaux,boxsize=1.3em}
\begin{ytableau}
\none &
\end{ytableau}\quad=\quad \ytableausetup
 {centertableaux, boxsize=1.3em}
N-1\left\{\begin{ytableau}
 \ \ & \ \ \\
 \ \ \\
 \none[\vdots]\\
 \ \
\end{ytableau}\right.\oplus \bf{1}
\eeq
We refer to the adjoint channel of scattering as $T$-Channel and the singlet channel as $S$-channel,
\beq
S_{\text{particle--anti-particle}} = S_T\left(\de_i^l \de_k^j -\f{\de_i^j \de_k^l}{N}\right)+S_S \f{\de_i^j \de_k^l}{N}.
\eeq
Writing $S = \mathbb{1} + i T$, to the leading order in large $N$ limit the amplitudes have the behavior
\beq
T_{U_s}\sim \mO\left(\f{1}{N}\right) \ ,  \ T_{U_a}\sim \mO\left(\f{1}{N}\right) \ , \ T_{T}\sim \mO\left(\f{1}{N}\right) \ , \ T_{S} \sim \mO\left(1\right).
\eeq
It follows from the above that the unitarity equation
$$ i (T^\dagger-T) = T T^\dagger$$
is a linear constraint for the symmetric, anti-symmetric and adjoint channels of scattering, while it is non-linear for the singlet channel. 

It is well known that the Chern-Simons gauge field attaches magnetic fluxes to the interacting matter particles. The $2\to 2$ scattering in Chern-Simons-matter theories can be viewed as Aharonov-Bohm scattering of quanta off a magnetic flux tube. Scattering of quanta in the representation $R_1$ and $R_2$ can be tensor decomposed into several channels of scattering
\beq
R_1 \otimes R_2 = \sum_m R_m,
\eeq
where $m$ refers to the channels of scattering.\footnote{For instance, when two fundamentals scatter, as explained above using Young tableauxs, the amplitude can be decomposed into symmetric and anti-symmetric channels. Similarly the scattering of a fundamental and an anti-fundamental can be decomposed into adjoint and singlet channels.}  Scattering of particles in the $m$th channel of scattering is equivalent to Aharonov-Bohm scattering of a $U(1)$ particle with the phase  \cite{Jain:2014nza}
\beq
2\pi \nu_m = \f{4\pi}{\ka} T_1^a T_2^a = \f{2\pi}{\ka} \left(C_2(R_m)- C_2(R_1)- C_2(R_2) \right),
\eeq
where $C_2(R)$ is the quadratic Casimir in representation $R$. The quadratic Casimirs take the following form 
\begin{align}
&C_2(\text{fundamental})=C_2(\text{anti-fundamental})= \f{N^2-1}{2N} ,\nn\\
&C_2\left(\text{symmetric}\right)=\f{N(N+1)-2}{N}\ , \ C_2\left(\text{anti-symmetric}\right)=\f{N(N-1)-2}{N},\nn\\
&C_2(\text{adjoint})= N\ , \  C_2(\text{singlet})=0.
\end{align}
In the limit $N\to \infty, \ka\to \infty, $ with the 't Hooft parameter $\la=\f{N}{\ka}$ held fixed, the anyonic phases take the form
\beq
\nu_{\text{symmetric}}\sim \nu_{\text{anti-symmetric}} \sim \nu_{\text{adjoint}}\sim \mO\left(\f{1}{N}\right) \ , \ \nu_{\text{singlet}} \sim \la.
\eeq
Thus, in the planar limit the anyonic effects in the symmetric, anti-symmetric and adjoint channels of scattering are suppressed to the leading order, and therefore it is reasonable to expect that the standard methods of quantum field theory are insensitive to the anyonic effects in these channels at the leading order. Hence, we collectively refer to the symmetric, anti-symmetric and adjoint channels of scattering as the ``non-anyonic channels." However, for the singlet channel, the scattering is effectively anyonic due to the non-vanishing Aharonov-Bohm phase, and one expects the standard methods of field theory to be in tension with anyonic statistics. 

Treating the scattering amplitude as an analytic function in the Mandelstam plane and performing a naive analytic continuation from any of the ``non-anyonic channels" to the ``anyonic channel" leads to conflicts with unitarity \cite{Jain:2014nza,Inbasekar:2015tsa}. The general form of the amplitude in the anyonic channel for $2\to 2$ scattering in any Chern-Simons-matter theory was conjectured in \cite{Jain:2014nza} to have the general structure
\beq\label{conj}
S(s,\te)= 8 i \sqrt{s} \cos(\pi\nu) \de(\te) + i\, \f{\sin(\pi\nu)}{\pi\nu} T^{naive}(s,\te) .
\eeq
The phase modification in the forward scattering amplitude is due to the constructive interference of the incoming wave packets, while the modified crossing factor is expected to arise from the analytic continuation of the ratio of Wilson lines between the channel of crossing to the singlet channel. The conjecture is consistent with unitarity, strong-weak bosonization duality, non-relativistic limit, and has been verified explicitly in Chern-Simons theories coupled to vector bosons/fermions \cite{Jain:2014nza}, and in $\mN=1,2$ supersymmetric Chern-Simons-matter theories \cite{Inbasekar:2015tsa}. However, unfortunately, no formal derivation of \eqref{conj} has been possible so far. 

In the rest of the paper we explicitly compute the $2\to2$ amplitudes for the $\mN=3$ theory in the \lq\lq non-anyonic" channels of scattering. Unitarity in these channels follows from Hermiticity. For the singlet channel of scattering we leave the verification of unitarity of the scattering amplitude using a manifestly $\mN=3$ on-shell formalism for a future work.

\subsection{$\mN=3$ supersymmetric Chern-Simons-matter theory in $\mN=1$ superspace}\label{lagr}
The maximal supersymmetric extension of Chern-Simons theory with fundamental matter is $\mathcal{N}=3$ \cite{Kapustin:1999ha,Kao:1995gf,Gaiotto:2007qi,Chang:2012kt}. The theory consists of a pair of fundamental and anti-fundamental chiral multiplets $(\Phi^+_i,\bar{\Phi}^{-i})$ coupled to a $SU(N)$ gauge superfield $\Ga_\al^a (T^a)_i^{\ j}$.
\footnote{The superfields $(\Phi^+_i,\Phi^-_i)$ map to the $\mathcal{N}=2$ chiral multiplets $(Q_i,\bar{\tilde{Q}}_i)$ in the  notation of \cite{Gaiotto:2007qi}, See \S \ref{conv} for our notations and conventions. } In $\mathcal{N}=1$ superspace the Euclidean action takes the form \cite{Inbasekar:2019azv} 
\begin{align}
\label{N3eucS}
 \mS^E_{\mN=3}=-\int & d^3x d^2\te\biggl[ \f{\ka}{4\pi}Tr\biggl(-\f{1}{4}D^\al \Ga^{\be} D_\be
\Ga_{\al}+\f{i}{6} D^\al \Ga^\be
\{\Ga_\al,\Ga_\be\}+\f{1}{24} \{\Ga^\al,\Ga^\be\}\{\Ga_\al,\Ga_\be\}\biggr) \nn\\
&-\f{1}{2}(D^\al\bar{\Ph}^++i \bar{\Ph}^+ \Ga^\al)(D_\al\Ph^+-i \Ga_\al
\Ph^+)-\f{1}{2}(D^\al\bar{\Ph}^-+i \bar{\Ph}^- \Ga^\al)(D_\al\Ph^--i \Ga_\al \Ph^-)\nn\\
&- \f{\pi}{\ka} \left(\bar\Phi^{+}\Phi^{+}\right)
\left(\bar\Phi^{+}\Phi^{+}\right)-\f{\pi}{\ka}
\left(\bar\Phi^{-}\Phi^{-}\right) \left(\bar\Phi^{-}\Phi^{-}\right)
+\f{4\pi}{\ka}\left(\bar\Phi^{+}\Phi^{+}\right) \left(\bar\Phi^{-}\Phi^{-}\right)\nn\\
& +\f{2\pi}{\ka}
\left(\bar\Phi^{+}\Phi^{-}\right) \left(\bar\Phi^{-}\Phi^{+}\right)-(m_0 \bPh^+\Ph^+ - m_0 \bPh^-\Ph^-)\biggr].
\end{align}
In the above, $\Phi^{\pm}$ are scalar superfields and $\Gamma^\al$ is a gauge superfield with the definitions
\begin{align}
 \Ph^\pm&= \ph^\pm +\te \ps^\pm -\te^2 F,\nn\\
 \Gamma^\al &=\ch^\al -\te^\al B +i\te^\be A_\be^{\ \al}-\te^2(2\la^\al -i \p^{\al\be} \ch_\be).
\end{align}
For the notation in $\mN=1$ superspace we follow the conventions stated in Appendix A of \cite{Inbasekar:2015tsa}. Since we formulate the $\mN=3$ theory in $\mN=1$ superspace, the $SO(2)_R$ subgroup of the full $SO(3)_R$ $R$-symmetry is manifest. Thus $\Phi^+$ and $\Ph^-$ transform under the two inequivalent one-dimensional representations of the residual $SO(2)_R$ symmetry. 

It is convenient to assign the $SO(2)_R$ charges $(+\f{1}{2},-\f{1}{2})$ for the superfields $(\Ph^+_i,\Phi^-_i)$ respectively. It follows that the complex conjugate fields $(\bar{\Ph}^{+i},\bar{\Ph}^{-i})$ have $SO(2)_R$ charges $(-\f{1}{2},+\f{1}{2})$ respectively. The mass deformation for the $\mN=3$ theory is unique and transforms as a triplet under the $R$-symmetry group \cite{Cordova:2016xhm}. Hence, it breaks the $R$-symmetry group from $SU(2)\to U(1)$. This becomes transparent in the component form written in the Wess-Zumino gauge \cite{Inbasekar:2019azv},
\begin{align}\label{N3WZE}
S_{\mN=3}^E= \int d^3x &\biggl[\text{Tr}\left(\f{i\ka}{4\pi}\ep^{\mu\nu\rh}\left(A_\mu\p_\nu
A_\rh-\f{2i}{3}A_\mu A_\nu A_\rh\right)\right)\nn\\
& - i
\bps^A\slashed{\mD}\ps_A-m_0\bps^A(\si^3)_A^{\
B}\ps_B+\mD^\mu\bph_A\mD_\mu\ph^A+m^2_0\bph_A\ph^A\nn\\
&+\f{4\pi^2}{\ka^2} (\bph_A\ph^B)(\bph_B\ph^C)(\bph_C\ph^A)-\f{4\pi}{\ka} 
(\bph_A\ph^B)(\bps^A\ps_B)-\f{2\pi}{\ka}
(\bps^A\ph_B)(\bph^B\ps_A)\nn\\
&+\f{4\pi}{\ka}(\bps^A\ph_A)(\bph^B\ps_B) +\f{2\pi}{\ka} (\bps^A\ph_A)(\bps^B\ph_B)+\f{2\pi}{\ka}
(\bph^A\ps_A)(\bph^B\ps_B) \nn\\
& -\f{4\pi m_0}{\ka}(\bph^A\ph_A)(\bph^C(\si_3)_C^{\ D}\ph_D)\biggr],
\end{align}
where $A,B$ are $SU(2)$ indices (see \S\ref{N3con} for the notation).\footnote{The gauge covariant derivatives are defined as
\begin{align}
&\mD^\mu\bph^\pm= \p^\mu \bph^\pm +i \bph^\pm A^\mu\ , \ \mD_\mu\ph^\pm= \p_\mu \ph^\pm -i  A_\mu\ph^\pm,\nn\\
&\slashed{\mD}\bps^\pm=\ga^\mu (\p_\mu \bps^\pm+i \bps^\pm A_\mu)\ , \ \slashed{\mD}\ps^\pm= \ga^\mu(\p_\mu \ps^\pm-i A_\mu) \ps^\pm, 
\end{align} with the trace conventions
\beq
\text{Tr} (T^a T^b)=\f{\de^{ab}}{2}\ , \ \sum_{a} (T^a)_i^{\ j} (T^a)_k^{\  l}=\f{\de_i^{\ l} \de_k^{\ j}}{2}.
\eeq
} Comparing the potential terms in \eqref{N3WZE} with the action in D.19 - D.22 of \cite{Chang:2012kt} we observe that both the operators of bare dimension 2 transform as triplet under the mass deformation and break the $R$-symmetry from $SU(2)\to U(1)$. We refer to \S\ref{N3susytransf} for the supersymmetry transformations of the Lorentzian $\mN=3$ action in the Wess-Zumino gauge. The mass deformation provides a manifestly supersymmetric IR regulator in the computation.  In fact, as we will see in \S\ref{exacpro}, there is no mass renormalization in this theory.\footnote{Similar to the $\mN=2$ theory, the point $m_0=0$ is a superconformal fixed point \cite{Gaiotto:2008sd,Avdeev:1992jt}.} Furthermore, we will also see that $m_0\to 0$ is a smooth limit of the scattering amplitude.

The theory \eqref{N3WZE} enjoys the strong-weak duality that is by now a standard feature of Chern-Simons-matter theories with $U(N), SU(N), SO(N)$ or $Sp(N)$ gauge groups. The statement of the duality for the $\mN=3$ theory is a statement of self-duality that is essentially the same as in the $\mN=2$ theory \cite{Jain:2013gza, Inbasekar:2015tsa}.\footnote{Starting from the $\mN=2$ theory, with a pair of fundamental and anti-fundamental chiral superfields, there exists a superpotential deformation that generates a duality preserving RG flow to the $\mN=3$ fixed point \cite{Gaiotto:2007qi}. At the fixed point the deformation parameter takes a fixed value that enhances the supersymmetry to $\mN=3$.} The theory \eqref{N3WZE} with rank $N$, level $\ka$, \rq t Hooft coupling $\la=\f{N}{\ka}$ and mass parameter $m$ is dual to the theory rank $N'$, level $\ka'$, \rq t Hooft coupling $\la'$ and mass parameter $m'$ under the map
\beq\label{dmap}
\quad \la'=\la-\text{Sign}(\la) \ , \ m'=-m \ , \ \ka'=-\ka
\eeq
in the limit $N\to\infty, \ka\to\infty$ with $\la=\f{N}{\ka}$ held fixed. Physical observables such as gauge invariant correlators, thermal partition functions and $S$-matrices would map under \eqref{dmap} as
\beq
\mathcal{O}^{\mN=3}(N',\ka',\la',m') \to \mathcal{O}^{\mN=3}(N,\ka,\la,m).
\eeq

\section{On-shell representation of $\mN=3$ theory in $\mN=1$ superspace}\label{onrepN3}
In this section, we formally outline the on-shell representation of $\mN=3$ theory in $\mN=1$ superspace. We begin in \S\ref{dir} from elementary properties of the Dirac equation in the mass deformed $\mN=3$ theory.
 
\subsection{Properties of the Dirac equation}\label{dir}
From the Lorentzian $\mN=3$ Lagrangian (see \eqref{N3WZL}) we find that the Dirac equation is of the form 
\beq
(i\slashed{\p} \ps_A +m (\si_3)_A^{\ B}\ps_B)=0.
\eeq
Let us break this in terms of the $SO(2)$ fields. We have the equations
\begin{align}
(i\slashed{\p}+m)\ps^-&=0,\nn\\
(i\slashed{\p}-m)\ps^+&=0.
\end{align}
The positive and negative energy wave functions $u_\al(\mbp,m), v_\al(\mbp,m)$ satisfy the equations \cite{Inbasekar:2015tsa}
\begin{equation}
\begin{split}
\label{uvde}
&(\slashed{p} - m)_\alpha^{\,~\beta} u_\beta(\vec[p],m) = 0 ,\\
&(\slashed{p} + m)_\alpha^{\,~\beta} v_\beta(\vec[p],m) = 0.
\end{split}
\end{equation}
The solutions are
\begin{align}\label{uvsol}
& u_\al(\mbp,m)=\begin{pmatrix}
         -\sqrt{p^0-p^1}\\
          \f{p^3+im}{\sqrt{p^0-p^1}}
        \end{pmatrix} \ , \
\bar{u}^\al(\mbp,m)=\begin{pmatrix}
               \f{i p^3 +m }{\sqrt{p^0-p^1}} & i \sqrt{p^0-p^1}
              \end{pmatrix}\ , \nn\\
& v_\al(\mbp,m)=\begin{pmatrix}
         \sqrt{p^0-p^1}\\
          \f{-p^3+im}{\sqrt{p^0-p^1}}
        \end{pmatrix} \ , \
\bar{v}^\al(\mbp,m)=\begin{pmatrix}
               \f{-i p^3 +m }{\sqrt{p^0-p^1}} & -i \sqrt{p^0-p^1}
              \end{pmatrix}\ ,
\end{align}
where we have defined
$$ p^0=+\sqrt{m^2+{\mbp}^2}\ .$$
Under the symmetry transformation $m\to-m$ we observe that
\beq
u_\al(\mbp,-m)= - v_\al(\mbp,m)\ , \ v_\al(\mbp,-m)= - u_\al(\mbp,m).
\eeq
The normalization of the spinors is chosen such that
\begin{equation}
\begin{split}
\label{uvnc}
&\bar{u}(\vec[p],m)\cdot u(\vec[p],m) = - 2m, \\
&\bar{v}(\vec[p],m)\cdot v(\vec[p],m) = 2m, \\
&u_\alpha(\vec[p],m) u^*_\beta(\vec[p],m) = - (\slashed{p} + m)_\alpha^{~\,\gamma} C_{\gamma\beta}, \\
&v_\alpha(\vec[p],m) v^*_\beta(\vec[p],m) = - (\slashed{p} - m)_\alpha^{~\,\gamma} C_{\gamma\beta}.
\end{split}
\end{equation}
With our conventions for the $\gamma$-matrices and the charge conjugation matrix $\mathcal{C}$, the spinors $(u_\alpha, v_\alpha)$ satisfy the properties
\begin{equation}
\begin{split}
\label{uvep}
u^*_\alpha(\vec[p],m) = - v_\alpha(\vec[p],m), \\
v^*_\alpha(\vec[p],m) = - u_\alpha(\vec[p],m).
\end{split}
\end{equation}
For the rest of the paper the notation $u_\al(\vec[p])$ stands for $u_\al(\vec[p],m)$, and similarly for $v_\al(\vec[p])$, i.e.\ unless specified the sign of $m$ is positive.

\subsection{Mass-deformed $\mN=3$ theory in $\mN=1$ superspace}
\label{fmon}
A scattering amplitude by definition is a matrix element which acts on an in-state to give an out-state. The in and out states are asymptotically free and obey free field equations. For the $\mN=3$ theory \eqref{N3eucS} the two superfields $\Phi^{\pm}(x,\theta)$ satisfy the free field equations of motion, given by
\beq
\label{osse1}
\left(D^2 \pm m\right) \Phi^\mp(x, \theta) = 0.
\eeq
The equations of motion \eqref{osse1} are met by the mode expansions
\begin{equation}
\begin{split}
\label{fmode1}
\Phi^-(x, \theta) = \int \frac{d^2p}{(2\pi)^2} \frac{1}{\sqrt{2 p^0}} \bigg[&\Big( a^-(\vec[p])(1+m\theta^2) + \theta^\alpha u_\alpha(\vec[p]) \alpha^-(\vec[p]) \Big) e^{ip\cdot x} \\
&+\Big( a^{-\dagger}_c(\vec[p])(1+m\theta^2) + \theta^\alpha v_\alpha(\vec[p]) \alpha^{-\dagger}_c(\vec[p]) \Big) e^{-ip\cdot x}\bigg],
\end{split}
\end{equation}
and
\begin{equation}
\begin{split}
\label{fmode2}
\Phi^+(x, \theta) = \int \frac{d^2p}{(2\pi)^2} \frac{1}{\sqrt{2 p^0}} \bigg[&\Big( a^+(\vec[p])(1-m\theta^2) + \theta^\alpha v_\alpha(\vec[p]) \alpha^+(\vec[p]) \Big) e^{ip\cdot x} \\
&+\Big( a^{+\dagger}_c(\vec[p])(1-m\theta^2) + \theta^\alpha u_\alpha(\vec[p]) \alpha^{+\dagger}_c(\vec[p]) \Big) e^{-ip\cdot x}\bigg].
\end{split}
\end{equation}
Here, $(a^\pm, a^{\pm\dagger})$ are annihilation and creation operators for bosons, $(a^\pm_c, a^{\pm\dagger}_c)$ are annihilation and creation operators for anti-bosons, $(\alpha^\pm, \alpha^{\pm\dagger})$ are annihilation and creation operators for fermions, and $(\alpha^\pm_c, \alpha^{\pm\dagger}_c)$ are annihilation and creation operators for anti-fermions. The bosonic operators obey the standard commutation relation
\beq
\label{stnc}
[a^\pm(\vec[p]), a^{\dagger \pm}(\vec[p]')] = (2\pi)^2 \delta^2(\vec[p] - \vec[p]'),
\eeq
and the fermionic operators obey the standard anti-commutation relation
\beq
\label{stnac}
\lbrace \alpha^\pm(\vec[p]), \alpha^{\pm\dagger}(\vec[p]') \rbrace = (2\pi)^2 \delta^2(\vec[p] - \vec[p]').
\eeq
Also, starting from the mode expansions for the fields $\Phi^\pm(x,\theta)$, \eqref{fmode1} and \eqref{fmode2}, the conjugate superfields $\bar\Phi^\pm(x,\theta)$ are given by
\begin{equation}
\begin{split}
\label{fbmode1}
\bar\Phi^-(x, \theta) = \int \frac{d^2p}{(2\pi)^2} \frac{1}{\sqrt{2 p^0}} \bigg[&\Big( a_c^-(\vec[p])(1+m\theta^2) - \theta^\alpha u_\alpha(\vec[p]) \alpha_c^-(\vec[p]) \Big) e^{ip\cdot x} \\
&+\Big( a^{-\dagger}(\vec[p])(1+m\theta^2) - \theta^\alpha v_\alpha(\vec[p]) \alpha^{-\dagger}(\vec[p]) \Big) e^{-ip\cdot x}\bigg],
\end{split}
\end{equation}
and
\begin{equation}
\begin{split}
\label{fbmode2}
\bar\Phi^+(x, \theta) = \int \frac{d^2p}{(2\pi)^2} \frac{1}{\sqrt{2 p^0}} \bigg[&\Big( a^+_c(\vec[p])(1-m\theta^2) - \theta^\alpha v_\alpha(\vec[p]) \alpha^+_c(\vec[p]) \Big) e^{ip\cdot x} \\
&+\Big( a^{+\dagger}(\vec[p])(1-m\theta^2) - \theta^\alpha u_\alpha(\vec[p]) \alpha^{+\dagger}(\vec[p]) \Big) e^{-ip\cdot x}\bigg].
\end{split}
\end{equation} 

\subsection{Asymptotic states and scattering amplitudes}
\label{astate}
In this section, we define the asymptotic states, classify them according to their $SO(2)_R$ charges in $\mN=1$ superspace, and define the $S$-matrices. The subject matter in this section may be viewed as a  recipe to extract various components of the $S$-matrix from the correlators computed in the $\mN=1$ superspace language. 

The scattering amplitude is a matrix element evaluated between asymptotic states. These asymptotic states are created by the creation and annihilation operators, and their mode expansions satisfy free field equations, as discussed in \S\ref{fmon}. The $SO(2)_R$ charges of the creation and annihilation operators follow from the $R$-charges of the corresponding superfields $\Phi^\pm$. It follows then that the asymptotic in/out states have definite $R$-charges. Since the $SO(2)_R$ symmetry is a global symmetry of the action \eqref{N3eucS} in $\mN=1$ superspace, it follows that the non-trivial $S$-matrices constructed in the $\mN=1$ language have to be invariant under this symmetry, i.e. they have zero net $SO(2)_R$ charge. We refer to the $S$-matrix in the $\mN=1$ language formally as 
\beq
S(R_1,R_2,R_3,R_4, \vec[p]_1,\vec[p]_2,\vec[p]_3,\vec[p]_4),
\eeq
where $R_i$ refers to $SO(2)_R$ charge of the $i^{th}$ particle, and $\vec[p]_i$ are the momenta. Unless stated otherwise, the incoming particles have momenta $p_1,p_2$, and outgoing particles have momenta $p_3,p_4$. The fact that the $S$-matrices are invariant under $SO(2)_R$ translates to the condition
\beq
\sum_i R_i=0.
\eeq
To proceed further, we define the operators
\begin{align}
U_B &= I+i L_B ,\nn\\
U_F &= I+i L_F .
\end{align}
We will encounter $L_B$ and $L_F$ at a later stage; for now we have used the same notation for convenience and readability. At this stage, $U_B$ and $U_F$ are unitary operators and $I$ is the identity operator formally defined as
\beq
I(\vec[p]_1,\vec[p]_2,\vec[p]_3,\vec[p]_4) = (2\pi)^2 \de^2(\vec[p]_1-\vec[p]_3) \de^2 (\vec[p]_2-\vec[p]_4).
\eeq

\subsubsection{Particle-particle scattering}\label{ppN3inN1}
From \S\ref{fmon}, we see that the two particle bosonic/fermionic asymptotic states are (recall that $a$, $\alpha$ denote the creation/annihilation operators for bosons and fermions respectively):
\beq\label{statepp}
\begin{array}{| c |c || c | c |}
\hline
SO(2)_R&\text{in-state} & \text{out-state}&SO(2)_R\\
\hline
-1 &a_+^{j\dag}(\vec[p]_2) a_+^{i\dag}(\vec[p]_1)|0\ran & \lan 0 | a_{n+}(\vec[p]_4)a_{m+}(\vec[p]_3) & +1 \\
+1 & a_-^{j\dag}(\vec[p]_2) a_-^{i\dag}(\vec[p]_1)|0\ran & \lan 0 | a_{n-}(\vec[p]_4)a_{m-}(\vec[p]_3) & -1 \\
0 & a_-^{j\dag}(\vec[p]_2) a_+^{i\dag}(\vec[p]_1)|0\ran & \lan 0 | a_{n+}(\vec[p]_4)a_{m-}(\vec[p]_3)& 0 \\
0 &a_+^{j\dag}(\vec[p]_2) a_-^{i\dag}(\vec[p]_1)|0\ran & \lan 0 | a_{n-}(\vec[p]_4)a_{m+}(\vec[p]_3)& 0 \\
\hline 
-1 &\al_+^{j\dag}(\vec[p]_2) \al_+^{i\dag}(\vec[p]_1)|0\ran & \lan 0 | \al_{n+}(\vec[p]_4)\al_{m+}(\vec[p]_3) & +1 \\
+1 & \al_-^{j\dag}(\vec[p]_2) \al_-^{i\dag}(\vec[p]_1)|0\ran & \lan 0 | \al_{n-}(\vec[p]_4)\al_{m-}(\vec[p]_3) & -1 \\
0 & \al_-^{j\dag}(\vec[p]_2) \al_+^{i\dag}(\vec[p]_1)|0\ran & \lan 0 | \al_{n+}(\vec[p]_4)\al_{m-}(\vec[p]_3)& 0 \\
0 &\al_+^{j\dag}(\vec[p]_2) \al_-^{i\dag}(\vec[p]_1)|0\ran & \lan 0 | \al_{n-}(\vec[p]_4)\al_{m+}(\vec[p]_3)& 0 \\
\hline
\end{array}
\eeq
The $R$-charge of the oscillators is same as that of the field in the mode expansion.  For instance, $\Phi^+$ has the $R$-charge $+\f{1}{2}$, so do $(a_+,\al_+)$ and similarly $((a_c)^\dagger_+,(\al_c)^\dagger_+)$. It is also clear from the commutation relations that $[a_+,a_-]=0$ (see \eqref{stnc} and \eqref{stnac}), and it follows that the non-trivial amplitudes are the ones that have vanishing $SO(2)_R$ charge. 
The bosonic $S$-matrices are defined as \footnote{In the $S$-matrices with zero $R$-charge for in and out states, it may naively appear that there are four possibilities. Of these only two are independent and are the third and fourth equations in \eqref{bosSUdUe}. }
\begin{small}
\begin{align}
\label{bosSUdUe}
S_B\left(- \f{1}{2},-\f{1}{2}, +\f{1}{2},+\f{1}{2}; \vec[p]_1,\vec[p]_2,\vec[p]_3,\vec[p]_4\right) & =\sqrt{(2p_1^0)(2p_2^0)(2p_3^0)(2p_4^0)} \lan 0 | a_{n+}(\vec[p]_4)a_{m+}(\vec[p]_3) U_B a_+^{j\dag}(\vec[p]_2) a_+^{i\dag}(\vec[p]_1)|0\ran,\nn\\
S_B\left(+ \f{1}{2},+\f{1}{2}, -\f{1}{2},-\f{1}{2}; \vec[p]_1,\vec[p]_2,\vec[p]_3,\vec[p]_4\right) & =\sqrt{(2p_1^0)(2p_2^0)(2p_3^0)(2p_4^0)} \lan 0 | a_{n-}(\vec[p]_4)a_{m-}(\vec[p]_3) U_B a_-^{j\dag}(\vec[p]_2) a_-^{i\dag}(\vec[p]_1)|0\nn\ran,\\
S_B\left( -\f{1}{2}, +  \f{1}{2}, + \f{1}{2}, - \f{1}{2}; \vec[p]_1,\vec[p]_2,\vec[p]_3,\vec[p]_4\right) &=\sqrt{(2p_1^0)(2p_2^0)(2p_3^0)(2p_4^0)} \lan 0 | a_{- n}(\vec[p]_4)a_{+ m}(\vec[p]_3) U_B a_-^{j\dag}(\vec[p]_2) a_+^{i\dag}(\vec[p]_1)|0\ran,\nn\\
S_B\left( -\f{1}{2}, +  \f{1}{2}, - \f{1}{2}, + \f{1}{2}; \vec[p]_1,\vec[p]_2,\vec[p]_3,\vec[p]_4\right) &=\sqrt{(2p_1^0)(2p_2^0)(2p_3^0)(2p_4^0)} \lan 0 | a_{+ n}(\vec[p]_4)a_{- m}(\vec[p]_3) U_B a_-^{j\dag}(\vec[p]_2) a_+^{i\dag}(\vec[p]_1)|0\ran.
\end{align}
\end{small}
As discussed in \S\ref{skin} the particle-particle scattering can be decomposed further into symmetric and anti-symmetric channels. Sometimes, it is convenient to use the direct and exchange channels, these are related to the symmetric and anti-symmetric channels via 
\begin{align}
\maS_{U_{s}}(\vec[p]_1,\vec[p]_2,\vec[p]_3,\vec[p]_4 )&=\maS_{U_{e}}(\vec[p]_1,\vec[p]_2,\vec[p]_4,\vec[p]_3) +\maS_{U_{d}}\vec[p]_1,\vec[p]_2,\vec[p]_3,\vec[p]_4 )\nn\\
\maS_{U_{a}}\vec[p]_1,\vec[p]_2,\vec[p]_3,\vec[p]_4 )&=\maS_{U_{e}}(\vec[p]_1,\vec[p]_2,\vec[p]_4,\vec[p]_3)-\maS_{U_{d}}(\vec[p]_1,\vec[p]_2,\vec[p]_4,\vec[p]_3 ).
\end{align}
Thus we have
\begin{small}
\begin{align}
S_B\left(\mp \f{1}{2},\mp\f{1}{2}, \pm\f{1}{2},\pm\f{1}{2}; \vec[p]_1,\vec[p]_2,\vec[p]_3,\vec[p]_4\right) &=  \de_m^i\de_n^j \left(I( \vec[p]_1,\vec[p]_2,\vec[p]_3,\vec[p]_4) + \TB^{U_d}(\mp \f{1}{2},\mp\f{1}{2}, \pm\f{1}{2},\pm\f{1}{2};  \vec[p]_1,\vec[p]_2,\vec[p]_3,\vec[p]_4)\right)\nn\\
&+ \de_m^j\de_n^i \left(I( \vec[p]_1,\vec[p]_2,\vec[p]_3,\vec[p]_3) + \TB^{U_e}(\mp \f{1}{2},\mp\f{1}{2}, \pm\f{1}{2},\pm\f{1}{2}; \vec[p]_1,\vec[p]_2,\vec[p]_4,\vec[p]_3)\right),\nn\\
S_B\left( - \f{1}{2}, +  \f{1}{2}, \pm \f{1}{2}, \mp \f{1}{2}; \vec[p]_1,\vec[p]_2,\vec[p]_3,\vec[p]_4\right)  &= \de_m^i\de_n^j \left(I( \vec[p]_1,\vec[p]_2,\vec[p]_3,\vec[p]_4) + \TB^{U_d}(  - \f{1}{2}, + \f{1}{2}, + \f{1}{2}, - \f{1}{2};\vec[p]_1,\vec[p]_2,\vec[p]_3,\vec[p]_4)\right)\nn\\
& +\de_m^j\de_n^i \left(I( \vec[p]_1,\vec[p]_2,\vec[p]_4,\vec[p]_3) + \TB^{U_e}(- \f{1}{2}, +  \f{1}{2}, - \f{1}{2}, + \f{1}{2}, \vec[p]_1,\vec[p]_2,\vec[p]_4,\vec[p]_3) \right).
\label{bosparteq1}
\end{align}
\end{small}
The fermionic $S$-matrices are defined similarly as
\begin{small}
\begin{align}
S_F\left(- \f{1}{2},-\f{1}{2}, +\f{1}{2},+\f{1}{2}; \vec[p]_1,\vec[p]_2,\vec[p]_3,\vec[p]_4\right) & =\sqrt{(2p_1^0)(2p_2^0)(2p_3^0)(2p_4^0)} \lan 0 | \al_{n+}(\vec[p]_4)\al_{m+}(\vec[p]_3) U_F \al_+^{j\dag}(\vec[p]_2) \al_+^{i\dag}(\vec[p]_1)|0\ran,\nn\\
S_F\left(+ \f{1}{2},+\f{1}{2}, -\f{1}{2},-\f{1}{2}; \vec[p]_1,\vec[p]_2,\vec[p]_3,\vec[p]_4\right) & =\sqrt{(2p_1^0)(2p_2^0)(2p_3^0)(2p_4^0)} \lan 0 | \al_{n-}(\vec[p]_4)\al_{m-}(\vec[p]_3) U_F \al_-^{j\dag}(\vec[p]_2) \al_-^{i\dag}(\vec[p]_1)|0\ran,\nn\\
S_F\left(- \f{1}{2},+\f{1}{2}, +\f{1}{2},-\f{1}{2}; \vec[p]_1,\vec[p]_2,\vec[p]_3,\vec[p]_4\right) & =\sqrt{(2p_1^0)(2p_2^0)(2p_3^0)(2p_4^0)} \lan 0 | \al_{n-}(\vec[p]_4)\al_{m+}(\vec[p]_3) U_F \al_-^{j\dag}(\vec[p]_2) \al_+^{i\dag}(\vec[p]_1)|0\ran,\nn\\
S_F\left(- \f{1}{2},+\f{1}{2}, -\f{1}{2},+\f{1}{2}; \vec[p]_1,\vec[p]_2,\vec[p]_3,\vec[p]_4\right) & =\sqrt{(2p_1^0)(2p_2^0)(2p_3^0)(2p_4^0)} \lan 0 | \al_{n+}(\vec[p]_4)\al_{m-}(\vec[p]_3) U_F \al_-^{j\dag}(\vec[p]_2) \al_+^{i\dag}(\vec[p]_1)|0\ran.
\end{align}
\end{small}
Decomposing the fermionic $S$-matrices further into direct and exchange channels of scattering we see that
\begin{small}
\begin{align}\label{ferUdUe}
S_F\left(\mp \f{1}{2},\mp\f{1}{2}, \pm\f{1}{2},\pm\f{1}{2}; \vec[p]_1,\vec[p]_2,\vec[p]_3,\vec[p]_4\right) = &- \de_m^i\de_n^j \left(I( \vec[p]_1,\vec[p]_2,\vec[p]_3,\vec[p]_4) + \TF^{U_d}(\mp \f{1}{2},\mp\f{1}{2}, \pm\f{1}{2},\pm\f{1}{2};  \vec[p]_1,\vec[p]_2,\vec[p]_3,\vec[p]_4)\right)\nn\\
&+ \de_m^j\de_n^i \left(I( \vec[p]_1,\vec[p]_2,\vec[p]_4,\vec[p]_3) + \TF^{U_e}(\mp \f{1}{2},\mp\f{1}{2}, \pm\f{1}{2},\pm\f{1}{2}; \vec[p]_1,\vec[p]_2,\vec[p]_4,\vec[p]_2)\right),\nn\\
S_F\left( - \f{1}{2}, +  \f{1}{2}, \pm \f{1}{2}, \mp \f{1}{2}; \vec[p]_1,\vec[p]_2,\vec[p]_3,\vec[p]_4\right)  = &-\de_m^i\de_n^j \left(I( \vec[p]_1,\vec[p]_2,\vec[p]_3,\vec[p]_4) + \TF^{U_d}(  - \f{1}{2}, + \f{1}{2}, + \f{1}{2}, - \f{1}{2};\vec[p]_1,\vec[p]_2,\vec[p]_3,\vec[p]_4)\right)\nn\\
+& \de_m^j\de_n^i \left(I( \vec[p]_1,\vec[p]_2,\vec[p]_4,\vec[p]_3) + \TF^{U_e}(- \f{1}{2}, +  \f{1}{2}, - \f{1}{2}, + \f{1}{2}, \vec[p]_1,\vec[p]_2,\vec[p]_4,\vec[p]_3)\right).
\end{align}
\end{small}

\subsubsection{Particle--anti-particle scattering}\label{papN3inN1}
For particle--anti-particle scattering the asymptotic states are defined as:
\beq\label{stateap}
\begin{array}{| c |c || c | c |}
\hline
SO(2)_R&\text{in-state} & \text{out-state}&SO(2)_R\\
\hline
0 &(a_c)_{+j}^{\dag}(\vec[p]_2) a_+^{i\dag}(\vec[p]_1)|0\ran & \lan 0 | a_{n+}(\vec[p]_4)(a_c)_{+}^m(\vec[p]_3) & 0 \\
0 & (a_c)_{-j}^{\dag}(\vec[p]_2) a_-^{i\dag}(\vec[p]_1)|0\ran & \lan 0 | a_{n-}(\vec[p]_4)(a_c)_{-}^m(\vec[p]_3) & 0 \\
-1 & (a_c)_{-j}^{\dag}(\vec[p]_2) a_+^{i\dag}(\vec[p]_1)|0\ran & \lan 0 | a_{n+}(\vec[p]_4)(a_c)_{-}^m(\vec[p]_3)& +1 \\
+1 &(a_c)_{+j}^{\dag}(\vec[p]_2) a_-^{i\dag}(\vec[p]_1)|0\ran & \lan 0 | a_{n-}(\vec[p]_4)(a_c)_{+}^m(\vec[p]_3)& -1 \\
\hline 
0 &(\al_c)_{+j}^{\dag}(\vec[p]_2) \al_+^{i\dag}(\vec[p]_1)|0\ran & \lan 0 | \al_{n+}(\vec[p]_4)(\al_c)_{+}^m(\vec[p]_3) & 0 \\
0 &(\al_c)_{-j}^{\dag}(\vec[p]_2) \al_-^{i\dag}(\vec[p]_1)|0\ran & \lan 0 | \al_{n-}(\vec[p]_4)(\al_c)_{-}^m(\vec[p]_3) & 0 \\
-1 & (\al_c)_{-j}^{\dag}(\vec[p]_2) \al_+^{i\dag}(\vec[p]_1)|0\ran & \lan 0 | \al_{n+}(\vec[p]_4)(\al_c)_{-}^m(\vec[p]_3)& +1 \\
+1 &(\al_c)_{+j}^{\dag}(\vec[p]_2) \al_-^{i\dag}(\vec[p]_1)|0\ran & \lan 0 | \al_{n-}(\vec[p]_4)(\al_c)_{+}^m(\vec[p]_3)& -1 \\
\hline
\end{array}
\eeq
It is clear from the commutation relations \eqref{stnc}, \eqref{stnac} that the non-trivial $2\to2$ $S$-matrices that one can construct have a net vanishing $SO(2)_R$ charge. The bosonic amplitudes are defined as
\begin{small}
\begin{align}
\tilde{S}_B(\mp\f{1}{2},\pm\f{1}{2},\mp\f{1}{2},\pm\f{1}{2};\vec[p]_1,\vec[p]_2,\vec[p]_3,\vec[p]_4) & =\sqrt{(2p_1^0)(2p_2^0)(2p_3^0)(2p_4^0)} \lan 0 | a_{n\pm}(\vec[p]_4)(a_c)_{\pm}^m(\vec[p]_3) U_B (a_c)_{\pm j}^{\dag}(\vec[p]_2) a_\pm^{i\dag}(\vec[p]_1)|0\ran,\nn\\
\tilde{S}_B(\mp\f{1}{2},\mp\f{1}{2},\pm\f{1}{2},\pm\f{1}{2};\vec[p]_1,\vec[p]_2,\vec[p]_3,\vec[p]_4) &=\sqrt{(2p_1^0)(2p_2^0)(2p_3^0)(2p_4^0)} \lan 0 | a_{n\pm}(\vec[p]_4)(a_c)_{\mp}^m(\vec[p]_3) U_B (a_c)_{\mp j}^{\dag}(\vec[p]_2) a_\pm^{i\dag}(\vec[p]_1)|0\ran.
\end{align}
\end{small}
Decomposing the $S$-matrix into adjoint and singlet channels gives
\begin{small}
\begin{align}
\label{bosTS}
\tilde{S}_B(\mp\f{1}{2},\pm\f{1}{2},\mp\f{1}{2},\pm\f{1}{2};\vec[p]_1,\vec[p]_2,\vec[p]_3,\vec[p]_4) =&  \left(\de_n^i \de_j^m - \f{\de_n^m \de_j^i }{N}\right) \left(I( \vec[p]_1,\vec[p]_2,\vec[p]_3,\vec[p]_4) + \TB^T(\mp\f{1}{2},\pm\f{1}{2},\mp\f{1}{2},\pm\f{1}{2};\vec[p]_1,\vec[p]_2,\vec[p]_3,\vec[p]_4)\right)\nn\\
& +\f{\de_n^m \de_j^i }{N} \left(I( \vec[p]_1,\vec[p]_2,\vec[p]_3,\vec[p]_4) + \TB^S(\mp\f{1}{2},\pm\f{1}{2},\mp\f{1}{2},\pm\f{1}{2};\vec[p]_1,\vec[p]_2,\vec[p]_3,\vec[p]_4)\right),\nn\\
\tilde{S}_B(\mp\f{1}{2},\mp\f{1}{2},\pm\f{1}{2},\pm\f{1}{2};\vec[p]_1,\vec[p]_2,\vec[p]_3,\vec[p]_4) =&  \left(\de_n^i \de_j^m - \f{\de_n^m \de_j^i }{N}\right) \left(I( \vec[p]_1,\vec[p]_2,\vec[p]_3,\vec[p]_4) + \TB^T(\mp\f{1}{2},\mp\f{1}{2},\pm\f{1}{2},\pm\f{1}{2};\vec[p]_1,\vec[p]_2,\vec[p]_3,\vec[p]_4)\right)\nn\\
& +\f{\de_n^m \de_j^i }{N} \left(I( \vec[p]_1,\vec[p]_2,\vec[p]_3,\vec[p]_4) + \TB^S(\mp\f{1}{2},\mp\f{1}{2},\pm\f{1}{2},\pm\f{1}{2};\vec[p]_1,\vec[p]_2,\vec[p]_3,\vec[p]_4)\right).
\end{align}
\end{small}

The fermionic $S$-matrices are defined similarly as
\begin{small}
\begin{align}
\tilde{S}_F(\mp\f{1}{2},\pm\f{1}{2},\mp\f{1}{2},\pm\f{1}{2};\vec[p]_1,\vec[p]_2,\vec[p]_3,\vec[p]_4) & =\sqrt{(2p_1^0)(2p_2^0)(2p_3^0)(2p_4^0)} \lan 0 | \al_{n\pm}(\vec[p]_4)(\al_c)_{\pm}^m(\vec[p]_3) U_F (\al_c)_{\pm j}^{\dag}(\vec[p]_2) \al_\pm^{i\dag}(\vec[p]_1)|0\ran,\nn\\
\tilde{S}_F(\mp\f{1}{2},\mp\f{1}{2},\pm\f{1}{2},\pm\f{1}{2};\vec[p]_1,\vec[p]_2,\vec[p]_3,\vec[p]_4) &=\sqrt{(2p_1^0)(2p_2^0)(2p_3^0)(2p_4^0)} \lan 0 | \al_{n\pm}(\vec[p]_4)(\al_c)_{\mp}^m(\vec[p]_3) U_F (\al_c)_{\mp j}^{\dag}(\vec[p]_2) \al_\pm^{i\dag}(\vec[p]_1)|0\ran.
\end{align}
\end{small}
Decomposing the $S$-matrix into adjoint and singlet channels gives
\begin{small}
\begin{align}
\label{fermTS}
\tilde{S}_F(\mp\f{1}{2},\pm\f{1}{2},\mp\f{1}{2},\pm\f{1}{2};\vec[p]_1,\vec[p]_2,\vec[p]_3,\vec[p]_4) =&  \left(\de_n^i \de_j^m - \f{\de_n^m \de_j^i }{N}\right) \left(I( \vec[p]_2,\vec[p]_1,\vec[p]_3,\vec[p]_4) + \TF(\mp\f{1}{2},\pm\f{1}{2},\mp\f{1}{2},\pm\f{1}{2};\vec[p]_1,\vec[p]_2,\vec[p]_3,\vec[p]_4)\right)\nn\\
& +\f{\de_n^m \de_j^i }{N} \left(I( \vec[p]_1,\vec[p]_2,\vec[p]_3,\vec[p]_4) + \TF(\mp\f{1}{2},\pm\f{1}{2},\mp\f{1}{2},\pm\f{1}{2};\vec[p]_1,\vec[p]_2,\vec[p]_3,\vec[p]_4)\right),\nn\\
\tilde{S}_F(\mp\f{1}{2},\mp\f{1}{2},\pm\f{1}{2},\pm\f{1}{2};\vec[p]_1,\vec[p]_2,\vec[p]_3,\vec[p]_4) =& \left(\de_n^i \de_j^m - \f{\de_n^m \de_j^i }{N}\right) \left(I( \vec[p]_1,\vec[p]_2,\vec[p]_3,\vec[p]_4) + \TF(\mp\f{1}{2},\mp\f{1}{2},\pm\f{1}{2},\pm\f{1}{2};\vec[p]_1,\vec[p]_2,\vec[p]_3,\vec[p]_4)\right)\nn\\
&+ \f{\de_n^m \de_j^i }{N} \left(I( \vec[p]_1,\vec[p]_2,\vec[p]_3,\vec[p]_4) + \TF(\mp\f{1}{2},\mp\f{1}{2},\pm\f{1}{2},\pm\f{1}{2};\vec[p]_1,\vec[p]_2,\vec[p]_3,\vec[p]_4)\right).
\end{align}
\end{small}
This concludes the formal definitions of the bosonic/fermionic $S$-matrices in $\mN=1$ superspace. 

\section{Manifestly $\mN=3$ covariant on-shell representation in the massless limit}\label{N3co}
Since the massive theory breaks $SU(2)_R$ symmetry, the on-shell representation was defined in a rather non-covariant way in \S\ref{fmon}. However, it turns out, as we will see later in \S\ref{onsmatrix}, that the massless limit of the amplitude is smooth and one recovers $SU(2)_R$ covariance. In this section, we construct the on-shell representation in the massless case.  
\subsection{Supersymmetry algebra}\label{N3susonsalgebra}
The free field equations that follow from \eqref{N3WZL} are given by
\beq
\Box\Ph^A=0 \ , \ i\ga^\mu\p_\mu\ps_A=0.
\eeq
They can be mode expanded as follows
\begin{align}
\label{modeexp}
\ph^A(x)&=\int \f{d^2 p}{\sqrt{2p^0}(2\pi)^2} (a^A(\vec[p]) e^{i p.x }+ b^{\dagger A}(\vec[p]) e^{-i p.x}),\nn\\
\bph^A(x)&=\int \f{d^2 p}{\sqrt{2p^0}(2\pi)^2} (a^{\dagger A}(\vec[p]) e^{-i p.x }+ b^{A}(\vec[p]) e^{i p.x}),\nn\\
\ps_{A\al}(x)&=\int \f{d^2 p}{\sqrt{2p^0}(2\pi)^2} (\al_A(\vec[p]) u_\al(\vec[p]) e^{i p.x }+ \be^{\dagger }_A(\vec[p]) v_\al(\vec[p]) e^{-i p.x}),\nn\\
\bps^A_\al(x)&=\int \f{d^2 p}{\sqrt{2p^0}(2\pi)^2} (\al^{\dagger A}(\vec[p]) u_\al(\vec[p])^* e^{-i p.x }+ \be^{A}(\vec[p]) v_\al(\vec[p])^* e^{i p.x}),
\end{align}
where $a^A/\alpha^A, a^{\dagger A}/\alpha^{\dagger A}$ create and annihilate bosonic/fermionic particles, while $b^A/\be^A, b^{\dagger A}/\be^{\dagger A}$ create and annihilate bosonic/fermionic anti-particles. These obey the canonical commutation rules
\begin{align}\label{covcomm}
 [a^A(\vec[p]), a_B^\dag(\vec[p]')] =(2\pi)^2\de_B^{\ A} \de^2(\vec[p]-\vec[p]')\ &, \  [b^A(\vec[p]), b_B^\dag(\vec[p]')] =(2\pi)^2\de_B^{\ A}\de^2(\vec[p]-\vec[p]'),\nn\\
 \{\al^A(\vec[p]), \al_B^\dag(\vec[p]')\} =(2\pi)^2\de_B^{\ A}\de^2(\vec[p]-\vec[p]')\ &, \  \{\be^A(\vec[p]), \be_B^\dag(\vec[p]')\} =(2\pi)^2\de_B^{\ A}\de^2(\vec[p]-\vec[p]').
\end{align}
The $SU(2)$ oscillators are related to \eqref{stnc}, \eqref{stnac} by
\begin{align}\label{SU2U1osc}
 a^A=\begin{pmatrix}
        a^+\\ a^-
       \end{pmatrix}\ , \ 
        b^{A\dag}=\begin{pmatrix}
        a_c^{+\dag}\\ a_c^{-\dag}
       \end{pmatrix}\ , \
        \al_A=\begin{pmatrix}
        \al^-\\ -\al^+
       \end{pmatrix}\ , \
       \be_A^\dag=\begin{pmatrix}
        \al_c^{-\dag}\\ -\al_c^{+\dag}
       \end{pmatrix},
\end{align}
with $(a_A)^\dag= a^{A\dag}$. The raising and lowering of the $SU(2)$ operators follow from \S\ref{N3con},
\begin{align}
 a^A = \ep^{AB} a_B \ , \ b^{\dagger A} =\ep^{AB} b^\dagger_B \ , \  \al^A = \ep^{AB} \al_B \ , \be^{\dagger A} =\ep^{AB} \be^\dagger_B,\nn\\
 a^{\dagger A} = a^\dagger_B\ep^{BA}  \ , \ b^{ A} = b_B\ep^{BA} \ , \  \al^{\dagger A} =  \al^\dagger_B\ep^{BA} \ , \be^{ A} = \be_B \ep^{BA}.
\end{align}
The free field $\mN=3$ supersymmetry transformations can be obtained from \S\ref{N3susytransf} and are given by
\begin{align}\label{frees}
 Q_{BC\al}\ph^A &= \ps_{\al (B}\de_{C)}^{\ A}\, ,\nn\\
 Q_{BC\al}\bph^A &= -\bps_{\al (B}\de_{C)}^{\ A}\, ,\nn\\
 Q_{BC\al}\ps_\be^A &= -i\p_{\al\be}\ph_{(B}\de_{C)}^{\ A}\, ,\nn\\
Q_{BC\al}\bps_\be^A &= i\p_{\al\be}\bph_{(B}\de_{C)}^{\ A}\, .
\end{align}
Using the mode expansions \eqref{modeexp} in \eqref{frees} we obtain the following representation for the $\mN=3$ supercharges,
\begin{align}\label{N3onsQ}
-i Q_{BC\al} &= u_\al (\al_{(B}\p_{a^{C)}}-a^\dag_{(B}\p_{\al^{\dag C)}})- u_\al^*(\al^\dag_{(B}\p_{a^{\dag C)}}+a_{(B}\p_{\al^{ C)}})\nn\\
& +v_\al( b_{(B}\p_{\be^{C)}}+ \be^\dag_{(B}\p_{b^{\dagger C)}})-v_\al^*( \be_{(B}\p_{b^{ C)}}- b^\dag_{(B}\p_{\be^{\dag C)}}).
\end{align}
Using \eqref{uvep} this can be expressed as
\begin{align}\label{N3onsQ2}
-i Q_{BC\al} &= u_\al\biggl( (\al_{(B}\p_{a^{C)}}-a^\dag_{(B}\p_{\al^{\dag C)}})-(\al^\dag_{(B}\p_{a^{\dag C)}}+a_{(B}\p_{\al^{ C)}})\nn\\
&\quad-( b_{(B}\p_{\be^{C)}}+ \be^\dag_{(B}\p_{b^{\dagger C)}})+( \be_{(B}\p_{b^{ C)}}- b^\dag_{(B}\p_{\be^{\dag C)}})\biggr).
\end{align}
It is straightforward to check that the supercharge \eqref{N3onsQ} closes to form the $\mN=3$ algebra,
\beq
\{Q_{BC\al},Q_{DE\be}\}\ph^A = -i\p_{\al\be} \biggl(\ph_{(B}\ep_{C)(D}\de_{E)}^{\ A}+\ph_{(D}\ep_{E)(B}\de_{C)}^{\ A}\biggr).
\eeq
The $\mN=3$ theory has $SU(2)_R$ symmetry and the $R$-symmetry algebra is
\beq\label{Rsym}
[R_{AB}, R_{CD}]=  \left(\ep_{AC}R_{BD} +\ep_{BC} R_{AD} +\ep_{AD} R_{BC} +\ep_{BD} R_{AC}\right).
\eeq
The $R$-symmetries act on the supersymmetry generators to form the algebra
\beq\label{QRsym}
[R_{AB}, Q_{CD}]=  \left(\ep_{AC}Q_{BD} +\ep_{BC} Q_{AD} +\ep_{AD} Q_{BC} +\ep_{BD} Q_{AC}\right).
\eeq
\subsection{Scattering amplitudes}\label{scmpa}
In this section, we conveniently organize all the states defined in \eqref{statepp} in a compact form. The definition of the $S$-matrix coincides, however, only in the massless limit since the formulation assumes manifest $SU(2)_R$ symmetry and the $S$-matrices are singlets under this global symmetry. In the subsections below we use the normalization for the asymptotic states such that
\begin{equation}
\f{1}{2}\f{\text{Tr}(N_A^{\ B})}{N}=1,
\end{equation}
where 
\begin{align}
&\text{Tr}(N_A^{ \ B}) =\text{Tr}\left(\lan 0| a_A a^{\dagger B}|0\ran\right) ,\nn\\
&N =\lan 0| a_+ a_+^\dagger|0\ran = \lan 0| a_- a_-^\dagger|0\ran.
\end{align}
\subsubsection{Particle-particle scattering}\label{N3covppons}
Consider two particle scattering,
\beq
P_i(p_1)+P_j(p_2)\to P_m(p_3) +P_n(p_4),
\eeq
where $i, j$ refer to the color indices of the particles. The two particle bosonic/fermionic asymptotic states are defined as 
\beq\label{stateppcov}
\begin{array}{| c | c |}
\hline
\text{in-state} & \text{out-state}\\
\hline
\f{1}{2} a^{i\dagger}_B(p_2)a^{i\dagger}_A(p_1)|0\ran & \f{1}{2}\lan 0 | a^D_n(p_4) a^C_n(p_3)\\
\f{1}{2}\al^{i\dagger}_B(p_2)\al^{i\dagger}_A(p_1)|0\ran & \f{1}{2}\lan 0 | \al^D_n(p_4) \al^C_n(p_3)\\
\hline
\end{array}
\eeq
Let us define super-creation/annihilation operators for particles as
\begin{align}\label{partosc}
 A_{ai}(\bar{x}_a,\bar{\eta}_a,\vec[p]_a) &= \bar{x}_{aA} a_{i}^A(\vec[p]_a) +\bar{\eta}_{aA} \al_{i}^A(\vec[p]_a),\nn\\
 A_a^{\dag i}(x_a,\eta_a,\vec[p]_a) &= a^{\dag i}_A (\vec[p]_a) x^{aA} +\al^{\dag i}_A(\vec[p]_a)\eta^{aA} ,
\end{align}
where $a=1,\ldots 4$ refers to the particle index, while $\bar{x}_A$ is a commuting $SU(2)$ valued Grassmann spinor variable whereas $\bar{\eta}_A$ is an anti-commuting $SU(2)$ valued Grassmann spinor variable.  We define the $S$-matrix for particle-particle scattering formally as
\begin{small}
\begin{align}\label{sppa}
\maS(x,\eta,\bar{x},\bar{\eta},\vec[p])  &=\sqrt{(2p_1^0)(2p_2^0)(2p_3^0)(2p_4^0)}\lan 0|A_{4n}(\bar{x}_4,\bar{\eta}_4,\vec[p]_4)  A_{3 m}(\bar{x}_3,\bar{\eta}_3,\vec[p]_3) (1+i L_4)  A_{2}^{\dag j}(x_2,\eta_2,\vec[p]_2) A_{1 }^{\dag i}(x_1,\eta_1,\vec[p]_1) |0\ran,
\end{align}
\end{small}
where $L_4$ is defined in \eqref{effon}. As discussed in \S\ref{skin} the amplitude can be decomposed into the symmetric and anti-symmetric channels of scattering,
\begin{align}
\maS(x,\eta,\bar{x},\bar{\eta},\vec[p])= \f{\de_m^i \de_n^j+\de_n^i \de_m^j }{2} \, \maS_{U_{s}}(x,\eta,\bar{x},\bar{\eta},\vec[p])+\f{\de_m^i \de_n^j-\de_n^i \de_m^j }{2} \, \maS_{U_{a}}(x,\eta,\bar{x},\bar{\eta},\vec[p]).
\end{align}
One can also rewrite these $S$-matrices in terms of the direct and exchange channel $S$-matrices as follows
\begin{align}
\maS_{U_{s}}(x,\eta,\bar{x},\bar{\eta},\vec[p])&=\maS_{U_{d}}(x,\eta,\bar{x},\bar{\eta},\vec[p])+\maS_{U_{e}}(x,\eta,\bar{x},\bar{\eta},\vec[p]),\nn\\
\maS_{U_{a}}(x,\eta,\bar{x},\bar{\eta},\vec[p])&=\maS_{U_{e}}(x,\eta,\bar{x},\bar{\eta},\vec[p])-\maS_{U_{d}}(x,\eta,\bar{x},\bar{\eta},\vec[p]).
\end{align}
The bosonic and fermionic $S$-matrices can be read off from \footnote{The brackets in $(x_1^A\bar{x}^3_A)$ etc.\ have been put for notational ease. } 
\begin{align}\label{sppax}
\maS_B^{U_d}(\vec[p]_1,\vec[p]_2,\vec[p]_3,\vec[p]_4) = (x_1^A \bar{x}^3_A) (x_2^B \bar{x}^4_B) \left(I(\vec[p]_1,\vec[p]_2,\vec[p]_3,\vec[p]_4) +i\TB^{U_d}\left(\vec[p]_1,\vec[p]_2,\vec[p]_3,\vec[p]_4\right)\right),\nn\\
\maS_B^{U_e}(\vec[p]_2,\vec[p]_1,\vec[p]_3,\vec[p]_4) = (x_1^A \bar{x}^3_B) (x_2^B \bar{x}^4_A) \left(I(\vec[p]_2,\vec[p]_1,\vec[p]_3,\vec[p]_4)+i\TB^{U_e}\left(\vec[p]_2,\vec[p]_1,\vec[p]_3,\vec[p]_4\right)\right),\nn\\
\maS_F^{U_d}(\vec[p]_1,\vec[p]_2,\vec[p]_3,\vec[p]_4) = (\et_1^A \bar{\et}^3_A) (\et_2^B \bar{\et}^4_B) \left(I(\vec[p]_1,\vec[p]_2,\vec[p]_3,\vec[p]_4) +i\TF^{U_d}\left(\vec[p]_1,\vec[p]_2,\vec[p]_3,\vec[p]_4 \right)\right),\nn\\
\maS_F^{U_e}(\vec[p]_2,\vec[p]_1,\vec[p]_3,\vec[p]_4) = (\et_1^A \bar{\et}^3_B) (\et_2^B \bar{\et}^4_A) \left(I(\vec[p]_2,\vec[p]_1,\vec[p]_3,\vec[p]_4)+i\TF^{U_e}\left(\vec[p]_2,\vec[p]_1,\vec[p]_3,\vec[p]_4\right)\right).
\end{align}
The polynomials in $x$ and $\eta$ encode the information on how the various component $S$-matrices discussed in \S\ref{astate} contribute to the covariant $S$-matrix as follows. Consider for instance the polynomial $(x_1^A \bar{x}^3_A) (x_2^B \bar{x}^4_B)$ in \eqref{sppax}.  The decompositions follow as
\begin{small}
\begin{align}
\label{bcovtononcovpp}
\TB^{U_d}\left(\vec[p]_1,\vec[p]_2,\vec[p]_3,\vec[p]_4\right) &= \f{1}{4} \biggl(\TB^{U_d}\left(+\f{1}{2},+\f{1}{2},-\f{1}{2},-\f{1}{2};\vec[p]_1,\vec[p]_2,\vec[p]_3,\vec[p]_4\right)+\TB^{U_d}\left(-\f{1}{2},-\f{1}{2},+\f{1}{2},+\f{1}{2};\vec[p]_1,\vec[p]_2,\vec[p]_3,\vec[p]_4\right)\nn\\
&+ \TB^{U_d}\left(-\f{1}{2},+\f{1}{2},+\f{1}{2},-\f{1}{2};\vec[p]_1,\vec[p]_2,\vec[p]_3,\vec[p]_4\right)+ \TB^{U_d}\left(+\f{1}{2},-\f{1}{2},-\f{1}{2},+\f{1}{2};\vec[p]_1,\vec[p]_2,\vec[p]_3,\vec[p]_4\right)\biggr),\nn\\
\TB^{U_e}\left(\vec[p]_1,\vec[p]_2,\vec[p]_3,\vec[p]_4\right) &= \f{1}{4} \biggl(\TB^{U_e}\left(+\f{1}{2},+\f{1}{2},-\f{1}{2},-\f{1}{2};\vec[p]_1,\vec[p]_2,\vec[p]_3,\vec[p]_4\right)+\TB^{U_e}\left(-\f{1}{2},-\f{1}{2},+\f{1}{2},+\f{1}{2};\vec[p]_1,\vec[p]_2,\vec[p]_3,\vec[p]_4\right)\nn\\
&+ \TB^{U_e}\left(-\f{1}{2},+\f{1}{2},-\f{1}{2},+\f{1}{2};\vec[p]_1,\vec[p]_2,\vec[p]_3,\vec[p]_4\right)+ \TB^{U_e}\left(+\f{1}{2},-\f{1}{2},+\f{1}{2},-\f{1}{2};\vec[p]_1,\vec[p]_2,\vec[p]_3,\vec[p]_4\right)\biggr).
\end{align}
\end{small}
It is easy to understand the decomposition \eqref{bcovtononcovpp}. 
Firstly, Using the definition of the S matrix \eqref{sppa} and the super oscillators \eqref{partosc}, we expands the $SU(2)$ oscillators in terms of the components using \eqref{SU2U1osc}. As explained in \S\ref{fmon} each of the component oscillators are associated with definite $SO(2)$ R charges, and thus \eqref{bcovtononcovpp} follows.

Similarly we have an identical decomposition for the fermionic amplitude,
\begin{small}
\begin{align}
\label{fcovtononcovpp}
\TF^{U_d}\left(\vec[p]_1,\vec[p]_2,\vec[p]_3,\vec[p]_4\right) &= \f{1}{4} \biggl(\TF^{U_d}\left(+\f{1}{2},+\f{1}{2},-\f{1}{2},-\f{1}{2};\vec[p]_1,\vec[p]_2,\vec[p]_3,\vec[p]_4\right)+\TF^{U_d}\left(-\f{1}{2},-\f{1}{2},+\f{1}{2},+\f{1}{2};\vec[p]_1,\vec[p]_2,\vec[p]_3,\vec[p]_4\right)\nn\\
&+ \TF^{U_d}\left(-\f{1}{2},+\f{1}{2},+\f{1}{2},-\f{1}{2};\vec[p]_1,\vec[p]_2,\vec[p]_3,\vec[p]_4\right)+ \TF^{U_d}\left(+\f{1}{2},-\f{1}{2},-\f{1}{2},+\f{1}{2};\vec[p]_1,\vec[p]_2,\vec[p]_3,\vec[p]_4\right)\biggr),\nn\\
\TF^{U_e}\left(\vec[p]_1,\vec[p]_2,\vec[p]_3,\vec[p]_4\right) &= \f{1}{4} \biggl(\TF^{U_e}\left(+\f{1}{2},+\f{1}{2},-\f{1}{2},-\f{1}{2};\vec[p]_1,\vec[p]_2,\vec[p]_3,\vec[p]_4\right)+\TF^{U_e}\left(-\f{1}{2},-\f{1}{2},+\f{1}{2},+\f{1}{2};\vec[p]_1,\vec[p]_2,\vec[p]_3,\vec[p]_4\right)\nn\\
&+ \TF^{U_e}\left(-\f{1}{2},+\f{1}{2},-\f{1}{2},+\f{1}{2};\vec[p]_1,\vec[p]_2,\vec[p]_3,\vec[p]_4\right)+ \TF^{U_e}\left(+\f{1}{2},-\f{1}{2},+\f{1}{2},-\f{1}{2};\vec[p]_1,\vec[p]_2,\vec[p]_3,\vec[p]_4\right)\biggr).
\end{align}
\end{small}

\subsubsection{Particle--anti-particle scattering}\label{N3covpapons}
Consider the scattering of a particle and an anti-particle,
\beq
P^i(p_1)+A_j(p_2)\to P_m(p_3) +A^n(p_4).
\eeq
The bosonic/fermionic asymptotic states are defined as
\beq\label{stateapcov}
\begin{array}{| c | c |}
\hline
\text{in-state} & \text{out-state}\\
\hline
\f{1}{2}b_j^{\dagger B}(p_2)a^{i\dagger}_A(p_1)|0\ran & \f{1}{2}\lan 0 | a^D_m(p_4) b_C^n(p_3)\\
\f{1}{2}\be^{\dagger B}_j(p_2)\al^{i\dagger}_A(p_1)|0\ran & \f{1}{2}\lan 0 | \al^D_m(p_4) \be^n_C(p_3)\\
\hline
\end{array}
\eeq
Let us define super-creation/annihilation operators for anti-particles as
\begin{align}
 B_{a}^i (x_a,\eta_a,\vec[p]_a) &= b^{i}_A (\vec[p]_a) x^{a A} +\be^{i }_A(\vec[p]_a)\eta^{aA},\\ 
 B_{a i}^\dag(\bar{x}_a,\bar{\eta}_a,\vec[p]_a) &= \bar{x}_{aA} b_{ i}^{\dag A}(\vec[p]_a) +\bar{\eta}_{aA} \be_{i}^{A\dag}(\vec[p]_a),
\end{align}
where $a=1,\ldots 4$.  We define the $S$-matrix for particle--anti-particle scattering formally as
\begin{small}
\begin{equation}
\tilde{S}(x,\eta,\bar{x},\bar{\eta},\vec[p])  =\sqrt{(2p_1^0)(2p_2^0)(2p_3^0)(2p_4^0)}\lan 0|B_{4}^n(x_4,\eta_4,\vec[p]_4)  A_{3 m}(\bar{x}_3,\bar{\eta}_3,\vec[p]_3)(1+i L_4)  B_{2 j}^{\dag}(\bar{x}_2,\bar{\eta}_2,\vec[p]_2) A_{1}^{\dag i}(x_1,\eta_1,\vec[p]_1) |0\ran,
\end{equation}
\end{small}
where $L_4$ is defined in \eqref{effon}. As discussed in \S\ref{skin}, the amplitude can be decomposed into the adjoint and singlet channels as
\begin{align}
\tilde{S}(x,\eta,\bar{x},\bar{\eta},\vec[p]) = \left(\de_m^i \de_j^n -\f{\de_j^i \de_m^n}{N}\right)S_T(x,\eta,\bar{x},\bar{\eta},\vec[p])+\f{\de_j^i \de_m^n}{N}\, S_S(x,\eta,\bar{x},\bar{\eta},\vec[p]).
\end{align}
The bosonic and fermionic $S$-matrices can be read off from
\begin{align}
\maS_B^{T}(\vec[p]_1,\vec[p]_2,\vec[p]_3,\vec[p]_4) = (x_1^A \bar{x}^3_A) (\bar{x}_{2B} x^{4 B}) \left(I(\vec[p]_1,\vec[p]_2,\vec[p]_3,\vec[p]_4) +i\TB^{T}\left(\vec[p]_1,\vec[p]_2,\vec[p]_3,\vec[p]_4\right)\right),\nn\\
\maS_B^{S}(\vec[p]_1,\vec[p]_2,\vec[p]_3,\vec[p]_4) = (x_1^A \bar{x}^3_A) (\bar{x}_{2B} x^{4 B}) \left(I(\vec[p]_2,\vec[p]_1,\vec[p]_3,\vec[p]_4)+i\TB^{S}\left(\vec[p]_1,\vec[p]_2,\vec[p]_3,\vec[p]_4\right)\right),\nn\\
\maS_F^{T}(\vec[p]_1,\vec[p]_2,\vec[p]_3,\vec[p]_4) = (\et_1^A \bar{\et}^3_A) (\bar{\et}_{2B} \et^{4 B}) \left(I(\vec[p]_1,\vec[p]_2,\vec[p]_3,\vec[p]_4) +i\TF^{T}\left(\vec[p]_1,\vec[p]_2,\vec[p]_3,\vec[p]_4\right)\right),\nn\\
\maS_F^{S}(\vec[p]_1,\vec[p]_2,\vec[p]_3,\vec[p]_4) = (\et_1^A \bar{\et}^3_A) (\bar{\et}_{2B} \et^{4 B})\left(I(\vec[p]_2,\vec[p]_1,\vec[p]_3,\vec[p]_4)+i\TF^{S}\left(\vec[p]_1,\vec[p]_2,\vec[p]_3,\vec[p]_4\right)\right).
\end{align}
Similar to our earlier discussion on particle-particle scattering, we can decompose the amplitude into the various component amplitudes discussed in \S\ref{astate} as
\begin{small}
\begin{align}
\label{bcovtononcovpa}
\TB^{T}\left(\vec[p]_1,\vec[p]_2,\vec[p]_3,\vec[p]_4\right)= \f{1}{4} \biggl(&\TB^{T}\left(-\f{1}{2},+\f{1}{2},+\f{1}{2},-\f{1}{2};\vec[p]_1,\vec[p]_2,\vec[p]_3,\vec[p]_4\right)+ \TB^{T}\left(+\f{1}{2},-\f{1}{2},-\f{1}{2},+\f{1}{2};\vec[p]_1,\vec[p]_2,\vec[p]_3,\vec[p]_4\right)\nn\\
&+ \TB^{T}\left(+\f{1}{2},+\f{1}{2},-\f{1}{2},-\f{1}{2};\vec[p]_1,\vec[p]_2,\vec[p]_3,\vec[p]_4\right)+\TB^{T}\left(-\f{1}{2},-\f{1}{2},+\f{1}{2},+\f{1}{2};\vec[p]_1,\vec[p]_2,\vec[p]_3,\vec[p]_4\right)\biggr),\nn\\
\TB^{S}\left(\vec[p]_1,\vec[p]_2,\vec[p]_3,\vec[p]_4\right)= \f{1}{4} \biggl(& \TB^{S}\left(-\f{1}{2},+\f{1}{2},+\f{1}{2},-\f{1}{2};\vec[p]_1,\vec[p]_2,\vec[p]_3,\vec[p]_4\right)+ \TB^{S}\left(+\f{1}{2},-\f{1}{2},-\f{1}{2},+\f{1}{2};\vec[p]_1,\vec[p]_2,\vec[p]_3,\vec[p]_4\right)\nn\\
&+\TB^{S}\left(+\f{1}{2},+\f{1}{2},-\f{1}{2},-\f{1}{2};\vec[p]_1,\vec[p]_2,\vec[p]_3,\vec[p]_4\right)+\TB^{S}\left(-\f{1}{2},-\f{1}{2},+\f{1}{2},+\f{1}{2};\vec[p]_1,\vec[p]_2,\vec[p]_3,\vec[p]_4\right)\biggr),
\end{align}
\end{small}
and an identical decomposition for the fermionic amplitudes,
\begin{small}
\begin{align}
\label{fcovtononcovpa}
\TF^{T}\left(\vec[p]_1,\vec[p]_2,\vec[p]_3,\vec[p]_4\right)= \f{1}{4} \biggl(& \TF^{T}\left(-\f{1}{2},+\f{1}{2},-\f{1}{2},+\f{1}{2};\vec[p]_1,\vec[p]_2,\vec[p]_3,\vec[p]_4\right)+ \TF^{T}\left(+\f{1}{2},-\f{1}{2},+\f{1}{2},-\f{1}{2};\vec[p]_1,\vec[p]_2,\vec[p]_3,\vec[p]_4\right)\nn\\
&+\TF^{T}\left(+\f{1}{2},+\f{1}{2},-\f{1}{2},-\f{1}{2};\vec[p]_1,\vec[p]_2,\vec[p]_3,\vec[p]_4\right)+\TF^{T}\left(-\f{1}{2},-\f{1}{2},+\f{1}{2},+\f{1}{2};\vec[p]_1,\vec[p]_2,\vec[p]_3,\vec[p]_4\right)\biggr),\nn\\
\TF^{S}\left(\vec[p]_1,\vec[p]_2,\vec[p]_3,\vec[p]_4\right)= \f{1}{4} \biggl(&\TF^{S}\left(-\f{1}{2},+\f{1}{2},-\f{1}{2},+\f{1}{2};\vec[p]_1,\vec[p]_2,\vec[p]_3,\vec[p]_4\right)+ \TF^{S}\left(+\f{1}{2},-\f{1}{2},+\f{1}{2},-\f{1}{2};\vec[p]_1,\vec[p]_2,\vec[p]_3,\vec[p]_4\right)\nn\\
&+ \TF^{S}\left(+\f{1}{2},+\f{1}{2},-\f{1}{2},-\f{1}{2};\vec[p]_1,\vec[p]_2,\vec[p]_3,\vec[p]_4\right)+\TF^{S}\left(-\f{1}{2},-\f{1}{2},+\f{1}{2},+\f{1}{2};\vec[p]_1,\vec[p]_2,\vec[p]_3,\vec[p]_4\right)\biggr).
\end{align}
\end{small}

\subsection{Unitarity}\label{unitN3cov}
As explained in \S\ref{skin}, the scattering amplitudes in the $U_d, U_e, T$ channels are $\mO\left(\f{1}{N}\right)$. Hence, Hermiticity of the amplitude is sufficient to ensure unitarity. More precisely, as a function of the momentum, the amplitudes must satisfy 
\begin{align}\label{unitaritycon}
\mathcal{T}_{B/F}^{U_d}(\vec[p]_1, \vec[p]_2,\vec[p]_3,\vec[p]_4) = \mathcal{T}_{B/F}^{U_d *}(\vec[p]_3, \vec[p]_4,\vec[p]_1,\vec[p]_2),\nn\\
\mathcal{T}_{B/F}^{U_e}(\vec[p]_1, \vec[p]_2,\vec[p]_3,\vec[p]_4) = \mathcal{T}_{B/F}^{U_e *}(\vec[p]_3, \vec[p]_4,\vec[p]_1,\vec[p]_2),\nn\\
\mathcal{T}_{B/F}^{T}(\vec[p]_1, \vec[p]_2,\vec[p]_3,\vec[p]_4) = \mathcal{T}_{B/F}^{T *}(\vec[p]_3, \vec[p]_4,\vec[p]_1,\vec[p]_2).
\end{align}
For the singlet channel, the unitarity equation is non-linear and requires an explicit analytic solution of the on-shell superspace constraints, especially the phase dependence and its relation to the sign of the mass. We address the unitarity for the singlet channel in a future work.

\section{Exact computation of the $S$-matrix to all orders}\label{exaccomp}
In this section, we present the computation of the exact propagator and the exact four point correlation function of the scalar superfield by setting up the Dyson-Schwinger equation in $\mN=1$ superspace. In \S\ref{onsmatrix} We will directly extract the S matrices in the symmetric, anti-symmetric and adjoint channels by explicitly taking the on-shell limit of the correlators constructed in this section. The four point correlators can also be used to construct correlation functions of conserved currents such as in the $\mN=1,2$ theories \cite{Aharony:2019mbc,Inbasekar:2019wdw}. 
\subsection{Supersymmetric light cone gauge}\label{slc}
For the rest of the paper, we employ the supersymmetric light cone gauge, given by the condition
\beq
\Ga_-=0.
\eeq
In components, this sets the gauge condition $A_-= A_1+i A_2=0$. The gauge is preserved under supersymmetry transformations and is referred to as the supersymmetric light cone gauge (\cite{Gates:1983nr}, also see Appendix F of 
\cite{Inbasekar:2015tsa}). The action \eqref{N3eucS} in the supersymmetric light cone gauge takes the form
\begin{small}
\begin{align}
\label{actionlcgauge1}
\mS^E_{\mN=3}=-&\int  d^3 x d^2\te \bigg[\f{\ka}{16\pi}
Tr(\Ga^-i\p_{--}\Ga^-)-\sum_{a=\pm}\f{1}{2}D^\al\bar{\Ph}^aD_\al\Ph^a-\f{i}{2}\Ga^-(\bar{\Ph}^a
D_-\Ph^a-D_-\bar{\Ph}^a\Ph^a)\nn\\
&- \f{\pi}{\ka} \left(\bar\Phi^{+}\Phi^{+}\right)
\left(\bar\Phi^{+}\Phi^{+}\right)-\f{\pi}{\ka}
\left(\bar\Phi^{-}\Phi^{-}\right) \left(\bar\Phi^{-}\Phi^{-}\right)
+\f{4\pi}{\ka}\left(\bar\Phi^{+}\Phi^{+}\right) \left(\bar\Phi^{-}\Phi^{-}\right)+\f{2\pi}{\ka}
\left(\bar\Phi^{+}\Phi^{-}\right) \left(\bar\Phi^{-}\Phi^{+}\right)\nn\\
& -(m_0 \bPh^+\Ph^+ - m_0 \bPh^-\Ph^-)\biggr],
\end{align}
\end{small}
where the $\pm$ indices on the scalar superfields indicate their $R$-charges, with the barred scalars carrying the opposite $R$-charge. The $\pm$ indices on the gauge superfield, super covariant derivatives etc.\ are two component spinor indices, see \S\ref{conv}. Note further that the triple gauge boson vertex vanishes in this gauge. 

\subsection{Exact propagators}
\label{exacpro}
In the $\mN=3$ theory the mass parameter is a central charge in the supersymmetry algebra, \eqref{sustn3md}, and is protected against quantum corrections. Thus the exact scalar superfield propagators in the $\mN=3$ theory are same as the bare propagators. On the other hand, the propagator for the Chern-Simons gauge field does not receive any quantum corrections to leading order in large $N$ \cite{Jain:2014nza,Inbasekar:2015tsa}. However, for pedagogical completeness, we construct the Dyson-Schwinger equations below and show that the exact scalar superfield propagators are identical to the bare propagators in the $\mN=3$ theory. 

\begin{figure}[h]
\begin{center}
\includegraphics[width=12.5cm,height=3.3cm]{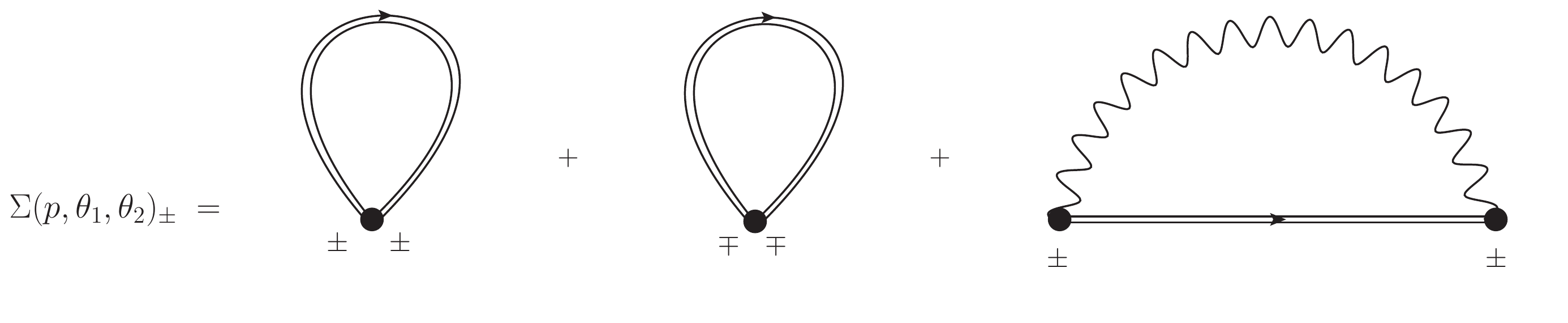}
\caption{\label{sigma3fig}Integral equation for self-energy. The double lines indicate the exact propagator. $\Si_\pm$ is the self-energy correction for the scalar superfield $\Ph^\pm$. The first diagram on the RHS is the
contribution from the vertex $(\bar{\Ph}^\pm \Ph^\pm)^2$, the second contribution is from the vertex $(\bPh^+\Ph^+)(\bPh^-\Ph^-)$, while the last diagram is the contribution from the gauge superfield exchange. The $\pm$ symbols denote the $R$-charge of the external superfields.}
\end{center}
\end{figure}

The bare propagators for the superfields that follow from the Lagrangian \eqref{actionlcgauge1} are
\begin{align}
 \lan\bar{\Ph}^\pm(\te_1,p)\Ph^\pm(\te_2,-p')\ran&=
\f{D^2_{\te_1,p}\pm m_0}{p^2+m_0^2}\de^2(\te_1-\te_2)
(2\pi)^3 \de^3(p-p'), \\
\lan\Ga^-(\te_1,p)\Ga^-(\te_2,-p')\ran &= -\f{8\pi}{\ka}\f{\de^2(\te_1-\te_2)}{p_{--}}(2\pi)^3
\de^3(p-p'), 
\end{align}
where $p_{--}=-(p_1+ip_2)=-p_-$. Note that upon $m\to -m$ the propagators for the two scalar superfields $\Ph^\pm$ map into one another. The exact 1PI quadratic effective action takes the form
\begin{align}
 S_2 = \int \f{d^3 p}{(2\pi)^3} \, d^2\te_1\, d^2\te_2 \,\bPh^\pm(-p,\te_1)\left( \exp(-\te_1^\al p_{\al\be}\te_2^\be) \pm m_0
\de^2(\te_1-\te_2)\right)\Ph^\pm(p,\te_2).
\end{align}
The general Grassmannian structure of the corrections in the effective action and in the two-point function are fixed by supersymmetric Ward identities (see \S 3.3 of \cite{Inbasekar:2015tsa}). The integral equation for the exact propagator for the $\Ph^\pm$ fields is schematically given in fig.\ \ref{sigma3fig}.
From the supersymmetric Ward identities for $\mN=1$ superspace, it follows that the exact propagator and the exact self-energy have the general expressions
\begin{align} \label{pform3}
P(\te_1,\te_2,p)_\pm&= \exp(-\te_1^\al p_{\al\be}\te_2^\be)\left(C_1(p^\mu)_\pm + C_2(p^\mu)_\pm
\de^2(\te_1-\te_2)\right),
\end{align}
\begin{equation}\label{fse3}
\Sigma(p, \theta_1, \theta_2)_\pm = C(p)_\pm\exp(-\te_1^\al p_{\al\be}\te_2^\be)+
D(p)_\pm \de^2(\te_1-\te_2).
\end{equation} 
Note that $C_1(p), C(p)$ have dimensions of $m^{-2}$, whereas both
$C_2(p), D(p)$ are of dimension $m^{-1}$.

The total contribution to the self-energy as illustrated in fig.\ \ref{sigma3fig} is 
\begin{align}
\Sigma(p, \theta_1, \theta_2)_\pm =& -2\pi\la  \int \f{d^3r}{(2\pi)^3} \de^2(\te_1-\te_2)P(r,
\theta_1,\theta_2)_\pm \nn\\
&+4\pi\la  \int \f{d^3r}{(2\pi)^3} \de^2(\te_1-\te_2)P(r, \theta_1,\theta_2)_\mp \nn\\
&  \  + 2\pi\la  \int \f{d^3r}{(2\pi)^3} \f{\de^2(\te_1-\te_2)}{(p-r)_{--}} D_-^{\te_1,r} D_-^{\te_2,-r} P(r,\te_1,\te_2)_{\pm}\, .
\end{align}
The first two lines are contributions from the two contact interactions in fig.\ \ref{sigma3fig}. The third line is the contribution from the gauge field exchange. The simplification of the last line is identical to that in equation 3.17 of \cite{Inbasekar:2015tsa}, and we will not repeat it here. Simplifying, we find that
\begin{align}\label{inteb3}
\Sigma(p, \theta_1, \theta_2)_\pm =& -2\pi\la  \int \f{d^3r}{(2\pi)^3} \de^2(\te_1-\te_2)P(r,
\theta_1,\theta_2)_\pm \nn\\
&+4\pi\la  \int \f{d^3r}{(2\pi)^3} \de^2(\te_1-\te_2)P(r, \theta_1,\theta_2)_\mp \nn\\
&  \  - 2\pi\la  \int \f{d^3r}{(2\pi)^3} \de^2(\te_1-\te_2)P(r, \theta_1,\theta_2)_\pm \ .
\end{align}
In the above, since the RHS is independent of $p$, it follows that
\beq
\Sigma(p, \theta_1, \theta_2)_\pm= (m \pm m_0)\,\delta^2(\theta_1-\theta_2),
\eeq
where $m$ is the renormalized mass. It follows that the exact propagator $P$ takes the form
of the tree-level propagator with $m_0$ replaced by $m$ i.e. 
\begin{equation}\label{ep3}
P(p, \theta_1, \theta_2)_\pm
= \frac{D^2\pm m }{ p^2 +  m^2} \, \de^2(\te_1- \te_2) .
\end{equation}
Substituting the expressions for the exact propagator \eqref{ep3} in \eqref{inteb3} and performing the Grassmann integrals we obtain
\begin{align} \label{deb3}
m\pm m_0&= -4\pi \la \int \f{d^3r}{(2\pi)^3} \f{1}{r^2+m^2} +4\pi \la \int
\f{d^3r}{(2\pi)^3} \f{1}{r^2+ m^2}.
\end{align}
The integrands cancel each other exactly and hence there is no mass renormalization of the matter fields in the $\mN=3$ theory as it should be. Thus the exact propagators are
\begin{align}\label{exactpropN3}
P_{\bPh^\pm\Ph^\pm}(\te_1,\te_2,p)\equiv \lan\bar{\Ph}^\pm(\te_1,p)\Ph^\pm(\te_2,-p')\ran&=
\f{D^2_{\te_1,p}\pm m}{p^2+m^2}\de^2(\te_1-\te_2)
(2\pi)^3 \de^3(p-p'), \\
P_{\Ga}(\te_1,\te_2,p)\equiv \lan\Ga^-(\te_1,p)\Ga^-(\te_2,-p')\ran &=
-\f{8\pi}{\ka}\f{\de^2(\te_1-\te_2)}{p_{--}}(2\pi)^3 \de^3(p-p').
\end{align}

\subsection{Organizing the Dyson-Schwinger equations for the exact four-point function}
\label{org}
\begin{figure}[h]
\begin{center}
\includegraphics[width=15cm,height=4cm]{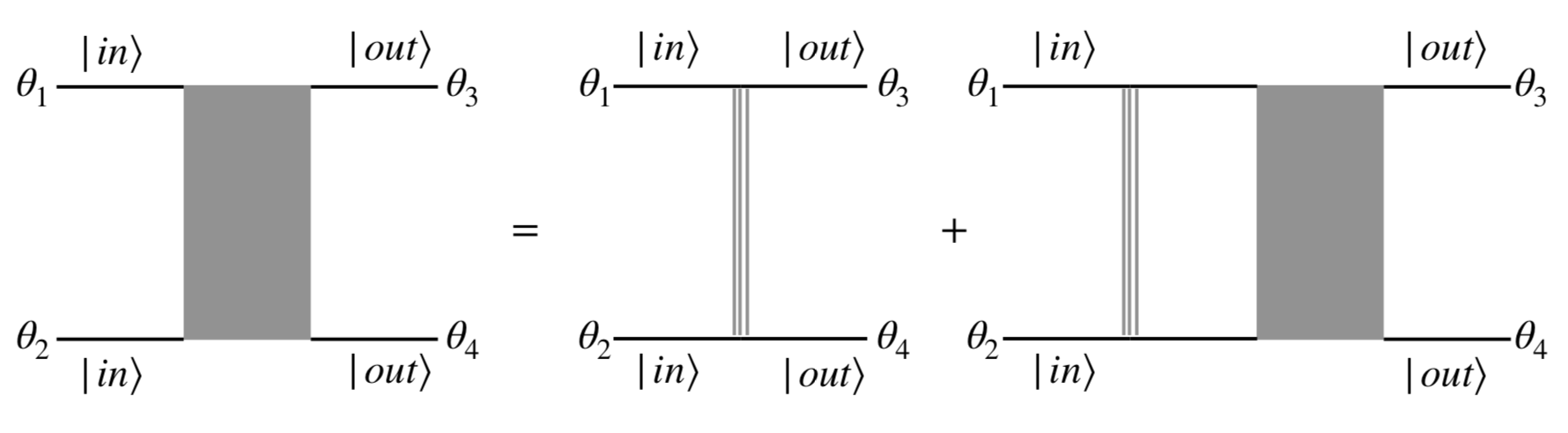}
  \caption{\label{DS} A schematic representation of the Dyson-Schwinger diagram for computing the planar all-loop four-point correlator. The striped diagrams are tree-level contributions. The solid blob is the contribution from all loops.}
\end{center}
\end{figure}

We now set up the Dyson-Schwinger integral equation for computing the exact four-point correlator to all orders in the \rq t Hooft coupling $\la$. First, we organize the various possible diagrams in terms of the $SO(2)_R$ charges of the in- and out-states that appear in the Dyson-Schwinger diagram of fig.\ \ref{DS}. We emphasize that these are \emph{not on-shell states}, the state terminology here simply refers to the $R$-charge carried by the external field in the diagrams for lack of a better word.

\begin{figure}[h]
\begin{center}
\includegraphics[width=15.2cm,height=7.6cm]{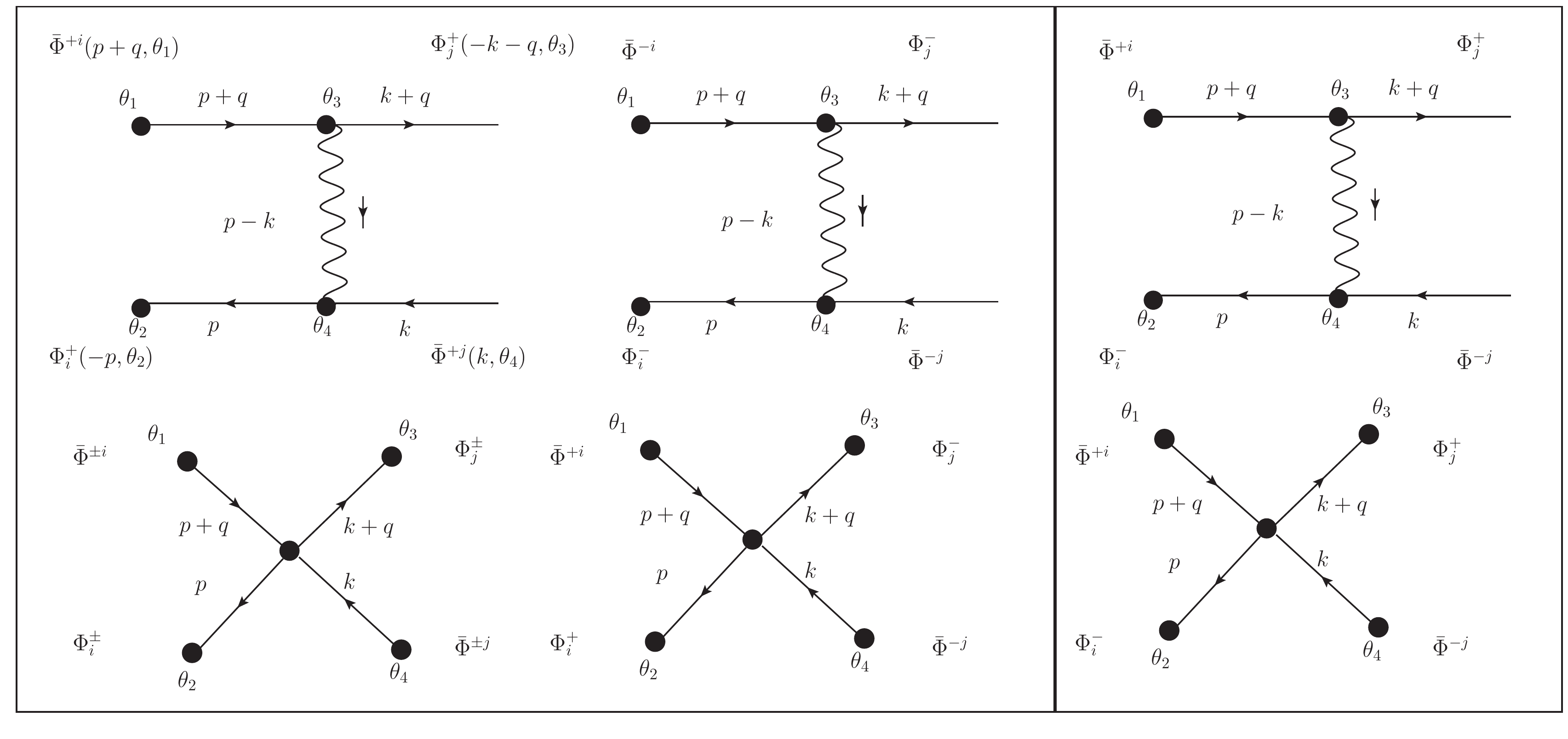}
\caption{\label{N3tree} In the above, the first block encloses the diagrams where the total $SO(2)_R$ charge of the in/out states respectively is zero. The second block constitutes diagrams that have in/out states with charge $\pm 1$. See \S \ref{N3con} for the notations and conventions. The color lines flow from anti-fundamentals to fundamentals. }
\end{center}
\end{figure}

In order to understand the classification it is instructive to observe all the tree-level diagrams that follow from \eqref{actionlcgauge1}. These are displayed in fig.\ \ref{N3tree}.
First, recall that the superfields $(\Ph^+,\Ph^-)$ have $R$-charges $(\f{1}{2},-\f{1}{2})$ respectively, and that the complex conjugates $(\bPh^+,\bPh^-)$ have $R$-charges $(-\f{1}{2},\f{1}{2})$ respectively. Using the $R$-charge assignments the in/out-states that appear in the Dyson-Schwinger series shown in fig.\ \ref{DS} can be divided into two independent sets of diagrams as shown in fig.\ \ref{N3tree}. Henceforth we have the following off-shell amplitudes with zero $R$-charge or charge neutral (in/out) states,
\begin{align}\label{corneut}
 M = \begin{pmatrix}
      \lan \bPh^+(\te_1)\Ph^+(\te_2)\Ph^+(\te_3)\bPh^+(\te_4)\ran & \lan \bPh^+(\te_1)\Ph^+(\te_2) \Ph^-(\te_3)\bPh^-(\te_4)\ran\\
      \lan \bPh^-(\te_1)\Ph^-(\te_2)\Ph^+(\te_3)\bPh^+(\te_4)\ran & \lan \bPh^-(\te_1)\Ph^-(\te_2)\Ph^-(\te_3)\bPh^-(\te_4)\ran
     \end{pmatrix}.
\end{align}
We refer to the above sector of correlators as the ``neutral sector''. The corresponding tree-level diagrams are in the first block of fig.\ \ref{N3tree}. The second set of diagrams have $\pm1$ $R$-charge for the (in/out) states. These correspond to the correlator
\beq\label{corchar}
\lan \bPh^-(\te_1)\Ph^+(\te_2)\Ph^-(\te_3)\bPh^+(\te_4)\ran
\eeq
and its complex conjugate. We refer to the diagrams that contribute to this correlator as the ``charged sector''. It is important to observe that due to the superfield interactions in \eqref{actionlcgauge1}, the diagrams between the charged and the neutral sector do not mix in the planar limit. This simple observation implies that the Dyson-Schwinger series is decoupled between the charged and neutral sectors. 

\subsection{Off-Shell four-point function}
\label{off4pt}
The $\mN=1$ Ward identity fixes the general form of any four-point correlator of a scalar superfield
\beq
  \lan \bPh(\te_1,p+q)\Ph(\te_2,-p)\Ph(\te_3,k)\bPh(\te_4,-k-q)\ran \equiv V(\te_1,\te_2,\te_3,\te_4,p,q,k)
\eeq
to be of the form \cite{Inbasekar:2015tsa}
\begin{align}
\label{totform}
&V(\te_1,\te_2,\te_3,\te_4,p,q,k)=\exp\bigg(\f{1}{4}X.(p.X_{12}+q.X_{13}+k.X_{43})\bigg)
F(X_{12},X_{13},X_{43},p,q,k),\nn\\
&F(X_{12},X_{13},X_{43},p,q,k)= X_{12}^+ X_{43}^+ \bigg( A(p,k,q) X_{12}^-
X_{43}^-X_{13}^+X_{13}^-+B(p,k,q) X_{12}^- X_{43}^-\nn\\ 
&\hspace{65mm}+ C(p,k,q)  X_{12}^- X_{13}^+ + D(p,k,q) X_{13}^+ X_{43}^-\bigg).
\end{align}
In the above we have used the notation
\beq
X=\sum_{i=1}^4 \te_i\ ,  \ X_{ij}=\te_i-\te_j.
\eeq
The form of $F$ was fixed in \cite{Inbasekar:2015tsa} by demanding associativity of the four-point function under multiplication. For a detailed discussion of closure of the structure \eqref{totform} under multiplication see Appendix H.3 of \cite{Inbasekar:2015tsa}. The $+,-$ on the difference variables $X_{ab}^\pm$ correspond to the two components of the Grassmann variable. For the $\mN=3$ theory, it is a straightforward exercise to compute the tree-level diagrams (see \S \ref{treeamp}) and check that they preserve the Grassmann structure \eqref{totform}.

As we discussed in \S\ref{org}, we will compute correlators in the charged and the neutral sectors. We formally define them below. In the ``neutral sector'' we have to evaluate the correlators
\begin{align}\label{neutcor}
\lan
\bar{\Ph}^+((p+q),\te_1)\Ph^+(-p,\te_2)\Ph^+(-(k+q),\te_3)\bar{\Ph}^+(k,\te_4) \ran
\equiv V_{\bar{\Ph}^+\Ph^+;\Ph^+\bar{\Ph}^+}(\te_1,\te_2,\te_3,\te_4,p,q,k),\nn\\
\lan
\bar{\Ph}^-((p+q),\te_1)\Ph^-(-p,\te_2)\Ph^-(-(k+q),\te_3)\bar{\Ph}^-(k,\te_4) \ran
\equiv V_{\bar{\Ph}^-\Ph^-;\Ph^-\bar{\Ph}^-}(\te_1,\te_2,\te_3,\te_4,p,q,k),\nn \\
\lan
\bar{\Ph}^+((p+q),\te_1)\Ph^+(-p,\te_2)\Ph^-(-(k+q),\te_3)\bar{\Ph}^-(k,\te_4) \ran
\equiv V_{\bar{\Ph}^+\Ph^+;\Ph^-\bar{\Ph}^-}(\te_1,\te_2,\te_3,\te_4,p,q,k), \nn\\
\lan\bar{\Ph}^-((p+q),\te_1)\Ph^-(-p,\te_2)\Ph^+(-(k+q),\te_3)\bar{\Ph}^+(k,\te_4) \ran
\equiv V_{\bar{\Ph}^-\Ph^-;\Ph^+\bar{\Ph}^+}(\te_1,\te_2,\te_3,\te_4,p,q,k),
\end{align}
and in the ``charged sector" we have to compute the  correlators
\begin{align}\label{chcor}
\lan
\bar{\Ph}^+((p+q),\te_1)\Ph^-(-p,\te_2)\Ph^+(-(k+q),\te_3)\bar{\Ph}^-(k,\te_4) \ran
\equiv V_{\bar{\Ph}^+\Ph^-;\Ph^+\bar{\Ph}^-}(\te_1,\te_2,\te_3,\te_4,p,q,k),\nn\\
\lan\bar{\Ph}^-((p+q),\te_1)\Ph^+(-p,\te_2)\Ph^-(-(k+q),\te_3)\bar{\Ph}^+(k,\te_4) \ran
\equiv V_{\bar{\Ph}^-\Ph^+;\Ph^-\bar{\Ph}^+}(\te_1,\te_2,\te_3,\te_4,p,q,k).
\end{align}
The most general 1PI action for both the ``charged'' and ``neutral'' sectors is given by
\begin{align}\label{eef1}
S_4&= \frac{1}{2} \int \frac{d^3 p}{(2 \pi)^3}  \frac{d^3 k}{(2 \pi)^3} 
\frac{d^3 q}{(2 \pi)^3} \, d^2 \theta_1 d^2 \theta_2 d^2 \theta_3 d^2 \theta_4 \\ 
&\biggl[\left(V_{\bar{\Ph}^+\Ph^+;\Ph^+\bar{\Ph}^+}(\te_1,\te_2,\te_3,\te_4,p,q,k)
\Phi^+_m(-(p+q), \theta_1) {\bar \Phi}^{m+}(p, \theta_2) {\bar \Phi}^{n+}(k+q, \theta_3)\Phi^+_n(-k,
\theta_4) \right) \nn\\
&+\left(V_{\bar{\Ph}^-\Ph^-;\Ph^-\bar{\Ph}^-}(\te_1,\te_2,\te_3,\te_4,p,q,k)
\Phi^-_m(-(p+q), \theta_1) {\bar \Phi}^{m-}(p, \theta_2) {\bar \Phi}^{n-}(k+q, \theta_3)\Phi^-_n(-k,
\theta_4) \right)\nn\\
&+\left(V_{\bar{\Ph}^+\Ph^+;\Ph^-\bar{\Ph}^-}(\te_1,\te_2,\te_3,\te_4,p,q,k)
\Phi^+_m(-(p+q), \theta_1) {\bar \Phi}^{m+}(p, \theta_2) {\bar \Phi}^{n-}(k+q,
\theta_3)\Phi^-_n(-k,\theta_4) \right)\nn\\
&+\left(V_{\bar{\Ph}^-\Ph^-;\Ph^+\bar{\Ph}^+}(\te_1,\te_2,\te_3,\te_4,p,q,k)
\Phi^-_m(-(p+q), \theta_1) {\bar \Phi}^{m-}(p, \theta_2) {\bar \Phi}^{n+}(k+q,
\theta_3)\Phi^+_n(-k,\theta_4) \right)\nn\\
&+\left(V_{\bar{\Ph}^+\Ph^-;\Ph^+\bar{\Ph}^-}(\te_1,\te_2,\te_3,\te_4,p,q,k)
\Phi^+_m(-(p+q), \theta_1) {\bar \Phi}^{m-}(p, \theta_2) {\bar \Phi}^{n+}(k+q,
\theta_3)\Phi^-_n(-k,\theta_4) \right)\nn\\
&+\left(V_{\bar{\Ph}^-\Ph^+;\Ph^-\bar{\Ph}^+}(\te_1,\te_2,\te_3,\te_4,p,q,k)
\Phi^-_m(-(p+q), \theta_1) {\bar \Phi}^{m+}(p, \theta_2) {\bar \Phi}^{n-}(k+q,
\theta_3)\Phi^+_n(-k,\theta_4) \right)\biggr].
\end{align}
We observe that the effective action is invariant under a $\mathbb{Z}_2$ symmetry transformation
\begin{align}\label{map}
& p\rightarrow k+q, k\rightarrow p+q ,q\rightarrow -q \ , \ \te_1\leftrightarrow \te_4 , \te_2\leftrightarrow \te_3,
\end{align}
which relates the following correlators under the action of \eqref{map},
\begin{align} \label{cormap}
V_{\bar{\Ph}^+\Ph^+;\Ph^-\bar{\Ph}^-}&\leftrightarrow V_{\bar{\Ph}^-\Ph^-;\Ph^+\bar{\Ph}^+},\nn\\
V_{\bar{\Ph}^+\Ph^-;\Ph^+\bar{\Ph}^-}&\leftrightarrow V_{\bar{\Ph}^-\Ph^+;\Ph^-\bar{\Ph}^+}.
\end{align}
This reduces the number of independent correlators to be computed in the neutral sector \eqref{neutcor} to three, and the number of independent correlators in the charged sector \eqref{chcor} to one.

\subsection{Integral equations for the four-point correlator}
\label{inteqall}
We now move on to compute the exact four-point correlators defined in \eqref{neutcor} and \eqref{chcor} to all orders in the \rq t Hooft coupling $\la$ by setting up a Dyson-Schwinger series. As explained in \S\ref{org} the ``charged'' and ``neutral'' sectors do not mix, hence we solve the integral equations in each case separately.

\subsubsection{Charged sector}\label{csec}
\begin{figure}[h]
\begin{center}
\includegraphics[width=15.2cm,height=6cm]{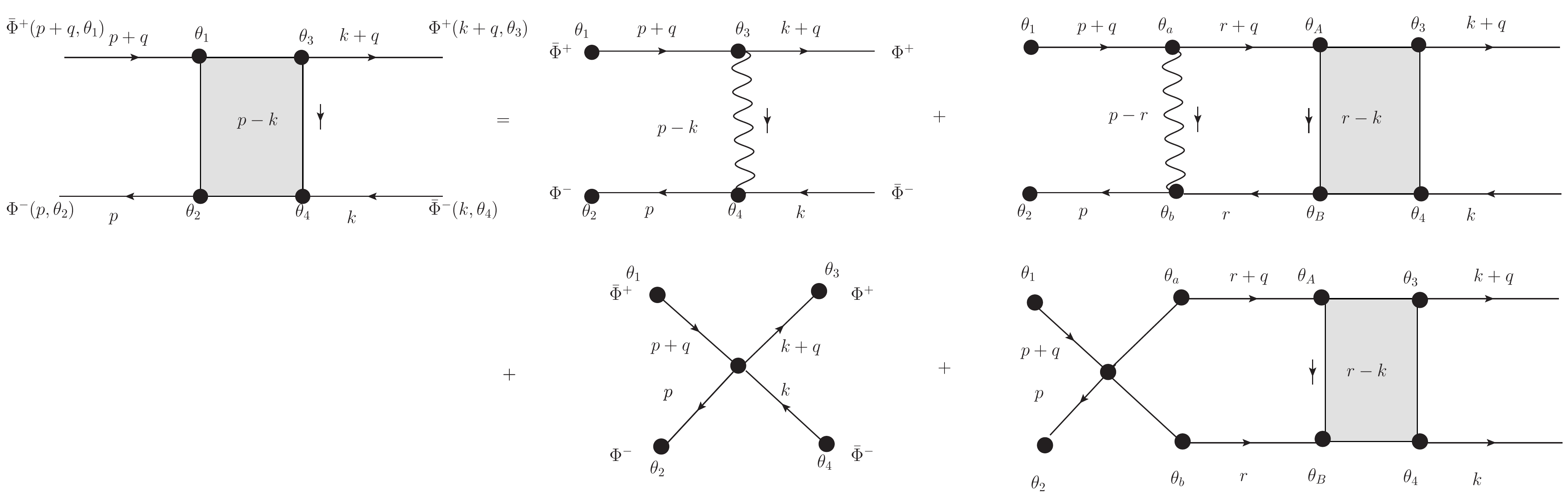}
\caption{\label{N3chDS} The Dyson-Schwinger series for the off-shell four-point function in the charged sector. The blob represents the off-shell four-point function to all orders in the \rq t Hooft coupling $\la$.}
\end{center}
\end{figure}

For the charged sector, the Dyson-Schwinger series is pictorially represented in fig.\ \ref{N3chDS}. We observe that the effective quartic coupling that enters the integral equation is identical to that of the 
$\mN=2$ theory \cite{Inbasekar:2015tsa}.\footnote{Note that the quartic coupling term in the $\mN=2$ theory is $\f{\pi}{\ka} (\bPh\Ph)(\bPh\Ph)$, and a symmetry factor of 2 enters the Dyson-Schwinger series due to Wick contractions. } One then expects that the correlation function in the charged sector would be identical to that of the $\mN=2$ theory. However, we caution the reader that due to the difference in signs for the mass terms in the propagators \eqref{exactpropN3} there are cancellations. Nevertheless, we find that the final result for the correlator in this sector indeed conforms to our expectations.
The integral equations that follow from fig.\ \ref{N3chDS} are of the form
\begin{align}\label{int4ptN3c}
&V_{\bar{\Ph}^+\Ph^-;\Ph^+\bar{\Ph}^-}(\te_1,\te_2,\te_3,\te_4,p,q,k)  =
V_{0:\bar{\Ph}^+\Ph^-;\Ph^+\bar{\Ph}^-}(\te_1,\te_2,\te_3,\te_4,p,q,k)\nn\\
&\qquad\qquad+ \int \f{d^3r}{(2\pi)^3} d^2\te_a d^2\te_b
d^2\te_A d^2\te_B \bigg(NV_{0:\bar{\Ph}^+\Ph^-;\Ph^+\bar{\Ph}^-}(\te_1,\te_2,\te_a,\te_b,p,q,r)\nn\\
&\qquad\qquad\qquad P_{\bPh^+\Ph^+}(r+q,\te_a,\te_A)
P_{\bPh^-\Ph^-}(r,\te_B,\te_b)V_{\bar{\Ph}^+\Ph^-;\Ph^+\bar{\Ph}^-}(\te_A,\te_B,\te_3,\te_4,r,q,
k)\bigg),
\end{align}
where $V_{0:\bar{\Ph}^+\Ph^-;\Ph^+\bar{\Ph}^-}(\te_1,\te_2,\te_3,\te_4,p,q,k)$ is the tree-level contribution \eqref{V0c}. The ansatz for the exact correlator takes the form
\begin{align}\label{totformcharged}
&V_{\bar{\Ph}^+\Ph^-;\Ph^+\bar{\Ph}^-}(\te_1,\te_2,\te_3,\te_4,p,q,k)=\exp\bigg(\f{1}{4}X.(p.X_{12}+q.X_{13}+k.X_{43})\bigg)
F_{\bar{\Ph}^+\Ph^-;\Ph^+\bar{\Ph}^-}(X_{12},X_{13},X_{43},p,q,k)\, ,\nn\\
&F_{\bar{\Ph}^+\Ph^-;\Ph^+\bar{\Ph}^-}(X_{12},X_{13},X_{43},p,q,k)= X_{12}^+ X_{43}^+ \bigg( A(p,k,q) X_{12}^-
X_{43}^-X_{13}^+X_{13}^-+B(p,k,q) X_{12}^- X_{43}^-\nn \\
&\hspace{80mm}+ C(p,k,q)  X_{12}^- X_{13}^+ + D(p,k,q) X_{13}^+ X_{43}^-\bigg).
\end{align}
Substituting the above in \eqref{int4ptN3c} and performing the Grassmann integrations, we find the component integral equations
\begin{small}
\begin{align}
\label{inteqch}
&A(p,k,q)+\f{2\pi i}{\ka}+i\pi\la \int \f{d^3r}{(2\pi)^3} \f{q_3\left(p_-(2 A-2B q_3-C k_-)+(4 A+2 C k_--3 D p_-+4 B q_3)r_-+2 D r_-^2\right) }{(r^2+m^2)((r+q)^2+m^2)(p-r)_-}=0,\nn\\
& B(p,k,q) + 2\pi i \la \int\f{d^3 r}{(2\pi)^3} \f{r_- (2 A-C k_-+2 B q_3+D r_-)}{(r^2+m^2)((r+q)^2+m^2)(p-r)_-}=0,\nn\\
&C(p,k,q)+\f{4\pi i}{\ka (k-p)_-} + 8 \pi i \la \int \f{d^3 r}{(2\pi)^3} \f{C q_3 r_-}{(r^2+m^2)((r+q)^2+m^2)(p-r)_-}=0,\nn\\
&D(p,k,q)+\f{4\pi i}{\ka (k-p)_-}+ 2 \pi i \la \int \f{d^3 r}{(2\pi)^3}\f{q_3 (-2 A+2 B q_3+C k_-+3 D r_-)}{(r^2+m^2)((r+q)^2+m^2)(p-r)_-}=0.
\end{align}
\end{small}
Note that these equations are the same as the equations 3.32 - 3.35 of \cite{Inbasekar:2015tsa} for the $\mN=2$ theory.\footnote{Upon substitution of $w=1$ (for the $\mN=2$ theory) in the integral equations 3.32 - 3.35 of \cite{Inbasekar:2015tsa}, the second and third equations match with \eqref{inteqch} exactly, while the first and fourth equations have an additional term proportional to
\beq
\int \f{d^3r}{(2\pi)^3}  \f{ 2 A(r,k,q) - C(r,k,q) k_- -2 B(r,k,q) q_3 + D(r,k,q) r_-}{(r^2+m^2)((r+q)^2+m^2)(p-r)_-}.
\eeq
However, it is easy to check that the integrand above vanishes on the solutions \eqref{solutionsN3ch}. } It follows then that the solutions of the unknown functions for the charged sector are the same as that for the $\mN=2$ theory. We found that the integral equations \eqref{inteqch} are solved in the kinematic regime $q_\pm=0$ by 
\begin{align}\label{solutionsN3ch}
A(p,k,q)=&-\f{2 i \pi  e^{2 i \la  \big(\tan ^{-1}\f{2
\sqrt{k_s^2+m^2}}{q_3}-\tan^{-1}\f{2
\sqrt{m^2+p_s^2}}{q_3}\big)}}{\ka }\ , \nn \\
B(p,k,q)=&\ 0 \ , \nn \\
C(p,k,q)=&-\f{4 i \pi  e^{2 i \la \big(\tan ^{-1}\f{2
\sqrt{k_s^2+m^2}}{q_3}-\tan ^{-1}\f{2
\sqrt{m^2+p_s^2}}{q_3}\big)}}{\ka(k-p)_-}\ ,
\nn\\
D(p,k,q)=&-\f{4 i \pi  e^{2 i \la  \big(\tan ^{-1}\f{2
\sqrt{k_s^2+m^2}}{q_3}-\tan ^{-1}\f{2
\sqrt{m^2+p_s^2}}{q_3}\big)}}{\ka(k-p)_-}\ .
\end{align}
Thus, the solutions \eqref{solutionsN3ch} above completely determine the correlators \eqref{chcor}.

\subsubsection{Neutral sector}
\label{nsec}
For the neutral sector, there are additional interactions which lead to a complicated set of coupled integral equations. For instance, it is clear from the first block of diagrams in fig.\ \ref{N3tree} that in the integral equation for $\lan(\bPh^+\Ph^+)(\bPh^+\Ph^+)\ran$ there will be intermediate contributions from $(\bPh^+\Ph^+ )(\bPh^-\Ph^-)$ and $(\bPh^-\Ph^-)(\bPh^-\Ph^-)$, each of which will have its own integral equation. Taking into account all the interactions, we write down the integral equations as follows. All the correlators defined in \eqref{neutcor} can be put in a matrix form given by
\begin{small}
\beq
M(\te_A,\te_B,\te_3,\te_4,p,q,k)=\begin{pmatrix}
   V_{\bar{\Ph}^+\Ph^+;\Ph^+\bar{\Ph}^+}(\te_A,\te_B,\te_3,\te_4,r,q,k) & V_{\bar{\Ph}^+\Ph^+;\Ph^-\bar{\Ph}^-}(\te_A,\te_B,\te_3,\te_4,r,q,k)\\
   V_{\bar{\Ph}^-\Ph^-;\Ph^+\bar{\Ph}^+}(\te_A,\te_B,\te_3,\te_4,r,q,k) & V_{\bar{\Ph}^-\Ph^-;\Ph^-\bar{\Ph}^-}(\te_A,\te_B,\te_3,\te_4,r,q,k)
  \end{pmatrix}.
\eeq
\end{small}
Similarly, we also write all the tree-level amplitudes \eqref{V0n} as
\begin{small}
\beq
M_0(\te_1,\te_2,\te_3,\te_4,p,q,k)=\begin{pmatrix}
   V_{0;\bar{\Ph}^+\Ph^+;\Ph^+\bar{\Ph}^+}(\te_1,\te_2,\te_3,\te_4,p,q,k) & 
   V_{0;\bar{\Ph}^+\Ph^+;\Ph^-\bar{\Ph}^-}(\te_1,\te_2,\te_3,\te_4,p,q,k)\\
   V_{0;\bar{\Ph}^-\Ph^-;\Ph^+\bar{\Ph}^+}(\te_1,\te_2,\te_3,\te_4,p,q,k) & 
   V_{0;\bar{\Ph}^-\Ph^-;\Ph^-\bar{\Ph}^-}(\te_1,\te_2,\te_3,\te_4,p,q,k)
  \end{pmatrix}.
\eeq
\end{small}
Using this notation, the integral equations in the neutral sector take the form
\begin{align}\label{intn}
 M(\te_1,\te_2,\te_3,\te_4,p,q,k)=&M_0(\te_1,\te_2,\te_3,\te_4,p,q,k)\nn\\
  &+\int \f{d^3r}{(2\pi)^3} \, d^2\te_a d^2\te_b d^2\te_A d^2\te_B \, N M_0(\te_1,\te_2,\te_a,\te_b,p,q,r) \nn\\
  &\quad \quad PP(\te_a,\te_b,\te_A,\te_B,r,q)M(\te_A,\te_B,\te_3,\te_4,r,q,k),
\end{align}
where (note that the exact propagators were defined in \eqref{exactpropN3}),
\begin{small}
\beq
 PP(\te_a,\te_b,\te_A,\te_B,r,q)=\begin{pmatrix}
 P_{\bPh^+\Ph^+}(r+q,\te_a,\te_A)
P_{\bPh^+\Ph^+}(r,\te_B,\te_b) & 0\\
 0 & P_{\bPh^-\Ph^-}(r+q,\te_a,\te_A)
P_{\bPh^-\Ph^-}(r,\te_B,\te_b)
\end{pmatrix}.
\eeq
\end{small}
For brevity, we introduce the notation
\begin{equation}
\begin{split}
&M_0= \begin{pmatrix}
V_0^{11} & V_0^{12}\\
V_0^{21} & V_0^{22}
\end{pmatrix}\ , \ M= \begin{pmatrix}
V^{11} & V^{12}\\
V^{21} & V^{22}
\end{pmatrix},\\
&PP=\begin{pmatrix}P_{11} & 0 \\ 0 & P_{22}\end{pmatrix} \ , \ \int d\mu = \int \f{d^3r}{(2\pi)^3}\, d^2\te_a d^2\te_b d^2\te_A d^2\te_B.
\end{split}
\end{equation}
Substituting the above in \eqref{intn}, we can read off the following coupled integral equations,
\begin{align}
V^{11} &= V_0^{11}+ \int d\mu  N V_0^{11} P_{11} V^{11} +\int d\mu NV_0^{12}P_{22} V^{21},\label{neut11}\\
V^{12} &= V_0^{12}+ \int d\mu  N V_0^{12} P_{22} V^{22} +\int d\mu NV_0^{11}P_{11} V^{12},\label{neut12}\\
V^{21} &= V_0^{21}+ \int d\mu  N V_0^{21} P_{11} V^{11} +\int d\mu NV_0^{22}P_{22} V^{21},\label{neut21}\\
V^{22} &= V_0^{22}+ \int d\mu  N V_0^{22} P_{22} V^{22} +\int d\mu NV_0^{21}P_{11} V^{12}.\label{neut22}
\end{align}
For the exact correlators we use
\begin{align}\label{totformneut}
&V^{ij}(\te_1,\te_2,\te_3,\te_4,p,q,k)=\exp\bigg(\f{1}{4}X.(p.X_{12}+q.X_{13}+k.X_{43})\bigg)
F^{ij}(X_{12},X_{13},X_{43},p,q,k),\nn\\
&F^{ij}(X_{12},X_{13},X_{43},p,q,k)= X_{12}^+ X_{43}^+ \bigg( A^{ij}(p,k,q) X_{12}^-X_{43}^-X_{13}^+X_{13}^-+B^{ij}(p,k,q) X_{12}^- X_{43}^-\nn \\
&\hspace{65mm}+ C^{ij}(p,k,q)  X_{12}^- X_{13}^+ + D^{ij}(p,k,q) X_{13}^+ X_{43}^-\bigg),
\end{align}
where $i,j=1,2$. Naively, there are sixteen unknown functions and sixteen coupled integral equations \eqref{neut11} - \eqref{neut22} (note that each equation is composed of four component equations). We simplify further by observing that $m\to-m$ is a symmetry of the integral equations, i.e.\ this operation maps \eqref{neut11} to \eqref{neut22}, \eqref{neut12} to \eqref{neut21}, and vice versa,
\beq\label{eqsym}
\begin{pmatrix}
A^{22} & B^{22} & C^{22} & D^{22}\\
A^{21} & B^{21} & C^{21} & D^{21}
\end{pmatrix}
\leftrightarrow
\begin{pmatrix}
A^{11} & B^{11} & C^{11} & D^{11}\\
A^{12} & B^{12} & C^{12} & D^{12}
\end{pmatrix} \text{under}\; m\to -m.
\eeq
Thus, we have effectively eight equations to solve. Obviously, this mapping is simply due to the fact that the propagators \eqref{exactpropN3} for the two superfields $\Phi^+$ and $\Ph^-$ are related by $m\to-m$. The final eight integral equations in components are
\begin{small}
\begin{align}
\label{inteqpp}
&A^{11}(p,k,q)-\f{2\pi i}{\ka}\nn\\
&\qquad+\pi \la \int \f{d^3 r}{(2\pi)^3} \biggl[\f{2 A^{11} (2 m (p-2 r)_-+i q_3 (p+2 r)_-)+(2 m+i q_3)(2 B^{11} p_- q_3 +C^{11} k_- p_- +D^{11} r_-  (p-2 r)_-)}{(p-r)_-(r^2+m^2)((r+q)^2+m^2)}\nn\\
&\hspace{72mm}+\f{2 (2 m (2 A^{21}+D^{21} r_-)-i q_3 (2 B^{21} q_3+C^{21} k_-+2 D^{21} r_-))}{(r^2+m^2)((r+q)^2+m^2)}\biggr]=0,\nn\\
&B^{11}(p,k,q)+\pi \la \int \f{d^3 r}{(2\pi)^3} \biggl[\f{2 i (2 A^{11} p_-+2 B^{11} q_3 r_--C^{11} k_- p_--D^{11} r_- (p-2 r)_-)}{(p-r)_-(r^2+m^2)((r+q)^2+m^2)}
\nn\\
&\hspace{40mm}+\f{2 (-2 i A^{21}+4 (B^{11}+B^{21}) m+i (C^{21} k_-+D^{21} r_-)) }{(r^2+m^2)((r+q)^2+m^2)}\biggr]=0,\nn\\
&C^{11}(p,k,q)+\f{4\pi i}{\ka} \f{1}{(k-p)_-} +4\pi \la \int \f{d^3r}{(2\pi)^3} \biggl[\f{i q_3 (p+r)_- C^{11}}{(p-r)_-(r^2+m^2)((r+q)^2+m^2)} \nn\\
&\hspace{70mm}+\f{ (2 m C^{11}+ C^{21}(2m-i q_3))}{(r^2+m^2)((r+q)^2+m^2)}\biggr]=0,\nn\\
& D^{11}(p,k,q)+\f{4\pi i}{\ka} \f{1}{(k-p)_-}+2\pi\la \int \f{d^3r}{(2\pi)^3}\biggl[\f{(2 m-i q_3)(2 A^{11} -2 B^{11} q_3 - k_- C^{11})+D^{11} r_- (2 m+3 i q_3)}{(p-r)_-(r^2+m^2)((r+q)^2+m^2)}\biggr]=0,
\end{align}
\end{small}
and
\begin{small}
\begin{align}
\label{inteqmp}
&A^{21}(p,k,q)+\f{4\pi i}{\ka} +\pi \la \int \f{d^3 r}{(2\pi)^3}\biggl[\f{-2 (2 m (2 A^{11}+D^{11} r_-)+i q_3 (2 B^{11} q_3+C^{11} k_-+2 D^{11} r_-))}{(r^2+m^2)((r+q)^2+m^2)}\nn\\
&\hspace{20mm}+\f{A^{21} (4 m (2 r-p)_-+2 i q_3 (p+2 r)_-)-(2 B^{21} p_- q_3 +C^{21} k_- p_- +D^{21} r_- (p-2 r)_-)(2 m-i q_3)}{(p-r)_-(r^2+m^2)((r+q)^2+m^2)}\bigg]=0,\nn\\
&B^{21}(p,k,q)+ 2 i \pi \la \int \f{d^3r}{(2\pi)^3}\biggl[\f{-2 A^{11}+4 i B^{11} m+C^{11} k_-+D^{11} r_-}{(r^2+m^2)((r+q)^2+m^2)}\nn\\
&\hspace{40mm}+\f{p_- (2 A^{21}+4 i B^{21} m-C^{21} k_--D^{21} r_-)+2 r_- (B^{21} (q_3-2i m)+D^{21} r_-)}{(p-r)_-((r^2+m^2)((r+q)^2+m^2)}\biggr]=0,\nn\\
&C^{21}(p,k,q)+4\pi\la \int \f{d^3 r}{(2\pi)^3} \biggl[ \f{ i C^{21} q_3 (p+r)_-}{(p-r)_-((r^2+m^2)((r+q)^2+m^2)} -\f{ (2 mC^{21}+C^{11} (2m+iq_3))}{((r^2+m^2)((r+q)^2+m^2)}\biggr]=0,\nn\\
&D^{21}(p,k,q)+2\pi \la \int \f{d^3r}{(2\pi)^3} \f{(2 m+i q_3)(-4 A^{21}+4 B^{21} q_3 +2 C^{21} k_-) +2 D^{21} r_- (-2 m+3 i q_3)}{{(p-r)_-(r^2+m^2)((r+q)^2+m^2)}}=0.
\end{align}
\end{small}
The other eight equations can of course be obtained by applying the rule \eqref{eqsym} to
the equations \eqref{inteqpp} and \eqref{inteqmp}. However, this still leaves us with eight complicated coupled integral equations to solve! Fortunately, the  $\mathbb{Z}_2$ symmetry \eqref{map} of the correlators lead to additional useful relations given by
\beq\label{consis}
\begin{pmatrix}
A^{ij}(p,k,q)\\
B^{ij}(p,k,q)\\
C^{ij}(p,k,q)\\
D^{ij}(p,k,q)
\end{pmatrix}=\begin{pmatrix}
1 & 0 & 0 & 0\\
0 & 1 & 0 & 0\\
0 & 0 & 0 & -1\\
0 & 0 & -1 & 0
\end{pmatrix}\begin{pmatrix}
A^{ji}(k,p,-q)\\
B^{ji}(k,p,-q)\\
C^{ji}(k,p,-q)\\
D^{ji}(k,p,-q)
\end{pmatrix}.
\eeq
Applying these relations to the integral equations above we are able to solve them in the kinematic regime $q_\pm=0$ for all the unknown functions, and express them in terms of the momenta $(p_s,k_s,q_3)$ and the coupling constant $\la$. The solutions are cumbersome, however, we present them below for the sake of completeness. They are
\begin{small}
\begin{align}
\label{solpp}
A^{11}&=\frac{\pi}{\kappa q_3} \Bigg((i q_3+2m)  \exp\left[{-2 i \lambda  \left(\tan ^{-1}\frac{2 |m|}{q_3}-\tan ^{-1}\frac{2 \sqrt{k_s^2+m^2}}{q_3}\right)}\right] \nn\\
&\qquad\qquad+(i q_3-2m) \exp\left[{2 i \lambda  \left(\tan ^{-1}\frac{2 |m|}{q_3}-\tan ^{-1}\frac{2 \sqrt{m^2+p_s^2}}{q_3}\right)}\right]\Bigg),\nn\\
B^{11}&=- \frac{2 i \pi }{\kappa  q_3} \left(\exp\left[{-2 i \lambda  \left(\tan ^{-1}\frac{2 |m|}{q_3}-\tan ^{-1}\frac{2 \sqrt{k_s^2+m^2}}{q_3}\right)}\right]-\exp\left[{2 i \lambda  \left(\tan ^{-1}\frac{2 |m|}{q_3}-\tan ^{-1}\frac{2 \sqrt{m^2+p_s^2}}{q_3}\right)}\right]\right),\nn\\
C^{11}&=\frac{2 \pi}{\kappa  k_- q_3}  \Bigg( \left(2 m-i q_3 \frac{(k+p)_-}{(k-p)_-}\right) \exp\left[{2 i \lambda  \left(\tan ^{-1}\frac{2 \sqrt{k_s^2+m^2}}{q_3}-\tan ^{-1}\frac{2 \sqrt{m^2+p_s^2}}{q_3}\right)}\right]\nn\\
&\qquad\qquad\qquad-(2 m+i q_3) \exp\left[{2 i \lambda  \left(\tan ^{-1}\frac{2 |m|}{q_3}-\tan ^{-1}\frac{2 \sqrt{m^2+p_s^2}}{q_3}\right)}\right]\Bigg),\nn\\
D^{11}&=\frac{2 \pi}{\kappa  p_- q_3}  \Bigg( \left(2 m-i q_3 \frac{(k+p)_-}{(k-p)_-}\right) \exp\left[{2 i \lambda  \left(\tan ^{-1}\frac{2 \sqrt{k_s^2+m^2}}{q_3}-\tan ^{-1}\frac{2 \sqrt{m^2+p_s^2}}{q_3}\right)}\right]\nn\\
&\qquad\qquad\qquad+(-2 m+i q_3) \exp\left[{-2 i \lambda  \left(\tan ^{-1}\frac{2 |m|}{q_3}-\tan ^{-1}\frac{2 \sqrt{m^2+k_s^2}}{q_3}\right)}\right]\Bigg),
\end{align}
\end{small}
and
\begin{small}
\begin{align}
\label{solmp}
A^{21}&=\frac{\pi  (2 m-i q_3) }{\kappa  q_3} {\left(\exp\left[{2 i \lambda  \left(\tan ^{-1}\frac{2 |m|}{q_3}-\tan ^{-1}\frac{2 \sqrt{m^2+p_s^2}}{q_3}\right)}\right]+\exp\left[{2 i \lambda \left( \tan ^{-1}\frac{2 \sqrt{k_s^2+m^2}}{q_3}-\tan ^{-1}\frac{2 |m|}{q_3}\right)}\right]\right)}\nn\\
&\quad- \frac{2 \pi  (2 m+i q_3)}{\kappa  q_3} \,{\exp\left[{2 i \lambda  \left(\tan ^{-1}\left(\frac{2 \sqrt{k_s^2+m^2}}{q_3}\right)-\tan ^{-1}\left(\frac{2 \sqrt{m^2+p_s^2}}{q_3}\right)\right)}\right]},\nn\\
B^{21}&= \frac{2 i \pi}{\kappa  q_3} {\left( \exp\left[-2 i \lambda  \left(\tan ^{-1}\frac{2 |m|}{q_3}-\tan ^{-1}\frac{2 \sqrt{k_s^2+m^2}}{q_3}\right)\right]-\exp\left[{2 i \lambda  \left(\tan ^{-1}\frac{2 |m|}{q_3}-\tan ^{-1}\frac{2 \sqrt{m^2+p_s^2}}{q_3}\right)}\right]\right)},\nn\\
C^{21}&=\frac{2 \pi  (2 m+i q_3)}{\kappa  k_- q_3} {\left(\exp\left[{2 i \lambda  \tan ^{-1}\frac{2 |m|}{q_3}}\right]-\exp\left[{2 i \lambda  \tan ^{-1}\frac{2 \sqrt{k_s^2+m^2}}{q_3}}\right]\right) \exp\left[{-2 i \lambda  \tan ^{-1}\frac{2 \sqrt{m^2+p_s^2}}{q_3}}\right]},\nn\\
D^{21}&=\frac{2 \pi  (2 m+i q_3) }{\kappa  p_- q_3} {\left(-\exp\left[{-2 i \lambda  \tan ^{-1}\frac{2 |m|}{q_3}}\right]+\exp\left[{-2 i \lambda  \tan ^{-1}\frac{2 \sqrt{p_s^2+m^2}}{q_3}}\right]\right) \exp\left[{2 i \lambda  \tan ^{-1}\frac{2 \sqrt{m^2+k_s^2}}{q_3}}\right]}.
\end{align}
\end{small}
The remaining functions can be obtained by using the map \eqref{eqsym}. We have checked explicitly that all the equations that arise from  \eqref{neut11} - \eqref{neut22} are indeed satisfied by the solutions above.
Thus, the solutions \eqref{solpp}, \eqref{solmp} completely determine the correlators defined in \eqref{neutcor} to all orders in the 't Hooft coupling $\la$. We therefore have completely determined all the off-shell four-point correlators defined in \S\ref{off4pt} for the $\mN=3$ theory in the planar limit. 

\subsection{On-shell limit and the $S$-matrix}
\label{onsmatrix}
The solutions in \S\ref{csec} and \S\ref{nsec} completely determine all the unknown $V$ functions in the quadratic 1PI effective action \eqref{eef1}, in the planar limit. The $S$-matrix is obtained, in the standard way, by substituting the on-shell superfield expansions and sandwiching the resultant expressions in between on-shell states. The on-shell superfield is defined as 
 \begin{align} \label{fmodeons}
2\pi \de(p^2+m^2)\Phi^\mp(p, \theta)|_{\text{on-shell}}&=2\pi \de(p^2+m^2)\biggl[(1\pm m\theta^2)\Big(\te(p^0) a^\mp(\vec[p]) +\te(-p^0) a^{\mp\dagger}_c(-\vec[p])\Big) \nn\\
&\quad\pm\te^\al\Big(\te(p^0)u_\alpha(\vec[p],\pm m) \alpha^\mp(\vec[p]) + \te(-p^0)v_\alpha(-\vec[p],\pm m) \alpha^{\mp\dagger}_c(-\vec[p]) \Big)\biggr],  
 \end{align}
with the complex conjugate field
 \begin{align} \label{fmodeons2}
2\pi \de(p^2+m^2)\bar{ \Phi}^\mp(p, \theta)|_{\text{on-shell}}&=2\pi \de(p^2+m^2)\biggl[(1\pm m\theta^2)\Big(\te(p^0) a^{\mp\dagger}(\vec[p]) +\te(-p^0) a^{\mp}_c(-\vec[p])\Big) \nn\\
&\quad\mp \te^\al\Big(\te(p^0)v_\alpha(\vec[p],\pm m) \alpha^{\mp\dagger}(\vec[p]) + \te(-p^0)u_\alpha(-\vec[p],\pm m) \alpha^{\mp}_c(-\vec[p]) \Big)\biggr].
 \end{align}
The creation and annihilation operators have their usual meaning and obey canonical commutation relations \eqref{stnc} and \eqref{stnac}. We refer to 
\S\ref{dir} for some basic properties of fermionic wave functions and 
\S\ref{fmon} for properties of the mode expansions in $\mN=1$ superspace. Substituting the on-shell mode expansion \eqref{fmodeons} into the effective action \eqref{eef1} and taking into account the total momentum conservation,\footnote{This accounts for some factors of $(2\pi)$ introduced in \eqref{fmodeons}.} we obtain a quartic polynomial in the creation/annihilation operators given by
\begin{align}\label{effon}
L_4&= \frac{1}{2} \int \frac{d^3 p}{(2 \pi)^3}  \frac{d^3 k}{(2 \pi)^3} 
\frac{dq_3}{(2 \pi)} d^2 \theta_1 d^2 \theta_2 d^2 \theta_3 d^2 \theta_4\de(p^2+m^2)\de(k^2+m^2) \de((p+q)^2+m^2)\de((k+q)^2+m^2)\nn\\ 
&\biggl[\left(V_{\bar{\Ph}^+\Ph^+;\Ph^+\bar{\Ph}^+}(\te_1,\te_2,\te_3,\te_4,p,q,k)
\Ph^+_m(-(p+q), \theta_1) {\bar \Ph}^{m+}(p, \theta_2) {\bar \Ph}^{n+}(k+q, \theta_3)\Ph^+_n(-k,
\theta_4) \right) \nn\\
&+\left(V_{\bar{\Ph}^-\Ph^-;\Ph^-\bar{\Ph}^-}(\te_1,\te_2,\te_3,\te_4,p,q,k)
\Ph^-_m(-(p+q), \theta_1) {\bar \Ph}^{m-}(p, \theta_2) {\bar \Ph}^{n-}(k+q, \theta_3)\Ph^-_n(-k,
\theta_4) \right)\nn\\
&+\left(V_{\bar{\Ph}^+\Ph^+;\Ph^-\bar{\Ph}^-}(\te_1,\te_2,\te_3,\te_4,p,q,k)
\Ph^+_m(-(p+q), \theta_1) {\bar \Ph}^{m+}(p, \theta_2) {\bar \Ph}^{n-}(k+q,
\theta_3)\Ph^-_n(-k,\theta_4) \right)\nn\\
&+\left(V_{\bar{\Ph}^-\Ph^-;\Ph^+\bar{\Ph}^+}(\te_1,\te_2,\te_3,\te_4,p,q,k)
\Ph^-_m(-(p+q), \theta_1) {\bar \Ph}^{m-}(p, \theta_2) {\bar \Ph}^{n+}(k+q,
\theta_3)\Ph^+_n(-k,\theta_4) \right)\nn\\
&+\left(V_{\bar{\Ph}^+\Ph^-;\Ph^+\bar{\Ph}^-}(\te_1,\te_2,\te_3,\te_4,p,q,k)
\Ph^+_m(-(p+q), \theta_1) {\bar \Ph}^{m-}(p, \theta_2) {\bar \Ph}^{n+}(k+q,
\theta_3)\Ph^-_n(-k,\theta_4) \right)\nn\\
&+\left(V_{\bar{\Ph}^-\Ph^+;\Ph^-\bar{\Ph}^+}(\te_1,\te_2,\te_3,\te_4,p,q,k)
\Ph^-_m(-(p+q), \theta_1) {\bar \Ph}^{m+}(p, \theta_2) {\bar \Ph}^{n-}(k+q,
\theta_3)\Ph^+_n(-k,\theta_4) \right)\biggr]\bigg|_{\text{on-shell}}. 
\end{align}
The functions $V$ that appear in \eqref{effon} are given by \eqref{totformcharged}, \eqref{totformneut} and \eqref{cormap}.  
The equation \eqref{effon} represents the most general effective action for the $2\to2$ amplitude at large $N$ in the $\mN=3$ theory. For our purposes, we limit the computation to that of the $2\to2$ bosonic and $2\to2$ fermionic amplitudes. This is easily accomplished by performing the Grassmann integrations and projecting out the purely bosonic/fermionic contributions. We write the bosonic and fermionic $2\to 2$ effective action as a sum of easily identifiable terms as\footnote{The first line in both the equations in \eqref{eefBF} is the contribution from the neutral sector \S\ref{nsec}, while the second line represents contributions from the charged sector \S\ref{csec}.}
\begin{small}
\begin{align}
\label{eefBF}
L_B &= L_B((\bph^+\ph^+) (\bph^+ \ph^+))+L_B((\bph^-\ph^-) (\bph^- \ph^-))+L_B((\bph^+\ph^+) (\bph^- \ph^-))+L_B((\bph^-\ph^-) (\bph^+ \ph^+))\nn\\
&+L_B((\bph^+\ph^-) (\bph^- \ph^+))+L_B((\bph^-\ph^+) (\bph^+ \ph^-)),\nn\\
L_F &= L_F((\bps^+\ps^+) (\bps^+ \ps^+))+L_F((\bps^-\ps^-) (\bps^- \ps^-))+L_F((\bps^+\ps^+) (\bps^- \ps^-))+L_F((\bps^-\ps^-) (\bps^+ \ps^+))\nn\\
&+L_F((\bps^+\ps^-) (\bps^- \ps^+))+L_F((\bps^-\ps^+) (\bps^+ \ps^-)).
\end{align}
\end{small}
These are quite cumbersome and not very illuminating, and we list all of them explicitly in the appendix \S\ref{termsineff}. For illustrative purposes we list the first terms in the bosonic and fermionic effective actions, which have the form
\begin{small}
\begin{align}
\label{bos1}
L_B((\bph^+\ph^+) (\bph^+ \ph^+))& = \frac{1}{2} \int \frac{d^3 p}{(2 \pi)^3}  \frac{d^3 k}{(2 \pi)^3} 
\frac{dq_3}{(2 \pi)}\de(p^2+m^2)\de(k^2+m^2) \de((p+q)^2+m^2)\de((k+q)^2+m^2)\nn\\ 
&\left(4 i A_{11}(p,k,q) m + (4m^2-q_3^2) B_{11}(p,k,q)-(C_{11}(p,k,q) k_-+ D_{11}(p,k,q) p_-) q_3\right)\nn\\
&\times \Big(\te(-p^0) a^+_m(-\vec[p]-\vec[q]) +\te(p^0) (a_c)^{+\dagger}_m(\vec[p]+\vec[q])\Big)\nn\\
&\times \Big(\te(p^0) a^{+m\dagger}(\vec[p]) +\te(-p^0) (a_c)^{+m}(-\vec[p])\Big)\nn\\
&\times  \Big(\te(k^0) a^{+n\dagger}(\vec[k]+\vec[q]) +\te(-k^0) (a_c)^{+n}(-\vec[k]-\vec[q])\Big)\nn\\
&\times \Big(\te(-k^0) a^+_n(-\vec[k]) +\te(k^0) (a_c)^{+\dagger}_n(\vec[k])\Big),
\end{align}
\end{small}
and
\begin{small}
\begin{align}
\label{fer1}
L_F((\bps^+\ps^+) (\bps^+ \ps^+))& = \frac{1}{2} \int \frac{d^3 p}{(2 \pi)^3}  \frac{d^3 k}{(2 \pi)^3} 
\frac{dq_3}{(2 \pi)}\de(p^2+m^2)\de(k^2+m^2) \de((p+q)^2+m^2)\de((k+q)^2+m^2)\nn\\ 
&\left((B_{11}(p,k,q) C^{\beta \alpha} C^{\delta \gamma}-i C_{11}(p,k,q)\  C^{\beta \alpha }C^{+\gamma }
C^{+\delta }+ i D_{11}(p,k,q) C^{\delta \gamma } C^{+\alpha } C^{+\beta })\right)\nn\\
&\times \Big(\te(p^0) v_\al(\vec[p],-m) \al^{+m\dagger}(\vec[p]) +\te(-p^0)u_\al(-\vec[p],-m)  (\al_c)^{+m}(-\vec[p])\Big)\nn\\
&\times \Big(\te(-p^0) u_\be(-\vec[p]-\vec[q],-m)\al^+_m(-\vec[p]-\vec[q]) +\te(p^0) v_\be(\vec[p]+\vec[q],-m)(\al_c)^{+\dagger}_m(\vec[p]+\vec[q])\Big)\nn\\
&\times  \Big(\te(k^0)v_\ga(\vec[k+q],-m)  \al^{+n\dagger}(\vec[k]+\vec[q]) +\te(-k^0)u_\ga(-\vec[k]-\vec[q],-m)  (\al_c)^{+n}(-\vec[k]-\vec[q])\Big)\nn\\
&\times \Big(\te(-k^0) u_\de(-\vec[k],-m)\al^+_n(-\vec[k]) +\te(k^0)v_\de(\vec[k],-m) (\al_c)^{+\dagger}_n(\vec[k])\Big).
\end{align}
\end{small}
Note that the functions determined in \S\ref{csec} and \S\ref{nsec} appear in specific combinations in the effective action \eqref{eefBF} and determine the $S$-matrix uniquely.

In principle, all the information needed for extracting the $S$-matrix in various channels of scattering is available in \eqref{eefBF}. However, we can extract the $S$-matrix only in the symmetric, anti-symmetric and adjoint channels of scattering as we explain now. First, observe that all the functions that appear in \eqref{eefBF} were derived in \S\ref{csec} and \S\ref{nsec} in the kinematic regime $q_\pm=0$. In the singlet channel, as one can see from \eqref{bos1} for instance, the incoming particle/anti-particle pair have momenta $p_1= p+q, p_2= -p$,\footnote{Note that setting $q_\pm=0$ is identical to $q_0=0, q_1=0$ and leaves only the space-like component $q_3$ non zero. This makes the center of mass energy space-like in the singlet channel. For the symmetric, anti-symmetric and adjoint channels, $q$ is just momentum transfer and it is completely legal to go to a frame where $q_\pm=0$ \cite{Jain:2014nza,Inbasekar:2015tsa}.} and thus the Mandelstam variable $s=-q^2$. It follows that $s<0$ whenever $q_\pm=0$, and thus the physical region $s\geq 4m^2$ of the singlet channel $S$-matrix cannot be extracted from our results in \S\ref{csec} and \S\ref{nsec}. This was already first observed in \cite{Jain:2014nza} and encountered in \cite{Inbasekar:2015tsa}.\footnote{Since by now we know and have checked that the four-point correlators in the $\mN=2,3$ theories are incredibly simple, it may be possible to compute them with some effort with arbitrary kinematics perhaps in the $\mN=2$ superspace. We hope to address this question in the near future.} 

For the rest of the section, we discuss the extraction of the direct, exchange (or symmetric and anti-symmetric, equivalently) and adjoint channels of scattering. The procedure is as follows. We first substitute the functions computed in \eqref{solutionsN3ch}, \eqref{solmp}, \eqref{solpp} and \eqref{eqsym} into \eqref{eefBF}. Simplifying further, the contributions to the bosonic and fermionic amplitudes can be formally rewritten as\footnote{We have the formal definitions \begin{align}
B_m^\pm(\vec[p]) &=\left(\te(p^0)a^\pm_m(\vec[p])+\te(-p^0) (a_c)_m^{\pm\dagger}(\vec[p]) \right)  \ , \  B^{n\pm\dagger}(\vec[p]) =\left(\te(p^0)a^{\pm n\dagger}(\vec[p])+\te(-p^0)(a_c)^{\pm n}(\vec[p]) \right),\nn\\
F_m^\pm(\vec[p]) &=\left(\te(p^0)\al^\pm_m(\vec[p])+\te(-p^0) (\al_c)_m^{\pm\dagger}(\vec[p]) \right) \ , \
F^{n\pm\dagger}(\vec[p]) =\left(\te(p^0)\al^{\pm n\dagger}(\vec[p])+\te(-p^0)(\al_c)^{\pm n}(\vec[p]) \right).
\end{align} }
\begin{align}\label{bosf}
L_B= \frac{1}{2} \int& \frac{d^3 p}{(2 \pi)^3}  \frac{d^3 k}{(2 \pi)^3} 
\frac{dq_3}{(2 \pi)}\de(p^2+m^2)\de(k^2+m^2) \de((p+q)^2+m^2)\de((k+q)^2+m^2)\nn\\ 
& \biggl(\mathcal{F}(p,q,k, - m) B_m^+(-\vec[p]-\vec[q])B^{m+\dagger}(\vec[p]) B^{n+\dagger}(\vec[k]+\vec[q])  B_n^+(-\vec[k])\nn\\
& +\mathcal{F}(p,q,k, + m) B_m^-(-\vec[p]-\vec[q])B^{m-\dagger}(\vec[p]) B^{n-\dagger}(\vec[k]+\vec[q])  B_n^-(-\vec[k])\nn\\
&+\mathcal{F}(p,q,k,0) B_m^+(-\vec[p]-\vec[q])B^{m-\dagger}(\vec[p]) B^{n+\dagger}(\vec[k]+\vec[q])  B_n^-(-\vec[k])\biggr),
\end{align}
\begin{align}\label{ferf}
L_F= \frac{1}{2} \int& \frac{d^3 p}{(2 \pi)^3}  \frac{d^3 k}{(2 \pi)^3} 
\frac{dq_3}{(2 \pi)}\de(p^2+m^2)\de(k^2+m^2) \de((p+q)^2+m^2)\de((k+q)^2+m^2)\nn\\ 
& \biggl(\mathcal{F}(p,q,k, - m) F^{m+\dagger}(\vec[p]) F_m^+(-\vec[p]-\vec[q]) F^{n+\dagger}(\vec[k]+\vec[q])  F_n^+(-\vec[k])\nn\\
&+\mathcal{F}(p,q,k,+ m) F^{m-\dagger}(\vec[p]) F_m^-(-\vec[p]-\vec[q]) F^{n-\dagger}(\vec[k]+\vec[q])  F_n^-(-\vec[k])\nn\\
&+\mathcal{F}(p,q,k,0) F^{m-\dagger}(\vec[p]) F_m^+(-\vec[p]-\vec[q]) F^{n+\dagger}(\vec[k]+\vec[q])  F_n^-(-\vec[k])\biggr),
\end{align}
where 
\beq\label{cova}
\mathcal{F}(p,q,k,m)= \f{4\pi i}{\ka} \ep_{\mu\nu\rh}\f{ q^\mu (p-k)^\nu (p+k)^\rh}{(p-k)^2} +  \f{8 m \pi}{\ka}.
\eeq
The equations \eqref{bosf} and \eqref{ferf} compactly capture the $2\to2$ bosonic/fermionic scattering amplitudes in the theory in the symmetric, anti-symmetric and adjoint channels of scattering. One simply sandwiches the expressions $L_B$ and $L_F$ in between appropriate on-shell states defined in \S\ref{astate} and reads off the amplitude in the relevant channel of scattering. 
We see that \eqref{cova} coincides with the tree-level amplitudes in the theory \S\ref{treeamp}.\footnote{Note that in the on-shell limit, the terms in \eqref{eefBF} $L_B((\bph^+\ph^+)(\bph^-\ph^-)), L_B((\bph^-\ph^-)(\bph^+\ph^+))$ and  $L_F((\bps^+\ps^+)(\bps^-\ps^-)), L_F((\bps^-\ps^-)(\bps^+\ph^+))$ (explicitly in \eqref{tb3}, \eqref{tb4}, \eqref{tf3}, \eqref{tf4} respectively)  have vanishing contributions in any channel and hence do not appear in \eqref{bosf} and \eqref{ferf}. Given that \eqref{cova} is tree-level exact, this is consistent with the fact that the corresponding tree-level amplitudes $\TB((\bph^+\ph^+) (\bph^-\ph^-))$ and $\TF((\bps^+\ps^+) (\bps^-\ps^-))$ vanish \eqref{mix}.}
Thus we have the remarkable result that the $2\to2$ bosonic/fermionic $S$-matrices computed to all loops in the \rq t Hooft coupling $\la$ are tree-level exact! The result is certainly not obvious, but not too surprising either, since it is known that four-point amplitude in the $\mN=2$ theory is tree-level exact to all orders in the planar limit \cite{Inbasekar:2015tsa,Inbasekar:2017sqp}. However, the $\mN=3$ theory has more complicated interactions and consequently the integral equations are more involved, as discussed in \S\ref{inteqall}. Nevertheless, the fact that despite the complications the final result is tree-level exact suggests deep symmetries underneath such results in all these theories. This also suggests perhaps it may be possible to undertake such computations for the $\mN\geq 4$ supersymmetric Chern-Simons-matter theories, perhaps by setting up the Dyson-Schwinger equations in a way similar to our approach for the $\mN=3$ theory in $\mN=1$ superspace. 

Before we conclude this section, we would like to point out that the computations in \S\ref{inteqall} were done in the kinematic frame $q_\pm=0$.  The result obtained by direct calculation reads as
\beq\label{noncova}
\mathcal{F}(p,q,k,\pm m)= \f{4\pi iq_3}{\ka} \f{(k+p)_-}{(k-p)_-}\pm \f{8 m \pi}{\ka}.
\eeq
We have extended the result \eqref{noncova} to arbitrary kinematics in \eqref{cova} by Lorentz covariantization. This is accurate for bosonic amplitudes. For fermionic amplitudes $\mathcal{F}(p,q,k)$ is accompanied by a relative phase factor that can be fixed by $\mN=3$ supersymmetry. For the massive case, it is also a function of the sign of the mass. Fixing this relative phase requires a very careful treatment of the $\mN=3$ on-shell superspace formalism that would enable us to express all component amplitudes in the superamplitude in terms of independent functions that we have computed. The relative phase is unobservable and irrelevant for the linear unitarity equations in the symmetric, anti-symmetric and adjoint channels of scattering that we considered in this paper (see \S\ref{skin}). For the singlet channel, fixing this relative phase in the fermionic amplitude is necessary to derive the correct form of the unitarity equation. We plan to address this in a future work. For instance, this phase is important for the correct cancellation of two boson going to two fermion contributions in the unitarity equation.

Before we conclude this section, we pause to note that the $m\to 0$ limit of the amplitude \eqref{bosf}, \eqref{ferf} and \eqref{cova} are smooth and free of singularities. The result is described by a single function
\beq\label{ampmless}
\mathcal{F}(p,q,k)= \f{4\pi i}{\ka} \ep_{\mu\nu\rh}\f{ q^\mu (p-k)^\nu (p+k)^\rh}{(p-k)^2}.
\eeq
In the next section, we discuss the covariantization of the amplitude in the massless limit.

\section{Amplitudes in various channels of scattering}\label{ampchannCov}
In this section we list the various amplitudes in different channels of scattering. We first write the non-covariant form in terms of the component $S$-matrices defined in \S\ref{astate}. In subsection \S\ref{covampform}, we take the massless limit, covariantize the amplitude and write it in terms of the Mandelstam variables. As discussed in \S\ref{skin}, for particle-particle scattering we consider
\beq
P_i(p_1)+P_j(p_2)\to P_k(p_3)+P_l(p_4),
\eeq
and for particle-anti particle scattering we consider
\beq
P_i(p_1)+A^j(p_2)\to P_k(p_3)+A^l(p_4)\, .
\eeq
\subsection{Component $S$-matrices}
\label{compo}
Applying the definitions of the $S$-matrices given in \S\ref{skin} on the expression \eqref{bosf} and \eqref{ferf}, we obtain the following component $S$- matrices. In particle-particle scattering we can read off the $U_d$ channel $S$-matrices as follows,
\begin{align}
\label{ampUd}
\mathcal{T}_{B/F}^{U_d}\left(-\f{1}{2},-\f{1}{2},+\f{1}{2},+\f{1}{2};\vec[p]_1,\vec[p]_2,\vec[p]_3,\vec[p]_4\right) & =\mF(\vec[p]_3, \vec[p]_1-\vec[p]_3, \vec[p]_2, - m), \nn\\
\mathcal{T}_{B/F}^{U_d}\left(+\f{1}{2},+\f{1}{2},-\f{1}{2},-\f{1}{2};\vec[p]_1,\vec[p]_2,\vec[p]_3,\vec[p]_4\right)& =\mF(\vec[p]_3, \vec[p]_1-\vec[p]_3, \vec[p]_2, + m),\nn\\
\mathcal{T}_{B/F}^{U_d}\left(-\f{1}{2},+\f{1}{2},+\f{1}{2},-\f{1}{2};\vec[p]_1,\vec[p]_2,\vec[p]_3,\vec[p]_4\right)& =\mF(\vec[p]_3, \vec[p]_1-\vec[p]_3, \vec[p]_2, 0),\nn\\
\mathcal{T}_{B/F}^{U_d}\left(+\f{1}{2},-\f{1}{2},-\f{1}{2},+\f{1}{2};\vec[p]_1,\vec[p]_2,\vec[p]_3,\vec[p]_4\right)& =\mF(\vec[p]_3, \vec[p]_1-\vec[p]_3, \vec[p]_2, 0).
\end{align}
Similarly, in the exchange channel $U_e$ we have
\begin{align}\label{ampUe}
\mathcal{T}_{B/F}^{U_e}\left(-\f{1}{2},-\f{1}{2},+\f{1}{2},+\f{1}{2};\vec[p]_1,\vec[p]_2,\vec[p]_4,\vec[p]_3\right) & =\mF(\vec[p]_4, \vec[p]_1-\vec[p]_4, \vec[p]_2, - m), \nn\\
\mathcal{T}_{B/F}^{U_e}\left(+\f{1}{2},+\f{1}{2},-\f{1}{2},-\f{1}{2};\vec[p]_1,\vec[p]_2,\vec[p]_4,\vec[p]_3\right)& =\mF(\vec[p]_4, \vec[p]_1-\vec[p]_4, \vec[p]_2, + m),\nn\\
\mathcal{T}_{B/F}^{U_e}\left(-\f{1}{2},+\f{1}{2},-\f{1}{2},+\f{1}{2};\vec[p]_1,\vec[p]_2,\vec[p]_4,\vec[p]_3\right)& =\mF(\vec[p]_4, \vec[p]_1-\vec[p]_4, \vec[p]_2, 0),\nn\\
\mathcal{T}_{B/F}^{U_e}\left(+\f{1}{2},-\f{1}{2},+\f{1}{2},-\f{1}{2};\vec[p]_1,\vec[p]_2,\vec[p]_4,\vec[p]_3\right)& =\mF(\vec[p]_4, \vec[p]_1-\vec[p]_4, \vec[p]_2, 0).
\end{align}
For the particle--anti-particle scattering, as we discussed earlier, we will be able to extract the amplitude only in the adjoint $(T)$ channel, which is given by 
\begin{align}\label{ampT}
\mathcal{T}_{B/F}^{T}\left(-\f{1}{2},+\f{1}{2},+\f{1}{2},-\f{1}{2};\vec[p]_1,\vec[p]_2,\vec[p]_3,\vec[p]_4\right) & =\mF(\vec[p]_3, \vec[p]_1-\vec[p]_3, -\vec[p]_4, - m) ,\nn\\
\mathcal{T}_{B/F}^{T}\left(+\f{1}{2},-\f{1}{2},-\f{1}{2},+\f{1}{2};\vec[p]_1,\vec[p]_2,\vec[p]_3,\vec[p]_4\right)& =\mF(\vec[p]_3, \vec[p]_1-\vec[p]_3, -\vec[p]_4, + m),\nn\\
\mathcal{T}_{B/F}^{T}\left(-\f{1}{2},+\f{1}{2},-\f{1}{2},+\f{1}{2};\vec[p]_1,\vec[p]_2,\vec[p]_3,\vec[p]_4\right)& =\mF(\vec[p]_3, \vec[p]_1-\vec[p]_3, -\vec[p]_4, 0),\nn\\
\mathcal{T}_{B/F}^{T}\left(+\f{1}{2},-\f{1}{2},+\f{1}{2},-\f{1}{2};\vec[p]_1,\vec[p]_2,\vec[p]_3,\vec[p]_4\right)& =\mF(\vec[p]_3, \vec[p]_1-\vec[p]_3, -\vec[p]_4, 0).
\end{align}
We note that in the massless limit, all the component $S$-matrices in each of \eqref{ampUd}, \eqref{ampUe} and \eqref{ampT} coincide. 

\subsection{Covariant form}\label{covampform}
\begin{figure}[h]
\begin{center}
\includegraphics[width=14cm,height=4.2cm]{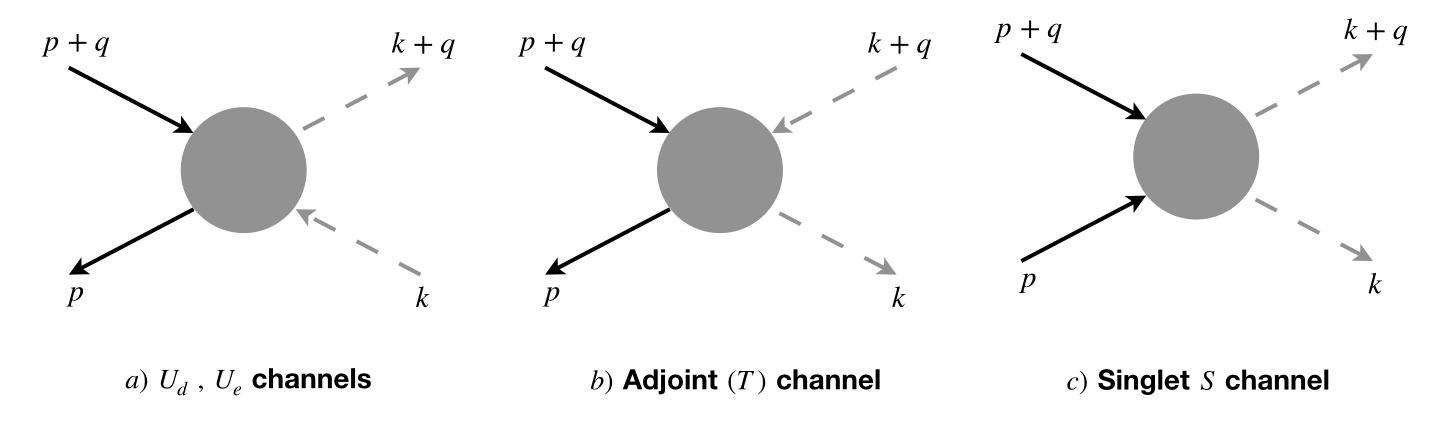}
\caption{\label{chann} Various channels of scattering as discussed in \S\ref{skin}. The blob indicates the off-shell four point correlator derived in \S\ref{csec} and \S\ref{nsec}. The dark continuous and grey striped lines indicate the color flow. The arrow indicates incoming/outgoing quanta. In the figure all the arrows are drawn for positive momenta. For an incoming/outgoing anti-particle the sign of the momentum is opposite to that of the direction of the arrow. Taking the on-shell limit of the external momenta, one reads of the amplitude in the relevant channel. As we repeatedly mentioned before, this is true only for the fig a) and fig b), the singlet channel is represented here only for the sake of completion and is not extractable due to the kinematic restriction $q_\pm=0$.}
\end{center}
\end{figure}

Let us now express the amplitudes in a manifestly $\mN=3$ covariant form.  We define the Mandelstam variables in the usual way as
\beq
s= -(p_1+p_2)^2\ , \ t= -(p_1-p_3)^2\ , \ u= -(p_1-p_4)^2 \ , \ s+t+u= 4m^2.
\eeq
In 2+1 dimensions, the $2\to2$ amplitude can be rewritten in terms of the Mandelstam variables up to a $Z_2$ invariant sign function of the participating momenta. This function takes the form
\beq
E(q,p-k,p+k)=\text{Sign}\left(\ep_{\mu\nu\rh}q^\mu (p-k)^\nu (p+k)^\rh\right).
\eeq
This issue has been discussed extensively in \cite{Jain:2014nza}, and the Lorentz invariant amplitudes are determined upto this sign function as in \cite{Jain:2014nza,Inbasekar:2015tsa}. 
The momentum assignments for the various channels of scattering can be read off from fig.\ \eqref{chann}. The $\mN=3$ covariance is manifest in the massless limit. In a practical note, this amounts to combining the various components listed in \S\ref{compo} as described in \eqref{bcovtononcovpp}, \eqref{fcovtononcovpp}, \eqref{bcovtononcovpa} and \eqref{fcovtononcovpa}.

\subsubsection{Amplitudes in the $U_d$, $U_e$ and $T$ channels}\label{resT}
In the \lq\lq direct\rq\rq ($U_d$) channel the momenta are
\beq
p_1=p+q \ , \ p_2= k \ , \ p_3= p \ , \ p_4=k+q.
\eeq
The bosonic/fermionic amplitudes in this channel are given by plugging in \eqref{ampUd} into \eqref{bcovtononcovpp}. Rewriting in terms of the Mandelstam variables we get
\beq\label{Udcov}
\mathcal{T}_{B/F}^{U_d}(s,t,u) =  E(q,p-k,p+k)\, \f{4 \pi i}{\ka} \sqrt{\f{t s}{u}} .
\eeq
In the \lq\lq exchange\rq\rq ($U_e$) channel the momenta are
\beq
p_1=p+q \ , \ p_2= k \ , \ p_3= k+q \ , \ p_4=p .
\eeq
The bosonic/fermionic amplitudes in this channel are given by plugging in \eqref{ampUe} into \eqref{bcovtononcovpp} and \eqref{fcovtononcovpp}. Rewriting in terms of the Mandelstam variables we get
\begin{align}
\label{Uecov}
\mathcal{T}_{B/F}^{U_e}(s,t,u) & = E(q,p-k,p+k) \,\f{4 \pi i}{\ka} \sqrt{\f{u s}{t}}.
\end{align}
In  particle--anti-particle scattering, as discussed in \S\ref{skin} we are able to read off the amplitude directly in the adjoint channel. The momenta take the form
\beq
p_1=p+q \ , \ p_2= -k-q \ , \ p_3= p \ , \ p_4=-k.
\eeq
The bosonic/fermionic amplitudes in this channel are given by plugging in \eqref{ampT} into \eqref{bcovtononcovpa} and \eqref{fcovtononcovpa}. Rewriting in terms of the Mandelstam variables we get
\begin{align}
\label{Tcov}
\mathcal{T}_{B/F}^{T}(s,t,u) & = E(q,p-k,p+k)\, \f{4 \pi i}{\ka} \sqrt{\f{t u}{s}} .
\end{align}
Under the duality map \eqref{dmap}, the $2\to 2$ bosonic amplitudes \eqref{Udcov}, \eqref{Uecov} and \eqref{Tcov} map to their corresponding 
$2\to 2$ fermionic counterparts trivially up to an overall physically unobservable phase. Furthermore, the $\mN=3$ covariant form of the amplitudes \eqref{Udcov}, \eqref{Uecov} and \eqref{Tcov} satisfy the unitarity conditions \eqref{unitaritycon}.

\subsubsection{Amplitude in the singlet channel}\label{ampsing}
As was discussed earlier in \S\ref{skin} and further in \S\ref{onsmatrix}, the off-shell correlation functions were computed in the kinematic regime $q_\pm=0$. As one can see from fig.\ \eqref{chann}, $q$ is the center of mass energy in the singlet channel and cannot be space-like. Thus we cannot read off the singlet channel from \eqref{bosf} and \eqref{ferf} by direct computation. This problem was first encountered in \cite{Jain:2014nza} and it was conjectured that the amplitude in the singlet channel takes the universal form \eqref{conj}. The veracity of this conjecture has been verified in \cite{Jain:2014nza,Inbasekar:2015tsa} in the non-supersymmetric boson/fermion coupled to Chern-Simons and $\mN=1,2$ supersymmetric Chern-Simons-matter theories by studying unitarity, duality and the non-relativistic limit. We believe that the conjecture should continue to hold for the $\mN=3$ theories as well. The obvious reason is that the the covariant form of the $\mN=3$ $2\to2$ bosonic/fermionic amplitude is identical to that of the $\mN=2$ theory in the $U_d,U_e,T$ channels of scattering. It follows then that consistency with $\mN=2$ unitarity would lead to the amplitude in the singlet channel to take the exact form as conjectured in \cite{Inbasekar:2015tsa}. However, it would be more satisfying to prove it from a manifestly $\mN=3$ on-shell formalism which we haven't been able to construct so far without ambiguities dependent on the mass sign.  In this paper we merely state the conjecture leaving a detailed proof of unitarity in the singlet channel to a later work. 

From fig.\ \eqref{chann} we have the momentum assignments of the singlet channel to be
\beq
p_1=p+q \ , \ p_2= -p \ , \ p_3= k+q \ , \ p_4=-k,
\eeq
and the amplitude with the naive analytic continuation from \eqref{ampmless} takes the form\footnote{Note that the singlet channel is multiplied by a factor of $N$ (see \S\ref{skin}). Furthermore, we used the definition of the \rq t Hooft coupling in the planar limit $\la= \f{N}{\ka}$. }
\begin{align}
\label{Scov}
\mathcal{T}_{B/F}^{S}(s,t,u) & = E(q,p-k,p+k) \,4 \pi i \la \sqrt{\f{s u}{t}}.
\end{align}
In the center of mass frame this takes the form\footnote{We define the center of mass coordinates \cite{Inbasekar:2015tsa} as
\beq
p_1= (\sqrt{p^2+m^2},p,0) \ , \ p_2=(\sqrt{p^2+m^2},-p,0) \ , \ p_3 =(\sqrt{p^2+m^2}, p\cos\te, p\sin\te)\ , \ p_4=(\sqrt{p^2+m^2},- p \cos\te,-p\sin\te),
\eeq
and the Mandelstam variables are expressed as
\begin{align}
s= 4(p^2+m^2) \ , \ t=-2p^2(1-\cos\te) \ , \ u=-2p^2 (1+\cos\te) \, .
\end{align}
}
\beq\label{snaive}
\mathcal{T}_{B/F}^{S; \text{naive}}(s,\te)  =4 \pi i \la \sqrt{s} \cot\f{\te}{2} .
\eeq
Following \cite{Jain:2014nza,Inbasekar:2015tsa}, we conjecture that the singlet channel $S$-matrix takes the form
\begin{align}
\label{Singcovfull}
\mathcal{S}_{B/F}^{S}(s,\te) & = 8\pi \sqrt{s} \cos(\pi\la) \de(\te) +i\, \f{\sin(\pi\la)}{\pi\la} \mathcal{T}_{B/F}^{S; \text{naive} }(s,\te).
\end{align}
The above result is identical to the $S$-matrix in the singlet channel of the $\mN=2$ theory and obviously obeys the $\mN=2$ unitarity equations discussed in \cite{Inbasekar:2015tsa} in the massless limit. \footnote{There is no mass renormalization in the $\mN=2$ theory and the unitarity equations eq.\ 4.14 and eq.\ 4.15 of \cite{Inbasekar:2015tsa} are true for any $m$. Naively, the term that could be singular when $m\to 0$ in eq.\ 4.14 and eq.\ 4.15 is $\f{(-s+4m^2)}{16 m^2 }(W_B-W_F)(W_B^*-W_F^*)  $. However, since $W_B$ and $W_F$ are proportional to $m$ upto a sign, the $m$ dependence cancels away as seen in eq.\ 4.21, and there is a smooth $m\to 0$ limit as expected. We would like to thank Xi Yin for a related discussion. } Under the duality \eqref{dmap} the conjectured bosonic/fermionic S matrix \eqref{Singcovfull} to the respective fermionic/bosonic S matrix in the dual theory up to an overall unobservable phase. We hope to set up a manifestly covariant $\mN=3$ formalism and demonstrate unitarity of the conjecture \eqref{Singcovfull} in a future work. 

\section{Discussion}\label{disco}
In this paper, we studied $2\to2$ amplitudes in $\mN=3$ supersymmetric Chern-Simons-matter theories, and computed them to all orders in the \rq t Hooft coupling $\la$ in the planar limit $N\to\infty, \ka \to \infty$ with $\la=\f{N}{\ka}$ held fixed.  We set up the Dyson-Schwinger equations for the exact four-point correlation function in \S\ref{org}  and solved it to all orders in the \rq t Hooft coupling (eqs.\ \eqref{solutionsN3ch}, \eqref{solpp}, \eqref{solmp}). We employed the supersymmetric light cone gauge and were able to solve for the off-shell correlators in the kinematic regime $q_\pm=0$. 

Taking the on-shell limit of the off-shell correlation function, we were able to extract the $2\to2$ bosonic/fermionic $S$-matrix in the symmetric, anti-symmetric and adjoint channels of scattering. Due to the kinematic restriction $q_\pm = 0$ the singlet channel amplitude is not extractable directly from our results. Our final result for the $2\to2$ $S$-matrix is remarkably simple and tree-level exact, with loop corrections vanishing to all orders. The conjectured singlet channel $S$-matrix continues to be simple yet minimally modified to be consistent with unitarity.
Such extraordinary ``non-renormalization" often indicates a powerful underlying symmetry. For instance, the $2\to2$ $S$-matrices in the $\mN=2$ theory are tree-level exact and it was shown in \cite{Inbasekar:2017sqp} that the amplitude enjoys dual superconformal symmetry exact to all loops. For the $\mN=3$ theory, the first step for some of these computations would be to come up with a simple on-shell representation of the $\mN=3$ superamplitude.  It is highly likely that the $2\to 2$ superamplitude in the $\mN=3$ theory would be dual superconformal and Yangian invariant \cite{Inbasekar:2017sqp,Yangian}. Following \cite{Inbasekar:2017ieo}, it may also be possible to compute arbitrary $n$-point tree-level amplitudes in the $\mN=3$ theory via BCFW recursions. Due to the high degree of supersymmetry, one may expect that the dual superconformal symmetry continues to be a symmetry for all tree-level amplitudes.

Going back to the conjectured amplitude \eqref{ampsingf}, the phase modification $\cos(\pi\la)$ of the forward scattering is physically motivated via the constructive interference of the Aharanov-Bohm phases of the incoming particles. The modified crossing factor $\f{\sin(\pi\la)}{\pi\la}$ essentially follows demanding unitarity. The authors of \cite{Jain:2014nza} physically motivated the crossing factor by relating it to the ratio of a Wilson loop in one of the non-anyonic channels and a Wilson loop in the singlet channel \cite{Witten:1988hf}. Unfortunately, so far there has not been any direct derivation of the conjecture. One way to derive this is the following. Formally one has to recompute the off-shell correlation function by attaching Wilson lines to make them gauge invariant. Assuming that one can set-up this computation and define a sensible on-shell limit, it is possible that the final result may factorize into an algebra of Wilson loops and the $S$-matrix computed in this paper in the large $N$ limit. Importantly, the exact result for Wilson loop algebra in $\mN=3$ theory has been computed using supersymmetric localization \cite{Kapustin:2013hpk}. Armed with this result, perhaps some clever argument with supersymmetry and large $N$ could assist in a formal derivation of the modified crossing factor. 

In this paper we have taken the first steps to compute amplitudes in higher supersymmetric theories in $\mN=1$ superspace. We observe that the $SO(\mN-1)_R$ invariance of $\mN$ supersymmetric theories on $\mN=1$ superspace plays an important role in the solution of the Dyson-Schwinger equation. However, for $\mN\geq 4$ supersymmetry matter transforms in the bi-fundamental representation and the diagrams are more complicated. Perhaps for $U(N)\times U(M)$ theories it may be possible to set up a $M/N$ expansion similar to the approach of \cite{Gurucharan:2014cva}. 

A peculiar feature in all the vector like Chern-Simons models that has puzzled us for a long time now is that all the exact amplitudes computed so far \cite{Jain:2014nza,Inbasekar:2015tsa} and together with the results of the current paper have been completely free from IR divergences. It is well known that the amplitudes in ABJM theory indeed have IR divergences \cite{Elvang:2013cua}. From the point of view of perturbation theory, this may be also related to the crossing factor though any computational effort has not been very fruitful so far. It is possible that there is something wrong with the amplitude computations in ABJM theory or that our Dyson-Schwinger methods are missing IR divergences consistently. It would be great to clarify this in a detailed manner since one might understand a great deal about the IR behavior in 2+1 dimensions. Another related direction to pursue is to reproduce our result - that the $2\to2$ amplitude is tree-level exact in the $\mN=2,3$ theories - using generalized unitarity approaches. 

\acknowledgments{
We would like to thank Sachin Jain, Shiraz Minwalla, Tarun Sharma and V.\ Umesh for helpful discussions. We would also like to thank Shiraz Minwalla, Sachin Jain, V. Vishal and V.\ Umesh for collaboration during the early stages of the project. We would like to thank Shiraz Minwalla, Shimon Yankielowicz and Xi Yin for comments on the draft.  We would like to acknowledge the use of the \texttt{Grassmann} package for superspace computations by Matt Headrick. The work of KI is supported by BSF grant number 2014707 at Ben Gurion University. KI would like to thank TIFR and Tel Aviv University for their hospitality during various stages of this project.  The work of LJ is supported by a KIAS Individual Grant PG077301 at Korea Institute for Advanced Study. The work of AS is supported by the Natural Sciences and Engineering Research Council of Canada.}

\appendix
\section{Notations and conventions}
\label{conv}
In this appendix we summarize the notations and conventions used in this paper. For the conventions in $\mN=1$ superspace we refer the reader to Appendix A of \cite{Inbasekar:2015tsa}. We list some relevant conventions for easy readability.
\subsection{$\mN=1$ superspace conventions}
The $\mN=1$ Grassmann spinor $\te^\al$ is a two-component spinor labeled by indices $\al=+,-$. It has the basic properties
\beq
\int d\te=0\ , \ \int d\te\,\te=1 \ , \  \int d^2\te\, \te^2= -1 .
\eeq
The raising and lowering of the spinor indices is done by
\beq
\te^\al = \mathcal{C}^{\al\be} \te_\be\ , \te_\al =\te^\be \mathcal{C}_{\be\al},
\eeq
where the anti-symmetric charge conjugation matrix is defined as
\beq
\mathcal{C}_{\al\be}=\begin{pmatrix}
            0 & -i\\
            i & 0
           \end{pmatrix}=- \mathcal{C}^{\al\be},
\eeq
with the property that $\mathcal{C}_{\al\ga} \mathcal{C}^{\ga\be}=-\de_\al^{\ \be}$. The $\mN=1$ supercharges are defined in momentum space as
\beq
Q_\al= i\left(\f{\p}{\p\te^\al}-\te^\be p_{\al\be}\right),
\eeq
and satisfy the algebra
\beq
\{Q_\al,Q_\be\}= 2 p_{\al\be}.
\eeq
The super covariant derivatives that appear in \eqref{actionlcgauge1} are defined as 
\beq
D_\al=\f{\p}{\p\te^\al}+\te^\be p_{\al\be},
\eeq
and satisfy the relations
\beq
\{D_\al,D_\be\}= 2 p_{\al\be}\ , \ \{Q_\al,D_\be\}= 0.
\eeq

\subsection{Other conventions}
In Lorentzian light cone coordinates a generic momentum matrix $p_{\al\be}\equiv p_\mu \ga^\mu_{\al\be}$ has the form 
\beq
p_{\al\be}=\begin{pmatrix}
            p_+ & p_3\\
            p_3 & -p_-
           \end{pmatrix},
\eeq
where $\{p_\mu=p_0,p_1,p_3\}$, and the light cone definitions
\beq
 p_+ =p_1+p_0 \ , \ p_{-}=p_1-p_0,
\eeq
such that $p_+ p_-=p_s^2=-p_0^2+p_1^2$. The Euclidean rotation is defined via $p_0\to -i p_2$. The gauge superfield is a spinor in $\mN=1$ superspace, and by choosing the supersymmetric light cone gauge (See Appendix F of \cite{Inbasekar:2015tsa} for more details)
\beq
\Ga_-=0
\eeq
sets in components the condition $A_-= A_1+i A_2=0$.

In the integral equations \eqref{inteqch}, \eqref{inteqpp}, \eqref{inteqmp} we use the following definition of the Euclidean measure
\beq
\int \f{d^3 r}{(2\pi)^3} = \f{1}{(2\pi)^3} \int r_s dr_s dr_3 d\te,
\eeq
where $r_\pm= r_s e^{\pm i \te}$ and $r^2= r_s^2+r_3^2= r_1^2+r_2^2+r_3^2$. We also make use of the following angular integrals (w.r.t. $r$),
\begin{align}
&\int \f{d\te}{(k-r)_-} =\f{2\pi}{k_-}\ \te(k_s-r_s),\nn\\
&\int \f{d\te r_-}{(k-r)_-} =-2\pi (1- \te(k_s-r_s)),\nn\\
&\int \f{d\te r_-^2}{(k-r)_-} =-2\pi k_- (1- \te(k_s-r_s)),\nn\\
&\int \f{d\te }{(k-r)_-(r-p)_-} = \f{2\pi}{(k-p)_-} \left( \f{\te(k_s-r_s)}{k_-}-\f{\te(p_s-r_s)}{p_-} \right),\nn\\
&\int \f{d\te r_-}{(k-r)_-(r-p)_-} = \f{2\pi}{(k-p)_-} \left( \te(k_s-r_s)-\te(p_s-r_s) \right),\nn\\
&\int \f{d\te r_-^2}{(k-r)_-(r-p)_-} = -\f{2\pi}{(k-p)_-} \left( k_-(1-\te(k_s-r_s))-p_-(1-\te(p_s-r_s)) \right),
\end{align}
and $r_3\in [-\infty,\infty]\ , \ r_s\in [0,\infty]$. 

\subsection{Conventions for $SU(2)_R$ and $SO(2)_R$}\label{N3con}
In this section, we list some relevant conventions for switching between $SU(2)_R$ and $SO(2)_R$ notations. For details see Appendix A of \cite{Inbasekar:2019azv}. The superfields $(\Ph^+,\Ph^-)$ have $R$-charges $(\f{1}{2},-\f{1}{2})$, 
\begin{align}
\Ph^\pm &=\ph^\pm+\te\ps^\pm-\te^2 F^\pm,\nn\\
\bPh^\pm &=\bph^\pm+\te\bps^\pm-\te^2 \bF^\pm\ .
\end{align}
Note that the complex conjugates $(\bPh^+,\bPh^-)$ have $R$-charges $(-\f{1}{2},\f{1}{2})$. 
The component fields have the $SO(2)_R$ charges carried by the superfield. We group the $SO(2)_R$ fields into $SU(2)_R$ doublets as discussed below.

We denote the $SU(2)_R$ spinor components labeled by $1,2$ as $\phi^1=\phi^+$ having $+$ charge and $\phi^2=\phi^-$ having $-$ charge under $SO(2)_R$. For the fermion we choose $\ps_1=\ps^-$ having $-$ charge and $\ps_2=\ps^+$ having $+$ charge under $SO(2)_R$. This is a consistent choice allowed by the supersymmetry transformation rules \eqref{sustn3md}. The $SO(2)_R$ spinors do not have raising or lowering and hence the location of the $\pm$ indices are pure convention.
\begin{eqnarray}\label{relations}
\phi^A = \left(
\begin{array}{c} 
\phi^+\\
\phi^-
 \end{array}
\right)
\begin{array}{c} 
+ \\
-
 \end{array}
\ , \
\phi_A = \left(
\begin{array}{c} 
-\phi^-\\
\phi^+
 \end{array}
\right)
\begin{array}{c} 
- \\
+
 \end{array}
\ , \
\psi^A = \left(
\begin{array}{c} 
\psi^+\\
-\psi^-
 \end{array}
\right)
\begin{array}{c} 
+ \\
-
 \end{array}
\ , \
\psi_A = \left(
\begin{array}{c} 
\psi^-\\
\psi^+
 \end{array}
\right)
\begin{array}{c} 
- \\
+
 \end{array}\nn\\
\bph_A = \left(
\begin{array}{c} 
\bph^+\\
\bph^-
 \end{array}
\right)
\begin{array}{c} 
- \\
+
 \end{array}\ , \
\bph^A = \left(
\begin{array}{c} 
-\bph^-\\
\bph^+
 \end{array}
\right)
\begin{array}{c} 
+ \\
-
 \end{array}
\ , \
\bps_A = \left(
\begin{array}{c} 
\bps^+\\
-\bps^-
 \end{array}
\right)
\begin{array}{c} 
- \\
+
 \end{array}
\ , \
\bps^A = \left(
\begin{array}{c} 
\bps^-\\
\bps^+
 \end{array}
\right)
\begin{array}{c} 
+\\
-
 \end{array}
\end{eqnarray}
where, $A,B=1,2$ are the $SU(2)_R$ symmetry indices. The anti-symmetric $SU(2)$ metric is used to raise/lower the indices 
\begin{align}
&\ph_A=\ph^B\ep_{BA}\ , \ \ps^A=\ep^{AB}\ps_B ,\nn\\
&\bph^A = \bph_B \ep^{BA} \ , \ \bph_A = \ep_{AB}\bph^B\, .
\end{align}
with $\ep_{12}=\ep^{12}=1$.  

\section{$\mN=3$ supersymmetry transformations}
\label{N3susytransf}
The mass-deformed Lorentzian $\mN=3$ action in Wess-Zumino gauge is given by \cite{Inbasekar:2019azv}
\begin{align}\label{N3WZL}
S_{\mN=3}^L= \int d^3x &\biggl[\text{Tr}\left(-\f{\ka}{4\pi}\ep^{\mu\nu\rh}\left(A_\mu\p_\nu
A_\rh-\f{2i}{3}A_\mu A_\nu A_\rh\right)\right)\nn\\
& + i \bps^A\slashed{\mD}\ps_A+m_0\bps^A(\si^3)_A^{\
B}\ps_B-\mD^\mu\bph_A\mD_\mu\ph^A-m^2_0\bph_A\ph^A\nn\\
&-\f{4\pi^2}{\ka^2} (\bph_A\ph^B)(\bph_B\ph^C)(\bph_C\ph^A)+\f{4\pi}{\ka} 
(\bph_A\ph^B)(\bps^A\ps_B)+\f{2\pi}{\ka}
(\bps^A\ph_B)(\bph^B\ps_A)\nn\\
&-\f{4\pi}{\ka}(\bps^A\ph_A)(\bph^B\ps_B) +\f{2\pi}{\ka} (\bps^A\ph_A)(\bps^B\ph_B)+\f{2\pi}{\ka}
(\bph^A\ps_A)(\bph^B\ps_B) \nn\\
& +\f{4\pi m_0}{\ka}(\bph^A\ph_A)(\bph^C(\si_3)_C^{\ D}\ph_D)\biggr].
\end{align}
The action is invariant under the supersymmetry transformations
\begin{align}\label{sustn3md}
&Q_{BC\alpha} \phi_A = \, \psi_{\alpha (B} \, \epsilon_{C)A} ,\nn \\
&Q_{BC\alpha} \bar\phi^A = -  \, \bar\psi_{\alpha (B} \, \delta_{C)}^{~~A} ,\nn\\
&Q_{BC\alpha} \psi_{\beta A} = - i  \, \mD_{\alpha\beta}\phi_{(B} \,
\epsilon_{C)A}+m_0\ch_1 C_{\al\be}\ph_{(B}(\si^3)_{C)A}\nn\\& \qquad\qquad+\frac{2\pi}{\kappa} \,
C_{\alpha\beta} (\bar\phi_A \phi_{(B}) \phi_{C)} +\frac{2\pi}{\kappa} \, C_{\alpha\beta}
(\bar\phi_{(B} \phi_{C)}) \phi_{A},\nn\\
&Q_{BC\alpha} \bar\psi^{\beta A} =  i \, \mD_{\alpha}^{~\,\beta} \bar\phi_{(B} \,
\delta_{C)}^{~~A}+m_0\ch_1 \de_\al^{\ \be}\bph_{(B}(\si_3)^{\ A}_{C)}\nn\\
&\qquad\qquad+\frac{2\pi}{\kappa} \,
\delta_{\alpha}^{~\,\beta}(\bar\phi_{(B} \phi^{A}) \bar\phi_{C)} - \frac{2\pi}{\kappa} \,
\delta_{\alpha}^{~\,\beta} (\bar\phi_{(B} \phi_{C)}) \bar\phi^{A},\nn\\
&Q_{BC\alpha} A^a_\mu = -\frac{4\pi}{\kappa} (\gamma_\mu)_\alpha^{~\,\beta} \bar\phi^i_{(B}
(T^a)_i^{~j} \psi_{C)\beta j} - \frac{4\pi}{\kappa} (\gamma_\mu)_\alpha^{~\,\beta}
\bar\psi^i_{\beta (B} (T^a)_i^{~j} \phi_{C)j}.
\end{align}

\section{Effective action}
\label{termsineff}
In this appendix we list all the terms in the bosonic and fermionic effective action given in \eqref{eefBF}. 
\begin{small}
\begin{align}\label{tb2}
L_B((\bph^-\ph^-) (\bph^- \ph^-)) = \frac{1}{2} \int& \frac{d^3 p}{(2 \pi)^3}  \frac{d^3 k}{(2 \pi)^3} 
\frac{dq_3}{(2 \pi)}\de(p^2+m^2)\de(k^2+m^2) \de((p+q)^2+m^2)\de((k+q)^2+m^2)\nn\\ 
&\left(-4 i A_{22}(p,k,q) m + (4m^2-q_3^2) B_{22}(p,k,q)-(C_{22}(p,k,q) k_-+ D_{22}(p,k,q) p_-) q_3\right)\nn\\
&\times \Big(\te(-p^0) a^-_m(-\vec[p]-\vec[q]) +\te(p^0) (a_c)^{-\dagger}_m(\vec[p]+\vec[q])\Big)\nn\\
&\times \Big(\te(p^0) a^{-m\dagger}(\vec[p]) +\te(-p^0) (a_c)^{-m}(-\vec[p])\Big)\nn\\
&\times  \Big(\te(k^0) a^{-n\dagger}(\vec[k]+\vec[q]) +\te(-k^0) (a_c)^{-n}(-\vec[k]-\vec[q])\Big)\nn\\
&\times \Big(\te(-k^0) a^-_n(-\vec[k]) +\te(k^0) (a_c)^{-\dagger}_n(\vec[k])\Big),
\end{align}
\end{small}
\begin{small}
\begin{align}\label{tb3}
L_B((\bph^+\ph^+) (\bph^- \ph^-)) = \frac{1}{2} \int& \frac{d^3 p}{(2 \pi)^3}  \frac{d^3 k}{(2 \pi)^3} 
\frac{dq_3}{(2 \pi)}\de(p^2+m^2)\de(k^2+m^2) \de((p+q)^2+m^2)\de((k+q)^2+m^2)\nn\\ 
&\left(i (2m+i q_3) (C_{12}(p,k,q) k_-+ D_{12}(p,k,q) p_- + B_{12}(p,k,q) (2 i m + q_3)) \right)\nn\\
&\times \Big(\te(-p^0) a^+_m(-\vec[p]-\vec[q]) +\te(p^0) (a_c)^{+\dagger}_m(\vec[p]+\vec[q])\Big)\nn\\
&\times \Big(\te(p^0) a^{+m\dagger}(\vec[p]) +\te(-p^0) (a_c)^{+m}(-\vec[p])\Big)\nn\\
&\times  \Big(\te(k^0) a^{-n\dagger}(\vec[k]+\vec[q]) +\te(-k^0) (a_c)^{-n}(-\vec[k]-\vec[q])\Big)\nn\\
&\times \Big(\te(-k^0) a^-_n(-\vec[k]) +\te(k^0) (a_c)^{-\dagger}_n(\vec[k])\Big),
\end{align}
\end{small}
\begin{small}
\begin{align}\label{tb4}
L_B((\bph^-\ph^-) (\bph^+ \ph^+)) = \frac{1}{2} \int& \frac{d^3 p}{(2 \pi)^3}  \frac{d^3 k}{(2 \pi)^3} 
\frac{dq_3}{(2 \pi)}\de(p^2+m^2)\de(k^2+m^2) \de((p+q)^2+m^2)\de((k+q)^2+m^2)\nn\\ 
&\left( i (i q_3-2m) (C_{21}(p,k,q) k_-+ D_{21}(p,k,q) p_- + B_{21}(p,k,q) (q_3-2i m))\right)\nn\\
&\times \Big(\te(-p^0) a^-_m(-\vec[p]-\vec[q]) +\te(p^0) (a_c)^{-\dagger}_m(\vec[p]+\vec[q])\Big)\nn\\
&\times \Big(\te(p^0) a^{-m\dagger}(\vec[p]) +\te(-p^0) (a_c)^{-m}(-\vec[p])\Big)\nn\\
&\times  \Big(\te(k^0) a^{+n\dagger}(\vec[k]+\vec[q]) +\te(-k^0) (a_c)^{+n}(-\vec[k]-\vec[q])\Big)\nn\\
&\times \Big(\te(-k^0) a^+_n(-\vec[k]) +\te(k^0) (a_c)^{+\dagger}_n(\vec[k])\Big),
\end{align}
\end{small}
\begin{small}
\begin{align}\label{tb5}
L_B((\bph^-\ph^+) (\bph^+ \ph^-)) = \frac{1}{2} \int& \frac{d^3 p}{(2 \pi)^3}  \frac{d^3 k}{(2 \pi)^3} 
\frac{dq_3}{(2 \pi)}\de(p^2+m^2)\de(k^2+m^2) \de((p+q)^2+m^2)\de((k+q)^2+m^2)\nn\\ 
&(-q_3 \left(B(p,k,q) q_3+ C(p,k,q) k_-+ D(p,k,q) p_-\right))\nn\\
&\times \Big(\te(-p^0) a^+_m(-\vec[p]-\vec[q]) +\te(p^0) (a_c)^{+\dagger}_m(\vec[p]+\vec[q])\Big)\nn\\
&\times \Big(\te(p^0) a^{-m\dagger}(\vec[p]) +\te(-p^0) (a_c)^{-m}(-\vec[p])\Big)\nn\\
&\times  \Big(\te(k^0) a^{+n\dagger}(\vec[k]+\vec[q]) +\te(-k^0) (a_c)^{+n}(-\vec[k]-\vec[q])\Big)\nn\\
&\times \Big(\te(-k^0) a^-_n(-\vec[k]) +\te(k^0) (a_c)^{-\dagger}_n(\vec[k])\Big),
\end{align}
\end{small}
\begin{small}
\begin{align}\label{tb6}
L_B((\bph^+\ph^-) (\bph^- \ph^+)) = \frac{1}{2} \int& \frac{d^3 p}{(2 \pi)^3}  \frac{d^3 k}{(2 \pi)^3} 
\frac{dq_3}{(2 \pi)}\de(p^2+m^2)\de(k^2+m^2) \de((p+q)^2+m^2)\de((k+q)^2+m^2)\nn\\ 
&(-q_3 \left(B(p,k,q) q_3+ C(p,k,q) k_-+ D(p,k,q) p_-\right))\nn\\
&\times \Big(\te(-p^0) a^-_m(-\vec[p]-\vec[q]) +\te(p^0) (a_c)^{-\dagger}_m(\vec[p]+\vec[q])\Big)\nn\\
&\times \Big(\te(p^0) a^{+m\dagger}(\vec[p]) +\te(-p^0) (a_c)^{+m}(-\vec[p])\Big)\nn\\
&\times  \Big(\te(k^0) a^{-n\dagger}(\vec[k]+\vec[q]) +\te(-k^0) (a_c)^{-n}(-\vec[k]-\vec[q])\Big)\nn\\
&\times \Big(\te(-k^0) a^+_n(-\vec[k]) +\te(k^0) (a_c)^{+\dagger}_n(\vec[k])\Big),
\end{align}
\end{small}
\begin{small}
\begin{align}\label{tf2}
L_F((\bps^-\ps^-) (\bps^- \ps^-)) = \frac{1}{2} \int& \frac{d^3 p}{(2 \pi)^3}  \frac{d^3 k}{(2 \pi)^3} 
\frac{dq_3}{(2 \pi)}\de(p^2+m^2)\de(k^2+m^2) \de((p+q)^2+m^2)\de((k+q)^2+m^2)\nn\\ 
&\left((B_{22}(p,k,q) C^{\beta \alpha} C^{\delta \gamma}-i C_{22}(p,k,q)\  C^{\beta \alpha }C^{+\gamma }
C^{+\delta }+ i D_{22}(p,k,q) C^{\delta \gamma } C^{+\alpha } C^{+\beta })\right)\nn\\
&\times \Big(\te(p^0) v_\al(\vec[p],m) \al^{-m\dagger}(\vec[p]) +\te(-p^0)u_\al(-\vec[p],m)  (\al_c)^{-m}(-\vec[p])\Big)\nn\\
&\times \Big(\te(-p^0) u_\be(-\vec[p]-\vec[q],m)\al^-_m(-\vec[p]-\vec[q]) +\te(p^0) v_\be(\vec[p]+\vec[q],m)(\al_c)^{-\dagger}_m(\vec[p]+\vec[q])\Big)\nn\\
&\times  \Big(\te(k^0)v_\ga(\vec[k+q],m)  \al^{-n\dagger}(\vec[k]+\vec[q]) +\te(-k^0)u_\ga(-\vec[k]-\vec[q],m)  (\al_c)^{-n}(-\vec[k]-\vec[q])\Big)\nn\\
&\times \Big(\te(-k^0) u_\de(-\vec[k],m)\al^-_n(-\vec[k]) +\te(k^0)v_\de(\vec[k],m) (\al_c)^{-\dagger}_n(\vec[k])\Big),
\end{align}
\end{small}
\begin{small}
\begin{align}\label{tf3}
L_F((\bps^+\ps^+) (\bps^- \ps^-)) = \frac{1}{2} \int& \frac{d^3 p}{(2 \pi)^3}  \frac{d^3 k}{(2 \pi)^3} 
\frac{dq_3}{(2 \pi)}\de(p^2+m^2)\de(k^2+m^2) \de((p+q)^2+m^2)\de((k+q)^2+m^2)\nn\\ 
&\left((B_{12}(p,k,q) C^{\beta \alpha} C^{\delta \gamma}-i C_{12}(p,k,q)\  C^{\beta \alpha }C^{+\gamma }
C^{+\delta }+ i D_{12}(p,k,q) C^{\delta \gamma } C^{+\alpha } C^{+\beta })\right)\nn\\
&\times \Big(\te(p^0) v_\al(\vec[p],-m) \al^{+m\dagger}(\vec[p]) +\te(-p^0)u_\al(-\vec[p],-m)  (\al_c)^{+m}(-\vec[p])\Big)\nn\\
&\times \Big(\te(-p^0) u_\be(-\vec[p]-\vec[q],-m)\al^+_m(-\vec[p]-\vec[q]) +\te(p^0) v_\be(\vec[p]+\vec[q],-m)(\al_c)^{+\dagger}_m(\vec[p]+\vec[q])\Big)\nn\\
&\times  \Big(\te(k^0)v_\ga(\vec[k+q],m)  \al^{-n\dagger}(\vec[k]+\vec[q]) +\te(-k^0)u_\ga(-\vec[k]-\vec[q],m)  (\al_c)^{-n}(-\vec[k]-\vec[q])\Big)\nn\\
&\times \Big(\te(-k^0) u_\de(-\vec[k],m)\al^-_n(-\vec[k]) +\te(k^0)v_\de(\vec[k],m) (\al_c)^{-\dagger}_n(\vec[k])\Big),
\end{align}
\end{small}
\begin{small}
\begin{align}\label{tf4}
L_F((\bps^-\ps^-) (\bps^+ \ps^+)) = \frac{1}{2} \int& \frac{d^3 p}{(2 \pi)^3}  \frac{d^3 k}{(2 \pi)^3} 
\frac{dq_3}{(2 \pi)}\de(p^2+m^2)\de(k^2+m^2) \de((p+q)^2+m^2)\de((k+q)^2+m^2)\nn\\ 
&\left((B_{21}(p,k,q) C^{\beta \alpha} C^{\delta \gamma}-i C_{21}(p,k,q)\  C^{\beta \alpha }C^{+\gamma }
C^{+\delta }+ i D_{21}(p,k,q) C^{\delta \gamma } C^{+\alpha } C^{+\beta })\right)\nn\\
&\times \Big(\te(p^0) v_\al(\vec[p],m) \al^{-m\dagger}(\vec[p]) +\te(-p^0)u_\al(-\vec[p],m)  (\al_c)^{-m}(-\vec[p])\Big)\nn\\
&\times \Big(\te(-p^0) u_\be(-\vec[p]-\vec[q],m)\al^-_m(-\vec[p]-\vec[q]) +\te(p^0) v_\be(\vec[p]+\vec[q],m)(\al_c)^{-\dagger}_m(\vec[p]+\vec[q])\Big)\nn\\
&\times  \Big(\te(k^0)v_\ga(\vec[k+q],-m)  \al^{+n\dagger}(\vec[k]+\vec[q]) +\te(-k^0)u_\ga(-\vec[k]-\vec[q],-m)  (\al_c)^{+n}(-\vec[k]-\vec[q])\Big)\nn\\
&\times \Big(\te(-k^0) u_\de(-\vec[k],-m)\al^+_n(-\vec[k]) +\te(k^0)v_\de(\vec[k],-m) (\al_c)^{+\dagger}_n(\vec[k])\Big),
\end{align}
\end{small}
\begin{small}
\begin{align}\label{tf5}
L_F((\bps^-\ps^+) (\bps^+ \ps^-)) = \frac{1}{2} \int& \frac{d^3 p}{(2 \pi)^3}  \frac{d^3 k}{(2 \pi)^3} 
\frac{dq_3}{(2 \pi)}\de(p^2+m^2)\de(k^2+m^2) \de((p+q)^2+m^2)\de((k+q)^2+m^2)\nn\\ 
&\left((B(p,k,q) \ C^{\beta \alpha} C^{\delta \gamma}-i C(p,k,q) \  C^{\beta \alpha }C^{+\gamma }
C^{+\delta }+ i D(p,k,q) C^{\delta \gamma } C^{+\alpha } C^{+\beta })\right)\nn\\
&\times \Big(\te(p^0) v_\al(\vec[p],m) \al^{-m\dagger}(\vec[p]) +\te(-p^0)u_\al(-\vec[p],m)  (\al_c)^{-m}(-\vec[p])\Big)\nn\\
&\times \Big(\te(-p^0) u_\be(-\vec[p]-\vec[q],-m)\al^+_m(-\vec[p]-\vec[q]) +\te(p^0) v_\be(\vec[p]+\vec[q],-m)(\al_c)^{+\dagger}_m(\vec[p]+\vec[q])\Big)\nn\\
&\times  \Big(\te(k^0)v_\ga(\vec[k+q],-m)  \al^{+n\dagger}(\vec[k]+\vec[q]) +\te(-k^0)u_\ga(-\vec[k]-\vec[q],-m)  (\al_c)^{+n}(-\vec[k]-\vec[q])\Big)\nn\\
&\times \Big(\te(-k^0) u_\de(-\vec[k],m)\al^-_n(-\vec[k]) +\te(k^0)v_\de(\vec[k],m) (\al_c)^{-\dagger}_n(\vec[k])\Big),
\end{align}
\end{small}
\begin{small}
\begin{align}\label{tf6}
L_F((\bps^+\ps^-) (\bps^- \ps^+)) = \frac{1}{2} \int& \frac{d^3 p}{(2 \pi)^3}  \frac{d^3 k}{(2 \pi)^3} 
\frac{dq_3}{(2 \pi)}\de(p^2+m^2)\de(k^2+m^2) \de((p+q)^2+m^2)\de((k+q)^2+m^2)\nn\\ 
&\left((B(p,k,q) C^{\beta \alpha} C^{\delta \gamma}-i C(p,k,q)\  C^{\beta \alpha }C^{+\gamma }
C^{+\delta }+ i D(p,k,q) C^{\delta \gamma } C^{+\alpha } C^{+\beta })\right)\nn\\
&\times \Big(\te(p^0) v_\al(\vec[p],-m) \al^{+m\dagger}(\vec[p]) +\te(-p^0)u_\al(-\vec[p],-m)  (\al_c)^{+m}(-\vec[p])\Big)\nn\\
&\times \Big(\te(-p^0) u_\be(-\vec[p]-\vec[q],m)\al^-_m(-\vec[p]-\vec[q]) +\te(p^0) v_\be(\vec[p]+\vec[q],m)(\al_c)^{-\dagger}_m(\vec[p]+\vec[q])\Big)\nn\\
&\times  \Big(\te(k^0)v_\ga(\vec[k+q],m)  \al^{-n\dagger}(\vec[k]+\vec[q]) +\te(-k^0)u_\ga(-\vec[k]-\vec[q],m)  (\al_c)^{-n}(-\vec[k]-\vec[q])\Big)\nn\\
&\times \Big(\te(-k^0) u_\de(-\vec[k],-m)\al^+_n(-\vec[k]) +\te(k^0)v_\de(\vec[k],-m) (\al_c)^{+\dagger}_n(\vec[k])\Big).
\end{align}
\end{small}

\section{Tree-level four-point amplitudes}\label{treeamp}
In this appendix, we list all the tree-level correlators in the ``charged sector'' and the ``neutral sector'' of the $\mN=3$ theory.
\subsection{Charged sector}
For the diagrams in the right hand side of the box in fig.\ \ref{N3tree} ($R$-charged $\pm 1$ in/out states) the tree-level amplitudes are of the form
\begin{align}
\label{V0c}
&V_{0:\bar{\Ph}^+\Ph^-;\Ph^+\bar{\Ph}^-}(\te_1,\te_2,\te_3,\te_4,p,q,k)=\exp\bigg(\f{1}{4}X.(p.X_{12} +q.X_{13}+k.X_{43} )\bigg) \times \nn\\
&\hspace{30mm}\biggl(-\f{2 \pi i }{\ka}X_{12}^-X_{12}^+X_{13}^-X_{13}^+X_{43}^-X_{43}^+ -\f{4\pi i}{\ka(p-k)_{--}}X_{12}^+X_{13}^+X_{43}^+ (X_{12}^-+X_{34}^-)\biggr)\\
&=V_{0:\bar{\Ph}^-\Ph^+;\Ph^-\bar{\Ph}^+}(\te_1,\te_2,\te_3,\te_4,p,q,k).
\end{align}
The first term is the contribution from the vertex  $(\bPh^+\Ph^-)(\bPh^-\Ph^+)$ and the second
term is the contribution from the gauge field ladder rung. The $+,-$ on the difference variables
$X_{ab}^\pm$ correspond to the two components of the Grassmann variable and have nothing to do with the $R$-charge. It is easy to project out the tree-level bosonic and fermionic $S$-matrices in this sector (in the $T$-channel), and they turn out to be
\beq
T_B^{tree}=T_F^{tree}=\f{4\pi i q_3}{\ka} \f{(k+p)_-}{(k-p)_-}.
\eeq

\subsection{Neutral sector}
For the neutral sector the relevant diagrams are in the first block of fig. \ref{N3tree}. The tree-level amplitudes are given by
\begin{align} \label{V0n}
&V_{0:\bar{\Ph}^+\Ph^+;\Ph^+\bar{\Ph}^+}(\te_1,\te_2,\te_3,\te_4,p,q,k)=\exp\bigg(\f{1}{4}X.(p.X_{12} +q.X_{13}+k.X_{43} )\bigg) \times \nn\\
&\hspace{30mm}\biggl(\f{ \pi i }{\ka}X_{12}^-X_{12}^+X_{13}^-X_{13}^+X_{43}^-X_{43}^+ -\f{4\pi i}{\ka(p-k)_{--}}X_{12}^+X_{13}^+X_{43}^+ (X_{12}^-+X_{34}^-)\biggr)\\
&= V_{0:\bar{\Ph}^-\Ph^-;\Ph^-\bar{\Ph}^-}(\te_1,\te_2,\te_3,\te_4,p,q,k)\, ,\\
&V_{0:\bar{\Ph}^+\Ph^+;\Ph^-\bar{\Ph}^-}(\te_1,\te_2,\te_3,\te_4,p,q,k)=\exp\bigg(\f{1}{4}X.(p.X_{12} +q.X_{13}+k.X_{43} )\bigg) \times\nn\\
&\hspace{68mm}\biggl(-\f{ 4 \pi i }{\ka}X_{12}^-X_{12}^+X_{13}^-X_{13}^+X_{43}^-X_{43}^+\biggr)\\
&=V_{0:\bar{\Ph}^-\Ph^-;\Ph^+\bar{\Ph}^+}(\te_1,\te_2,\te_3,\te_4,p,q,k).
\end{align}
It is obvious that the tree-level amplitudes all fit in the structure of \eqref{totform}. We also define the tree-level matrix
\beq\label{M0nmat}
M_0(\te_1,\te_2,\te_3,\te_4,p,q,k)=\begin{pmatrix}
   V_{0;\bar{\Ph}^+\Ph^+;\Ph^+\bar{\Ph}^+}(\te_1,\te_2,\te_3,\te_4,p,q,k) & 
   V_{0;\bar{\Ph}^+\Ph^+;\Ph^-\bar{\Ph}^-}(\te_1,\te_2,\te_3,\te_4,p,q,k)\\
   V_{0;\bar{\Ph}^-\Ph^-;\Ph^+\bar{\Ph}^+}(\te_1,\te_2,\te_3,\te_4,p,q,k) & 
   V_{0;\bar{\Ph}^-\Ph^-;\Ph^-\bar{\Ph}^-}(\te_1,\te_2,\te_3,\te_4,p,q,k)
  \end{pmatrix}.
\eeq

\noindent The bosonic and fermionic tree-level $S$-matrices for the process $\bar{\Ph}^+\Ph^+\to\bar{\Ph}^+\Ph^+$ are given by
\begin{align}
T_B^{tree}&=\f{4\pi i q_3}{\ka} \f{(k+p)_-}{(k-p)_-}- \f{8 m \pi}{\ka},\nn\\
T_F^{tree}&=\f{4\pi i q_3}{\ka} \f{(k+p)_-}{(k-p)_-}-\f{8 m \pi}{\ka}.
\end{align}
The bosonic and fermionic tree-level $S$-matrices for the process $\bar{\Ph}^-\Ph^-\to\bar{\Ph}^-\Ph^-$ are given by
\begin{align}
T_B^{tree}&=\f{4\pi i q_3}{\ka} \f{(k+p)_-}{(k-p)_-}+ \f{8 m \pi}{\ka},\nn\\
T_F^{tree}&=\f{4\pi i q_3}{\ka} \f{(k+p)_-}{(k-p)_-}+\f{8 m \pi}{\ka}.
\end{align}
The bosonic and fermionic tree-level $S$-matrices for the process $\bar{\Ph}^+\Ph^+\to\bar{\Ph}^-\Ph^-$ are given by
\beq\label{mix}
T_B^{tree}=T_F^{tree}=0.
\eeq

\bibliographystyle{JHEP}
\bibliography{massh.bib}
\end{document}